\documentclass[11pt,a4paper]{article}
\pdfoutput=1
\usepackage{jheppub}
\usepackage{extarrows}
\usepackage{caption}
\allowdisplaybreaks
\usepackage{subfigure}
\usepackage{slashed}
 \usepackage{setspace} 
 \usepackage{amsmath}
\usepackage{caption}
\usepackage{xcolor}  
\usepackage{color}
\allowdisplaybreaks
\def\bea{\begin{eqnarray}}
\def\eea{\end{eqnarray}}
\def\be{\begin{equation}}
\def\ee{\end{equation}}
\def\nn{\nonumber}

\def\nn{\nonumber}

\usepackage[utf8]{inputenc}

\DeclareUnicodeCharacter{2212}{-}
\usepackage{comment}

\usepackage[T1]{fontenc}

\def\nue{{\nu_e}}
\def\anue{{\bar{\nu}_e}}
\def\numu{{\nu_{\mu}}}
\def\anumu{{\bar{\nu}_{\mu}}}

\newcommand{\eg}{{\it e.g.}}
\newcommand{\ie}{{\it i.e.}}

\newcommand{\beq}{\begin{equation}}
\newcommand{\eeq}{\end{equation}}
\newcommand{\beqa}{\begin{eqnarray}}
\newcommand{\eeqa}{\end{eqnarray}}

\def\nn{\nonumber}

\usepackage{cancel}
\usepackage{soul}

\vspace*{-2cm}

\title{A $17$ MeV pseudoscalar and the LSND, MiniBooNE and ATOMKI anomalies\footnote{\large This work is dedicated to the memory of Boris J Kayser.}}
  \author[a]{Waleed Abdallah,}
  \author[b]{Raj Gandhi,}
  \author[b]{Tathagata Ghosh,}
  \author[c]{Najimuddin Khan,}
  \author[d]{Samiran Roy,}
  \author[e]{and Subhojit Roy}
\affiliation[a]{Department of Mathematics, Faculty of Science, Cairo University, Giza 12613, Egypt}
\affiliation[b]{Harish-Chandra Research Institute, Chhatnag Road, Jhunsi, Allahabad 211019, India}
\affiliation[c]{Department of Physics, Aligarh Muslim University, Aligarh-202002, India}
\affiliation[d]{School of Physics, University of Hyderabad, Hyderabad - 500046, India}
\affiliation[e]{HEP Division, Argonne National Laboratory, 9700 Cass Ave., Argonne, IL 60439, USA}

    \emailAdd{awaleed@sci.cu.edu.eg} \emailAdd{raj@hri.res.in}
    \emailAdd{tathagataghosh@hri.res.in}
    \emailAdd{nkhan.ph@amu.ac.in}
    \emailAdd{samiranroy.hri@gmail.com}
    \emailAdd{sroy@anl.gov}
    \preprint{HRI-RECAPP-2024-02} 
  \abstract{In the absence of any new physics signals at the Large Hadron Collider (LHC), anomalous results at low energy experiments have become the subject of increased attention. We focus on three such 
results from the LSND, MiniBooNE (MB), and ATOMKI experiments. A 17~MeV pseudoscalar mediator ($a'$) can account for two ($^8$Be and $^4$He) out of the three cases in which  excess events have been seen in  pair creation transitions in ATOMKI. We incorporate this mediator in a gauge invariant extension of the Standard Model (SM) with a second Higgs doublet and three singlet (seesaw)  neutrinos ($N_i, i=1,2,3$). $N_{1,2}$ participate in an interaction in MB and LSND which, with $a'$ as mediator, leads to the production of $e^+ e^-$ pairs. The $N_i$ also lead to mass-squared differences for SM neutrinos in agreement with global oscillation data. We first show that such a model offers a natural joint solution to the MB and LSND excesses, providing excellent fits to their data. Next, using the values of the couplings to the quarks and electrons which are required to explain pair creation nuclear transition data for $^8$Be and $^4$He in  ATOMKI, we show that these values still lead to  fits for MB and LSND data. However, once ATOMKI is incorporated, we find that strong constraints  from the decays $K^+ \rightarrow \pi^+ a' \, (a'\rightarrow e^+e^-)$ and $\pi^+ \rightarrow $ $ e^+ ~\nu_e ~e^+ e^- $ come into play. While our solution is in conformity with the bounds on the former decay, it remains in tension with $90\%$
 CL bounds on the latter. We also discuss other constraints  from both collider and non-collider experiments and from electroweak precision data, stability and unitarity.  We  compute the contributions to the electron and muon $g-2$ up to two loops for our model.  We discuss tests of the model in upcoming experiments.
}

\begin{document}
\maketitle
\flushbottom 
\section{Introduction}
The search for new physics beyond the Standard Model (SM) over the past couple of decades has, to a significant degree, focused on the exploration of  scales beyond the electroweak, extending to energies above a TeV. Strong theoretical motivations for these searches have been provided by Supersymmetry (see~\cite{Martin:1997ns,ParticleDataGroup:2022pthAA} and references therein) and Compositeness~\cite{ParticleDataGroup:2022pthAA, Eichten:1983hw}. The lack of new physics signals at the Large Hadron Collider (LHC), as predicted by these and other high-scale theories  has, however, led to a renewed interest in the search for new, weakly coupled physics at low energy scales. 

Such  physics may manifest itself at experiments already optimized to pick up signals of weak (as in feeble) interactions at low energies, like  neutrino and dark matter detectors, as well as at those looking for rare decays. Anomalies in such experiments could thus be an important signpost to the existence of new physics. In addition, we believe that attempts to understand such anomalies should also examine whether the same new physics may underlie more than one of them, providing a common explanation. Our work in this paper pursues this line of thought.

Two of the most statistically significant and long-standing low energy anomalies are the excesses in electron-like events at the Liquid Scintillator Neutrino Detector (LSND)~\cite{LSND:2001aii} and MiniBooNE (MB)~\cite{MiniBooNE:2007uho,MiniBooNE:2008yuf,MiniBooNE:2013uba}, both of which are short-baseline  liquid scintillation detectors with incident $\numu$ and/or $\anumu$ beams with average energies below 1~GeV. The significances of the LSND and MB excesses are $3.8\sigma$ and  $4.8\sigma$, respectively. Their combined significance stands at $6.1\sigma$. The results are backed by careful checks and studies of possible SM backgrounds~\cite{LSND:1996jxj, Katori:2020tvv, Dasgupta:2021ies, Brdar:2021ysi, Alvarez-Ruso:2021dna, MicroBooNE:2021zai} in order to eliminate SM physics explanations.

In recent years the ATOMKI collaboration has studied rare nuclear transitions for a number of nuclei. In particular, it has focussed on  Internal Pair Creation (IPC), where the nucleus emits a virtual photon which then
decays to an $e^+ e^-$ pair for  excited $^8{\rm Be}$~\cite{Krasznahorkay:2015iga}, $^4{\rm He}$~\cite{Krasznahorkay:2019lyl,Krasznahorkay:2021joi} and $^{12}{\rm C}$~\cite{Krasznahorkay:2022pxs} nuclei. The collaboration has  reported unexpected measurements in all of these decays. Specifically, it reported
anomalous bumps for both the invariant mass and the angular opening
of the $e^+ e^-$ pairs with statistical significance above $6\sigma$. In its analyses of these results, the collaboration has stated that if one assumes that they signal new physics, they can be
interpreted as due to the on-shell emission of a new boson $X$ from the excited nuclei, which subsequently decays to an $e^+ e^-$ pair. They estimate the best fit mass
for this hypothetical new particle to be $\sim 17$~MeV.

Previous work on understanding the LSND and MB excesses in terms of physics beyond the SM has focused, to a significant degree,  on the existence of one or more light ($m^2\sim$eV$^2$) sterile neutrinos. These states  induce additional $\numu\rightarrow \nue$ or $\anumu \rightarrow \anue$ oscillations which are then used to explain the signals in these two detectors. The sterile hypothesis has also been linked to other low-energy anomalies~\cite{GALLEX:1994rym, GALLEX:1997lja, Kaether:2010ag, Abdurashitov:1996dp, SAGE:1998fvr, Abdurashitov:2005tb, SAGE:2009eeu} involving electron neutrino appearance and disappearance at short baselines. It is, however,  in strong tension with cosmology~\cite{Hamann:2011ge, Archidiacono:2013xxa, Hagstotz:2020ukm}, which limits the number of relativistic degrees of freedom in thermal equilibrium prior to neutrino decoupling at $T\sim 1$~MeV as well as the amount of hot dark matter in the universe. Additionally, it exhibits significant tension with $\numu$ disappearance results~\cite{MINOS:2020iqj, IceCube:2020phf,IceCube:2020tka}, which is manifest in global  analyses and fits~\cite{Dentler:2018sju, Diaz:2019fwt, Boser:2019rta, Dasgupta:2021ies, Acero:2022wqg} of all neutrino data.

These tensions have, in part, motivated a significant amount of work on other new physics explanations of the MB and LSND anomalies, which typically involve, in various combinations, the introduction of new mediators and/or heavy neutral leptons and transition magnetic moments. A recent review may be found in~\cite{Abdullahi:2023ejc}. In some cases~\cite{Moss:2017pur,Moulai:2019gpi, Akhmedov:2011zza, Bramante:2011uu, Karagiorgi:2012kw, Asaadi:2017bhx, Smirnov:2021zgn, Alves:2022vgn, Palomares-Ruiz:2005zbh, Bai:2015ztj, deGouvea:2019qre, Dentler:2019dhz, Hostert:2020oui,Chang:2021myh} these ideas combine sterile oscillations and decay with new physics of the kind mentioned above. Other ideas focus on new, non-oscillatory interactions using these elements of new physics, which produce electron-like signals inside the detectors~\cite{Gninenko:2009ks, Gninenko:2010pr, Gninenko:2012rw, Masip:2012ke, Radionov:2013mca, Magill:2018jla, Bertuzzo:2018itn, Ballett:2018ynz, Ballett:2019pyw, Datta:2020auq,Dutta:2020scq, Abdallah:2020biq, Abdullahi:2020nyr, Abdallah:2020vgg, Schwetz:2020xra, Vergani:2021tgc, Hammad:2021mpl, Dutta:2021cip, Alvarez-Ruso:2021dna,Abdallah:2022grs, Kamp:2022bpt, Bansal:2022zpi,Ghosh:2023dgk}.

In this work, we re-visit one of the solutions~\cite{Abdallah:2020vgg} proposed in previous work. It involved a $CP$-even  scalar mediator of mass $\sim 750$~MeV from a second Higgs doublet and  a real ($CP$-even) dark singlet $h'$  with $m_{h'} \simeq 17$~MeV. That model provides a very good fit to both the LSND and MB energy and angular distributions~\cite{Abdallah:2020vgg}. Proposals invoking new mediators producing electron-like signals are subject to multiple constraints, from near detectors
in neutrino experiments, meson decay data, high and ultra-high energy neutrino
experiments, colliders, active-sterile mixings, beam dump results and dark photon searches. Discussions and references on these constraints may be found in~\cite{Abdallah:2020vgg, Magill:2018jla, Brdar:2020tle, Atre:2009rg, McKeen:2010rx, ISTRA:2011bgc, Drewes:2015iva, Drewes:2015iva, deGouvea:2015euy, Coloma:2017ppo, MiniBooNEDM:2018cxm, Jordan:2018qiy, Arguelles:2018mtc, Bryman:2019ssi, Coloma:2019qqj,Bryman:2019bjg}. The model in~\cite{Abdallah:2020vgg} conforms to the constraints in the above references.
In light of recent Higgs data, however, in this work  we test this model against more stringent collider constraints not considered earlier. Specifically, we study the effects of the new scalars on the Higgs decay width and the Higgs di-photon channel. Depending on the charged Higgs masses of the second Higgs doublet, mild violations of the 1$\sigma$ limit
for  the measured di-photon signal strength, $1.04^{+0.10}_{-0.09}$~\cite{ATLAS:2022tnm} occur or, alternatively, a violation of 
  the constraint on the Higgs decay width~\cite{ParticleDataGroup:2022pthAA} is present. However, if the real 17~MeV singlet scalar is replaced by a pseudoscalar of the same mass,  we find that the model conforms to these and other constraints. 
  
  The switch to a $17$~MeV pseudoscalar also obviates the need in~\cite{Abdallah:2020vgg} for one of the $CP$-even scalars of the second Higgs doublet to be much lighter ($\ie \sim 750$~MeV)  than its charged and pseudoscalar partners. As we show in a later section, due to the nature of its spin-dependant couplings and the absence of a dominant coherent contribution\footnote{A 17~MeV real scalar, on the other hand, when used to fit MB angular distributions,  provides events which are predominantly forward, necessitating a companion heavier real scalar ($\sim 750$~MeV) which helps provide events in non-forward directions, as discussed in~\cite{Abdallah:2022grs}.}, the pseudoscalar alone provides a very good fit to MB and LSND data. This modification to the solution proposed in~\cite{Abdallah:2020vgg} thus allows for a natural hierarchy in the 2HDM + $a'$ model, where the singlet is light while all the members of the second doublet stay heavy, with masses in the several hundred GeV range.

  Since one of the proposed solutions to the ATOMKI anomaly~\cite{Ellwanger:2016wfe} has a 17 MeV pseudoscalar as a candidate mediator\footnote{It is appropriate to note that the pseudoscalar solution to ATOMKI is not without its caveats, because it does not provide a solution for all possible decays of $^8{\rm Be}$, $^4{\rm He}$ and $^{12}{\rm C}$ which could in principle contribute to the excess. This apparent limitation, when combined with the MB and LSND results, could however, be indicative of the presence of a 17 MeV \textit{complex} scalar, as we discuss later in the paper.}, it is thus natural to examine if there is a common parameter space of quark and electron couplings that allows a common solution to all three anomalies under consideration. We show that this is indeed possible if the effective couplings of the pseudoscalar to nucleons are significantly higher than those necessary to understand MB and LSND alone. This enhancement, while staying safe from other constraints, leads to tension with bounds from meson decay. Specifically, the decays 
  $K^+\rightarrow \pi^+ + a'$ and $\pi^+ \rightarrow $ $ e^+ ~\nu_e ~e^+ e^- $ require special attention in the light of strong existing bounds. We thus devote significant effort to finding solutions which respect these constraints. However, as we show, our solution remains in tension with bounds on  the second decay above.

  We supplement the constraints on a model such as ours discussed in~\cite{Abdallah:2020vgg, Magill:2018jla, Brdar:2020tle, Atre:2009rg, McKeen:2010rx, ISTRA:2011bgc, Drewes:2015iva, Drewes:2015iva, deGouvea:2015euy, Coloma:2017ppo, MiniBooNEDM:2018cxm, Jordan:2018qiy, Arguelles:2018mtc, Bryman:2019ssi, Coloma:2019qqj,Bryman:2019bjg} by a fuller discussion of those which also arise from $i)$ the LEP measurements of the $Z$ decay width, $ii)$ the LHC measurements of the Higgs decay width and its couplings to fermions, $iii)$ the vacuum stability of the scalar potential, $iv)$ the unitarity of its $S$-matrix, $v$) heavy Higgs searches at LHC,  $vi)$  electroweak precision measurements.

New physics such as that introduced here is expected to affect charged lepton anomalous magnetic moments; specifically those of the muon and the electron, both of which are the subjects of current experimental measurements. They are denoted as $a_{\mu,e}$,  defined by
$a_{\mu,e} \equiv \left( g_{\mu,e} - 2 \right)/2$ for the muon and electron, respectively, where $g_{\mu,e}$ is the Lande $g{\text{-}\rm factor}$. We calculate the effects of our model on $a_{\mu,e}$ up to two-loop level.
We find that  it is not possible to explain the discrepancy observed in $a_{\mu}$ using the new physics ingredients of this model if, at the same time, we wish to explain MB and LSND. It may, however, be possible to understand  the observed discrepancy in $a_{e}$ within its context.

This paper is organized as follows: In section~\ref{sec2} we describe our model and its constituents. Section~\ref{sec3} 
 focuses on the interaction in MB and LSND which leads to the electron-like signal.
 In section~\ref{sec4} we demonstrate that this model provides very good fits to MB and LSND.  The benchmark parameters used are shown in table~\ref{tab} in this section. In section~\ref{sec5} we use the ATOMKI results to derive the required couplings of the pseudoscalar to nucleons  and to electrons in order to explain the observed excess in that experiment. 
 In section~\ref{sec6} we use the couplings obtained from ATOMKI results in the previous section to obtain fits to LSND and MB also. The relevant benchmark parameters used are shown in table~\ref{tabNeW} of this section.
 Section~\ref{sec7} contains a detailed discussion of collider and non-collider constraints on the model for the benchmark values of table~\ref{tab} and table~\ref{tabNeW}.
 Section~\ref{sec8} discusses the contributions to $a_{\mu, e}$. Tests of the model in upcoming experiments are discussed in section~\ref{sec9}, while the final section summarizes the work and presents our conclusions.
\section{The Model}
\label{sec2}
We extend the scalar sector of the SM by incorporating a second Higgs doublet, and also add a singlet pseudoscalar $\phi_{h'}=i \, A_3^0/\sqrt{2}$. Additionally, three right-handed neutrinos help generate neutrino masses via the seesaw mechanism and participate in the interaction which generates electron-like signals in MB and LSND. We can write the scalar potential $V$ as
\begin{eqnarray}
    V=V_{\rm 2HDM}+V_{h'},
\end{eqnarray}
where $V_{\rm 2HDM}$ and $V_{h'}$ are given in the Higgs basis $(\phi_h,\phi_H,\phi_{h'})$, with $\lambda_i$ denoting the usual set of quartic couplings:
\begin{eqnarray}
V_{\rm 2HDM}&=& \mu_1 |\phi_{h}|^2 +\mu_2 |\phi_{H}|^2 + \frac{\lambda_1}{2} |\phi_{h}|^4+ \frac{\lambda_2}{2} |\phi_{H}|^4 +\lambda_3 |\phi_{H}|^2|\phi_{h}|^2+\lambda_4 (\phi_{h}^\dagger\phi_{H})(\phi_{H}^\dagger\phi_{h})\nonumber\\
&&+\frac{\lambda_5}{2} \big{\{}(\phi_{h}^\dagger\phi_{H})^2 +h.c\big{\}}+ (\lambda_6 |\phi_{h}|^2 + \lambda_7  |\phi_{H}|^2)\, (\phi_{h}^\dagger\phi_{H}+ \phi_{H}^\dagger\phi_{h})\nonumber,\\  
V_{h'}&=&\mu' |\phi_{h'}|^2 + \lambda'_2 |\phi_{h'}|^4+\lambda'_3 |\phi_{h}|^2 |\phi_{h'}|^2+\lambda'_4|\phi_{H}|^2 |\phi_{h'}|^2   +\big{\{}( \lambda'_5 |\phi_{h'}|^2 \,-\mu_3)(\phi_{h}^\dagger\phi_{H}) \nonumber\\
&&+(m_1|\phi_{h}|^2+m_2|\phi_{H}|^2+m_3\phi_{h}^\dagger\phi_{H}- m_s \phi_{h'})\phi_{h'}+h.c. \big{\}}\nonumber.
\end{eqnarray}
Here,
\begin{eqnarray}
\phi_{h}&=&\left( \begin{array}{c}
G^+ \\
\frac{v+H^0_1+i G^0}{\sqrt{2}} \\
\end{array} \right),~~~
\phi_{H}=\left( \begin{array}{c}
H^+_2 \\
\frac{H^0_2+i A_2^0}{\sqrt{2}} \\
\end{array} \right),~~~
\phi_{h'}=i\, A_3^0/\sqrt{2}\,.
\end{eqnarray}
We consider the vacuum expectation values (VEV) $\langle\phi_h\rangle=v(\equiv v_{SM})\simeq  246$~GeV and $\langle \phi_H\rangle\!=\!0\!=\!\langle \phi_{h'}\rangle$. Here, $G^+, G^0$ are the Goldstone modes, which give the gauge bosons mass after the electroweak symmetry is spontaneously broken.

The mass matrix of the neutral $CP$-even Higgses in the basis $\left(H_1^0, H_2^0\right)$ is given by
\begin{equation} \label{$CP$-Even-MM}
{\cal M}^2_{\cal H} = \left( 
\begin{array}{ccc}
\lambda_1 v^2 &\lambda_6 v^2  \\ 
\lambda_6 v^2 & \mu_H 
 \end{array} 
\right), 
 \end{equation} 
where $\mu_H = \mu_2 +(\lambda_3 + \lambda_4 + \lambda_5)v^2/2$. Here, we have minimized the scalar potential $V$ using the following conditions:
\begin{equation}
\mu_1 = -\frac{1}{2}\lambda_1 v^2,~~\mu_{3} = \frac{1}{2}\lambda_6 v^2.
\label{eq:conds}
\end{equation} 
The matrix in eq.~\eqref{$CP$-Even-MM},   $m^2_{\cal H}$, is diagonalized by $Z^{\cal H}$ as follows:
\begin{equation} 
Z^{\cal H} {\cal M}^2_{\cal H} (Z^{\cal H})^T = ({\cal M}^{2}_{\cal H})^{\rm diag}\,, 
~~{\rm with}~~
H^0_i = \sum_{j}Z_{{j i}}^{\cal H}h_{{j}}\,,
\end{equation} 
\begin{equation}
{Z^{\cal H}}=\left(
\begin{array}{ccc}
  c_\alpha & -s_\alpha \\
  s_\alpha & c_\alpha
\end{array}
\right),~~\tan{2 \alpha}=\frac{ 2\lambda_6 v^2 }{\lambda_1 v^2 -\mu_H},
\end{equation}
where $s_\alpha\equiv \sin \alpha$, $c_\alpha \equiv \cos \alpha$. In the alignment limit ($\ie$, $\lambda_6\sim0$), the SM-like Higgs is $H^0_1\approx h$ with $M^2_h\simeq \lambda_1 v^2$ and $H^0_2\approx H$ having mass $M_H^2\simeq\mu_H$.
The mass matrix of the neutral $CP$-odd Higgses in the basis $\left(A_2^0, A_3^0\right)$, satisfying the  conditions in eq.~\eqref{eq:conds}, is given by
 \begin{equation} \label{$CP$-Odd-MM}
{\cal M}^2_{\cal A} = \left( 
\begin{array}{ccc}
 \mu_A & -m_{3} v/\sqrt{2}  \\ 
 -m_{3} v/\sqrt{2}  &\mu_{a'}
 \end{array} 
\right), 
 \end{equation} 
where, $\mu_A = \mu_H- \lambda_5 v^2$ and $\mu_{a'}=\mu'+ 2\,m_s +\lambda'_3 v^2/2$. The  masses of the  $CP$-odd physical Higgs states $(A, a')$ are given by
\begin{equation}
M^{2}_{A,a'}\simeq\frac{1}{2}\left[\mu_A+\mu_{a'}\!\pm\! \sqrt{(\mu_A-\mu_{a'})^2 + 2 m_{3}^2 v^2 }\right]\!.
\label{eq:masscpeven2}
\end{equation}
 The  corresponding mixing matrix and angle $\xi$ is given by
\begin{equation}
{Z^{\cal A}}=\left(
\begin{array}{ccc}
c_\xi & -s_\xi \\
s_\xi & c_\xi
\end{array}
\right),~~\tan{2 \xi}=\frac{\sqrt{2}\, m_{3} v}{\mu_A-\mu_{a'}}.
\end{equation}
 The charged Higgs mass is given by 
\begin{eqnarray}
M^2_{H^\pm}&=&\mu_2+\lambda_3 v^2/2\,.
\label{eq:mass2}
\end{eqnarray}
In the Higgs basis the relevant Lagrangian ${\cal L}$ can be written as follows
\begin{eqnarray}
{\cal L}&=&\sqrt{2}\,\Big[(X^u_{ij} \tilde\phi_h+\bar{X}^u_{ij} \tilde\phi_H) \bar{Q}_L^i  u^j_R
+(X^d_{ij}  \phi_h+\bar{X}^d_{ij} \phi_H) \bar{Q}_L^i  d^j_R+(X^e_{ij}  \phi_h+\bar{X}^e_{ij} \phi_H) \bar{L}_L^i  e^j_R\nonumber\\
&+&(X^\nu_{ij}  \tilde\phi_h +\bar{X}^\nu_{ij} \tilde\phi_H) \bar{L}_L^i  \nu_{R_j}+\frac{1}{\sqrt{8}}{m_{ij} \bar{\nu}^c_{R_i}{\nu_{R_j}}}+\lambda^N_{ij}\bar{\nu}^c_{R_i}\phi_{h'}\nu_{R_j}+h.c.\Big],
\label{eq:YkN}
\end{eqnarray}
here, $X_{ij}$ stands for SM Yukawa couplings depending on SM charged fermion mass, whereas $\bar{X}_{ij}$ (with $i,j=1,2,3$) are independent Yukawa matrices. The singlet sector Yukawa coupling matrices denoted by $\lambda^N_{ij}$ define the couplings at the neutrino vertices. The fermion masses receive contributions only from $X^f_{ij}~(f=u,d,e)$, since  we work in the Higgs basis, i.e., only $\phi_h$ acquires a non-zero VEV while $\langle\phi_H\rangle$ = $\langle\phi'_h\rangle=0$, leading to $X^f=\frac{M_f}{v}$, where $M_f$ are the SM charged fermion mass matrices.
$\bar{X}_{ij}$ are free parameters and non-diagonal matrices. We  work in a basis in which charged fermion mass matrices are real and diagonal, where $U_k M_f V_k^\dagger=m_f^{\rm diag}=\frac{y^h_f}{\sqrt{2}} \, v$ are the requisite bi-unitary transformations. The additional independent doublet sector Yukawa couplings $\bar{X}_{ij}$ are henceforth considered in a diagonal basis (as $y_f$) for simplicity. We will consider constraints on them from collider searches in a later section.

For neutrinos, we rotate the right-handed fields as $n_{R_i}= (U_{\nu_R})_{ij} \, \nu_{R_j}$, and the  left-handed SM neutrino fields as $n_{L_i} = (U_{\nu_L})_{ij} \, \nu_{L_j}$. To  generate  neutrino mass,  the mass matrices $\mathcal{M}_\nu= X^\nu_{ij} \, v$ and $m_{ij}$ can be diagonalized as,
\begin{eqnarray}
    U_{\nu_L} \mathcal{M}_\nu U_{\nu_R}^\dagger = m_D^{\rm diag}, ~~~ U_{\nu_R} m_{ij} U_{\nu_R}^\dagger  = m_{\nu_R}^{\rm diag} 
    \label{eq:rotanferm1} \, .
\end{eqnarray}
Similarly, we get,
\begin{eqnarray}
    U_{\nu_L} X^\nu_{ij} U_{\nu_R}^\dagger = y_\nu, ~~~ U_{\nu_R} \lambda^N_{ij} U_{\nu_R}^\dagger  = \lambda_n \, .
    \label{eq:rotanferm2}
\end{eqnarray}
The neutrino mass matrix in $(n_L,\, n_R^c)$ basis can now be written as
\begin{eqnarray}
m_\nu  = \left(  \begin{array}{c}
\overline{n^c}_L\, ~\overline{n}_R
\end{array} \right) \,
\left(  \begin{array}{ccc}
0 & m_D^{\rm diag} \\
m_D^{\rm diag} & m_{\nu_R}^{\rm diag}
\end{array} \right) \,
\left( \begin{array}{c}
n_L \\
n_R^c \\ 
\end{array} \right) \, ,
    \label{eq:numass}
\end{eqnarray}
which could be diagonalized by $m_\nu^{\rm diag} = \mathcal{N}\,m_\nu\,\mathcal{N}^\dagger$, where $m_\nu^{\rm diag}  \simeq {\rm diag}\{ -\frac{m_{D_i}^2}{m_{\nu_{R_i}}},  m_{\nu_{R_i}}\} $. As in~\cite{Abdallah:2020vgg}   the physical neutrino mass eigenstates are
\begin{eqnarray}
\left(  \begin{array}{c}
\nu \\ N
\end{array} \right) \,
\simeq \left(  \begin{array}{ccc}
I-\frac{\Theta^2}{2} & \Theta \\
\Theta & I-\frac{\Theta^2}{2}
\end{array} \right) \,
\left( \begin{array}{c}
n_L \\
n_R^c \\
\end{array} \right).
\end{eqnarray}
Here the neutrino mass matrix is diagonalized up to corrections of $\Theta_i=\frac{m_{D_i}}{m_{\nu_{R_i}}}$. For normal ordering, i.e., $m_{\nu_1} < m_{\nu_2} < m_{\nu_3}$, the two mass squared differences from  neutrino oscillation data~\cite{Esteban:2020cvm} are: $\Delta m_{21}^2=(6.82-8.04)\times 10^{-5}~{\rm eV^2}$ and $\Delta m_{31}^2=(2.435-2.598)\times 10^{-3}~{\rm eV^2}$.

Finally, after a rotation of the scalar fields, one finds the following coupling strengths of the scalars $h$, $H$,  $A$, and $a'$ with fermions, respectively:
\begin{eqnarray}
y^h_f= \frac{\sqrt{2} \, m_f}{v},  ~ y^H_f= y_f , ~ y^A_f = \, y_f  \, c_\xi,~ y^{a'}_f =  \, y_f  \, s_\xi,  ~ \lambda^A_{N_{ij}} = \,\lambda^N_{ij} \, s_\xi, ~{\rm and }~ \lambda^{a'}_{N_{ij} }= \, \lambda^N_{ij}  \, c_\xi\,. \, \, \, \, \, \, \,
\label{eq:fermcouplingstrength1}
\end{eqnarray}

\begin{figure}[t!]
\center
\includegraphics[width=0.6\textwidth]{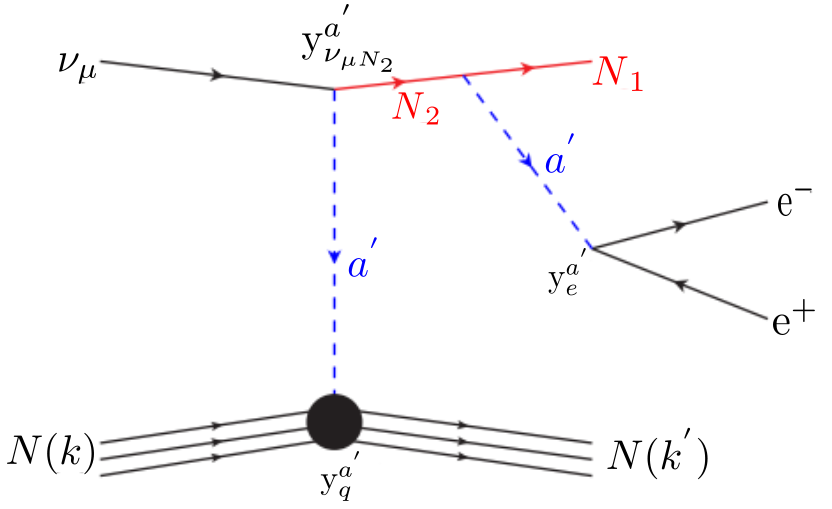}
\caption{Feynman diagram of the scattering process involving $a'$ which leads to the excess in MB and LSND.}
\label{FD-SP-LSND-MB}
\end{figure}
\section{The interaction in MB and LSND}
\label{sec3}
The Feynman diagram of the process that occurs in MB and LSND in our model, leading to an electron-like signal, is shown in figure~\ref{FD-SP-LSND-MB}. The heavy neutral lepton $N_2$ is generated via the up-scattering of  muon-type neutrinos in the beam. Following its production, $N_2$ promptly decays into a lighter neutral lepton $N_1$ and a light ($17$~MeV) pseudoscalar $a'$, which also decays promptly to a collimated $e^+e^-$ pair and produces the observed visible light signal. The heavy scalar ($H$) and pseudoscalar ($A$) can also mediate the process that occurs between the incoming neutrino and the nucleon at the primary vertex. However, if their contribution dominates, the angular distribution will lose the necessary forward character, with more events crowding the non-forward bins~\cite{Abdallah:2022grs}. The relative contribution of $a'$ and heavy particles ($H,\,A$) depends on the mass of heavy mediators and the mixing angle $ \xi$. The contribution of $H$ and $A$ to the MB excess for our chosen benchmark values is less than $5\%$. Increasing (decreasing) the masses of $A,H$ would reduce (increase) this contribution, whereas increasing (decreasing) $ \xi$ would decrease (increase) it. 

Our numerical calculations employ the cross-section for the interaction and the model outlined in section~\ref{sec2}. Fits to LSND and MB, as well as those for a common understanding of all three anomalies, which we take up in a later section)  depend crucially on the couplings of $a'$ to $u,d$ quarks and to electrons. The quark couplings are then used to obtain effective nucleon couplings.

The Yukawa interaction of $a'$ with quarks is given by
\bea
\mathcal{L}_{a'qq} = y^{a'}_q \, a' \bar{q} \, i\gamma_5 \, q \, .
\label{eq:sudolag}
\eea 
In our model, when seeking a solution to MB and LSND alone, the $a'$ predominantly couples to the first generation of quarks ($u$ and $d$) and has negligible and much smaller couplings to other families\footnote{This results in a relatively small allowed parameter space. We have thus used a benchmark point as a representative solution. It is expected that a parameter scan would yield a region in its neighbourhood. This changes when we seek a solution to all three anomalies and relax this assumption, as may be seen in table~\ref{tabNeW}.}. The effective coupling $(F_N)$ of $a'$ to a nucleon $(N)$ can be written as~\cite{Cheng:2012qr,PhysRevLett.114.011301}, 
\bea
F_N = \frac{m_N}{m_q} \, \sum_{q=u,d,s} \Delta_q^{(N)} \, \left( y^{a'}_q-   \sum_{q'=u,..,t}  y^{a'}_q \frac{\overline{m}}{m_{q'}}\right),
\label{eq:ncoup}
\eea
where $\Delta_q^{(N)}$ are the quark spin components of the nucleon $N$,
\be
\frac{1}{\overline{m}} = \frac{1}{m_u} + \frac{1}{m_d} + \frac{1}{m_s},
\label{effmass}
\ee
$\Delta_u^{(p)}=0.84$, $\Delta_d^{(p)}=-0.44$, $\Delta_s^{(p)}=-0.03$, $\Delta_u^{(n)}=-0.44$, $\Delta_d^{(n)}=0.84$, $\Delta_s^{(n)}=-0.03$~\cite{Cheng:2012qr}. Here $m_N$ is the nucleon mass.  All the relevant quark masses are taken from~\cite{ParticleDataGroup:2022pth}.

We note that for the  carbon nucleus, which is the primary target in MB and LSND, the total spin of the nucleus is zero. Since any pseudoscalar mediated contribution to the coherent production depends on the spin of the target, we need only  consider  the incoherent production of $N_2$ in MB.  The total differential cross-section, for the target in MB,  $\ie$, CH$_2$, is thus given by
{\selectfont\fontsize{9.5}{9.5}{
\begin{equation*}
\Big[\dfrac{d\sigma}{dE_{N_2}}\Big]_{{\rm CH_{2}}}\!=\!\Big[\underbrace{(8F^2_{p}\!+\!6F^2_{n})}_{\textrm{\footnotesize{{incoherent}}}}
\Big]\dfrac{d\sigma}{dE_{N_2}}.
\label{tot_xsec}
\end{equation*}}}
The number of events is given by
\begin{equation}
{\rm N}_{\rm{events}} \! =\! \eta\!\! \int\!\! dE_{\nu} dE_{N_2}\dfrac{d\Phi^{\rm{\nu}}}{dE_{\nu}} \!\,  \dfrac{d\sigma}{dE_{N_2}}\! \times\! {\rm BR}(N_2 \!\rightarrow\! N_1 a'),
\end{equation}
 $E_{a'}$, which represents the visible energy that manifests itself as light subsequent to $a'$ decay, $\in [E_{a'},E_{a'}+\Delta E_{a'}]$. Here 
 $\Delta E_{a'}$ is essentially the bin size in our plots for the two detectors. $\eta$
 encompasses all detector-related information, such as efficiencies, POT, etc., while $\Phi^{\rm{\nu}}$
 represents the incoming muon neutrino flux. The calculations for LSND and MB alone are carried out using the benchmark values in table~\ref{tab}, while when trying to obtain a common solution to all three anomalies, we use the benchmarks given in table~\ref{tabNeW}. The quark and neutrino vertex couplings are quite different for  the two tables. We note that the  LSND and MB fits depend on the product of the  effective couplings at the neutrino and nucleon vertex, for given masses of $N_2$ and $a'$.  The resolution of the  ATOMKI anomaly demands higher values of $F_p$ and $F_n$ which translate into higher (lower) values of quark (neutrino vertex) couplings\footnote{Although the fit to MB and LSND consequently remains unaffected, the enhanced values of the quark couplings emerging from  ATOMKI do impact the resolution of certain constraints, as we discuss in a later section.} in table~\ref{tabNeW} compared to table~\ref{tab}.
\section{Results for MB and LSND}
\label{sec4}
This section gives our fits for MB and LSND alone using a light 17~MeV pseudoscalar, without any input from the ATOMKI results.

\begin{figure}[h!]
	\begin{center}
			{\includegraphics[scale=0.530]{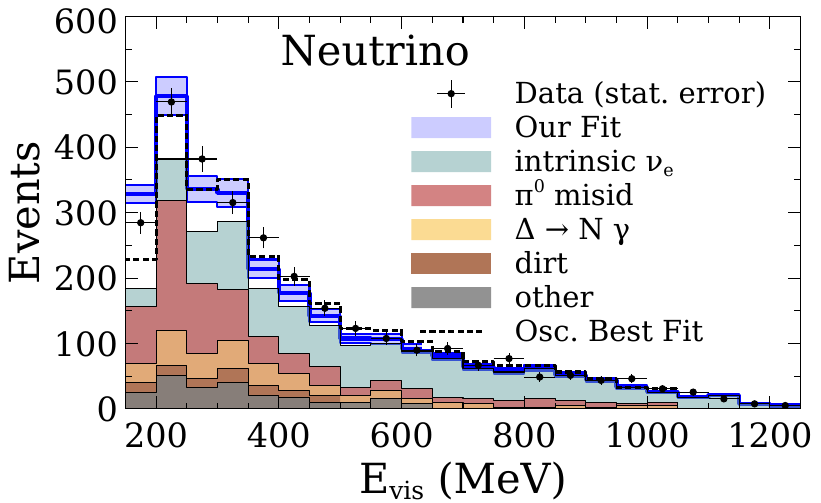}}	
			{\includegraphics[scale=0.530]{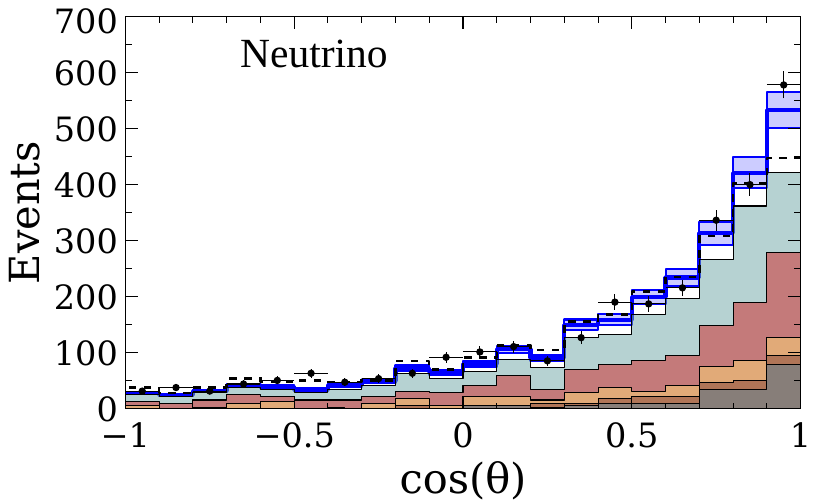}}
			{\includegraphics[scale=0.530]{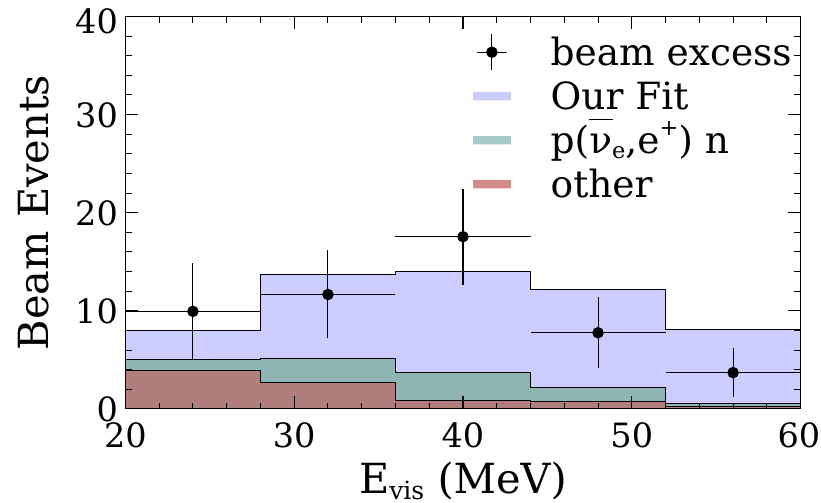}}	
			{\includegraphics[scale=0.530]{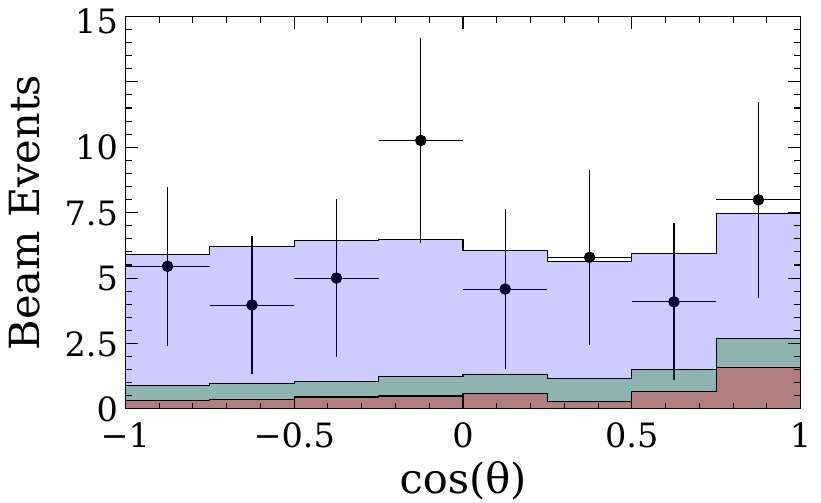}}
	\end{center}
\caption{ Top panels: These show, for neutrino runs, the MB electron-like events (backgrounds and signal) sourced from~\cite{MiniBooNE:2020pnu}. They are  plotted against the visible energy E$_{\rm vis}$ (left panel) and  the emitted angle of the light (right panel). The blue solid line denotes the prediction of excess events in our model.
Bottom panels: These show, for $R_\gamma> 10$, the energy distribution (left panel) of the LSND data~\cite{Aguilar:2001ty} and, for $R_\gamma > 1$, the angular distribution (right panel) of the light originating from the electron-like final state. In both panels, the shaded blue region represents our fit, and other shaded regions denote the backgrounds.}
\label{fig:MB_LSND_Plot}
\end{figure}

In Figure~\ref{fig:MB_LSND_Plot} (top panels) we present  the SM backgrounds, the MB data points,  and the predictions from our model (depicted by the blue solid line) within each bin. Additionally, the oscillation best fit is represented by the black dashed line. Our fit utilizes the most recent data set for the neutrino mode, corresponding to $18.75 \times 10^{20}$ POT~\cite{MiniBooNE:2020pnu}. The left panel shows $E_{vis}$, the measured visible energy, plotted against the neutrino events. Note that  $E_{vis}$ corresponds to $E_{a'}$ in our model. Meanwhile, the right panel displays the angular distributions for the emitted light. The fit is conducted using benchmark parameter values outlined in table~\ref{tab}. We have utilized fluxes, efficiencies and other pertinent details from~\cite{Aguilar-Arevalo:2018gpe} and the references therein to generate these plots.  Excellent fits to the data are achieved for both the angular and the energy distributions. The data points  reflect statistical uncertainties only. We have incorporated a systematic uncertainty of $15\%$ in our calculations, and this tolerance is represented by the blue bands in the figures.

 In addition to detecting the $2.2$~MeV photon resulting from coincident neutron capture on hydrogen,  LSND  measures the visible energy stemming from  Cerenkov and scintillation light radiated by the  electron-like event. Figure~\ref{FD-SP-LSND-MB} shows how this process takes place in the context of our model, via  the scattering off a target neutron within the Carbon nucleus. All requisite information concerning fluxes\footnote{We note that only the decay-in-flight (DIF) flux for LSND has been used, since it is energetic enough to produce $N_2$, while the decay-at-rest (DAR) flux is not.}, efficiencies, POT, etc., for LSND has been sourced from~\cite{Aguilar:2001ty} and the references therein.

Figure~\ref{fig:MB_LSND_Plot} (bottom-left panel) illustrates our prediction compared  to the LSND data for $R_\gamma >10$.  Here $R_\gamma$ is a measure of the likelihood ratio that the observed  $2.2$~MeV photon, which signals the presence of an emitted  neutron, was correlated rather than accidental. It  has been   defined by the LSND Collaboration in \eg ~\cite{Aguilar:2001ty}. This panel shows the excess events observed and their  energy distribution, as well as those resulting from our model employing the same benchmark parameters as are used to generate the MB fit. For this specific $R_\gamma$, we obtain 39.4 total excess events from our model, while LSND reports 32 events. 


Figure~\ref{fig:MB_LSND_Plot} (bottom-right panel) illustrates, for $R_\gamma >1$, the angular distribution of light stemming from the observed  electron-like final state. The visible energies shown lie  within the range\footnote{The bottom panels employ different ranges of $R_\gamma$ and ${\rm E}{\rm_{vis}}$. This choice has been made  to align our results with those typically presented by the LSND Collaboration, which employs distinct $R_\gamma$ and ${\rm E}{\rm_{vis}}$ ranges for the energy and angular distributions.} $36~{\rm MeV} < {\rm E_{vis}}<60$~MeV. The blue shaded region in both panels represents the prediction of our model, juxtaposed with backgrounds and data. For our benchmark values, we get  $\chi^2/{\rm dof} =1.9$.
 \begin{table}[h!]
\vspace{0.5cm}
\begin{center}\scalebox{0.9}{
 \begin{tabular}{|c|c|c|c|c|c|c|c|c|}
  \hline
  $m_{N_1}$& $m_{N_2}$ & $m_{N_3}$&$y_{u}^{a'}\!\!\times\!\! 10^{6}$ &$y_{e}^{a'}\!\!\times\!\! 10^{5}$&$y_{\mu}^{a'}\!\!\times\!\! 10^{5}$ & $M_{H^{\pm}}$&$y_{c}^{a'}$& $y_{t}^{a'}$\\
  \hline
   70\,MeV  & $120$\,MeV & $10$\,GeV&$4.34$ &$2.3$&$1$ & 305~GeV & 0 & 0\\
  \hline
    \hline
  $M_{a'}$& $M_{H}$ &$\sin\xi$&  $y_{d}^{a'}\!\!\times\!\! 10^{6}$&$y_{\nu_{\mu N_2}}^{a'}\!\!\times\!\! 10^{2}$&$\lambda_{N_{\!12}}^{a'}$  & $M_{A}$ & $y_{s}^{a'}$ & $y_{b}^{a'}$\\[0.05 cm]
  \hline
   17\,MeV  & 300\,GeV &$0.01$& $4.0$&$3.15$&$0.1$ &400~GeV & 0 & 0\\
  \hline
 \end{tabular}}
 \caption{Benchmark parameter values used to generate the event spectrum in LSND and~MB.}
\label{tab}
\end{center}
\end{table}

\section{Derivation of couplings of the pseudoscalar from ATOMKI results}
\label{sec5}
As mentioned above, the ATOMKI collaboration has reported anomalous
measurements in the IPC  decays of excited $^8$Be~\cite{Krasznahorkay:2015iga, Krasznahorkay:2018snd},
$^4$He~\cite{Krasznahorkay:2019lyl, Krasznahorkay:2021joi} and, more recently, $^{12}$C~\cite{Krasznahorkay:2022pxs} nuclei. We note that observations were
conducted for two excited states of $^8$Be, specifically $^8$Be(18.15) and
$^8$Be(17.64). Similarly, for $^4$He, decays of $^4$He(21.01) and $^4$He(20.21)
were studied.\footnote{For a review, see Ref.~\cite{Barducci:2022lqd}.} The experiment observed unexpected bumps for both
the invariant mass and the angular opening of the $e^+e^-$ pairs with high
statistical significance, well above $6\sigma$.  As in~\cite{Ellwanger:2016wfe}, we derive the values of the couplings of the pseudoscalar to quarks, electrons and nucleons appropriate 
 to a resolution of this anomaly.

Noting that in the SM, the decay of an excited nucleus to a lower state with an equal
 number of protons and neutrons can happen only via EM processes, the
allowed channels are:
\begin{itemize}
    \item the emission of a real photon, due to the decay of the nucleus.
    
    \item The emission of a virtual photon by the nucleus, which
decays to an $e^+e^-$ pair, (Internal Pair Creation (IPC)),  i.e.,
\begin{equation}
    p+A \rightarrow N^* \rightarrow N + e^+ + e^-\,.
\end{equation}
\end{itemize}
Here $p$ denotes a beam proton and $A$ is a target nucleus. The anomaly appears only in the IPC events. We consider, as an example, the Be case. The experiment used a beam of protons with kinetic energies tuned to the resonance energy of 1.03~MeV, which were made to  collide with Li nuclei, in order  to form the resonant state $^8$Be$^*$.  A small
percentage of these decayed via $^8{\!\,\rm Be}^*\rightarrow \,^8{\!\,\rm Be}+e^+e^-$.
We note that $^8{\!\,\rm Be}^*$ decays to $^7{\!\,\rm Li} + p$ most of the time, but it also has  electromagnetic transitions
with branching fractions ${\rm BR}(^8{\!\,\rm Be}^* \rightarrow\,^8{\!\,\rm Be}\, \gamma) \simeq 1.4 \times 10^{-5}$~\cite{Tilley:2004zz}
and ${\rm BR}(^8{\!\,\rm Be}^* \rightarrow\,^8{\!\,\rm Be} + e^+ e^-) \simeq 3.9 \times 10^{-3}~{\rm BR}(^8{\!\,\rm Be}^* \rightarrow\,^8{\!\,\rm Be}\,\gamma)$~\cite{Schluter:1981cjo,Rose:1949zz}.

The experiment measured both the electron and positron energies, as well as the opening angle of the $e^+ e^-$ pairs, $\theta$, in order to  determine the invariant mass ($m_{ee}$)  and angular distributions.  It did not observe the behaviour predicted by the  SM
 \ie  ~that the $\theta$ and  $m_{ee}$ distributions should fall monotonically.  The $\theta$
distribution exhibited a high-statistics bump that peaked at $\theta \simeq 140^{\circ}$ before it 
returned close  to the SM prediction at  $\theta \simeq 170^{\circ}$.

Such a bump at large opening angles is expected from the kinematics  of  a massive particle that is  is produced with low velocity
in the $^8{\!\,\rm Be}^*$ decay and which then subsequently decays to $e^+ e^-$ pairs. The hypothesis of  a new particle $X$ and the
two-step decay ${\rm BR}(^8{\!\,\rm Be}^* \rightarrow\, ^8{\!\,\rm Be} + X)$ followed by $X \rightarrow e^+ e^- $ thus emerges  as a natural resolution to the anomaly. With the assumption of a fixed background, Krasznahorkay \emph{et al.}~\cite{Krasznahorkay:2015iga} give the best fit mass and branching fraction as
\bea
M_X = 16.7 \pm 0.35 ({\rm stat}) \pm 0.5 ({\rm sys}) ~~ {\rm MeV},&&\nn\\[0.5cm]
\frac{{\rm BR}(^8{\!\,\rm Be}^* \rightarrow\, ^8{\!\,\rm Be}\, X) \times {\rm BR}(X \rightarrow e^+ e^-)}{{\rm BR}(^8{\!\,\rm Be}^* \rightarrow\, ^8{\!\,\rm Be}\, \gamma)} =  &&5.8\times 10^{-6}\,.   \nn
\eea
These values correspond to a statistical significance in excess of $6.8\sigma$.
Possible types of particles which may provide candidate solutions of the
anomaly are a pseudoscalar, a vector or an axial-vector~\cite{Barducci:2022lqd}.
Assuming parity and angular momentum conservation in the decay, the
possibility of $X$ being a real scalar is ruled out. Additionally, very strong
coupling constraints based on anomaly cancellations apply to a light vector or
axial vector that couples to SM fermions via an additional $U(1)$~\cite{Dror:2017ehi,Dror:2017nsg,Dror:2018wfl}. For specificity, we have thus chosen to focus on a pseudoscalar as our calculational choice for $X$.

 On a more general footing, however, we note that 
for both the Be excited states, a pseudoscalar, which has intrinsic parity $(-1)$
can be emitted in a state of orbital angular momentum 1 with respect to the
ground state $^8{\!\,\rm Be}$, thus conserving both angular momentum and parity. In the
case of $^4{\rm He}$, however, only the decay from the excited state
$^4{\rm He}$ (21.01), which has 0 angular momentum and parity $-1$, will be allowed for a pseudoscalar, while in the case of $^4{\rm He}$ (20.21), if it contributes to the observed excess, only a real scalar is allowed.  Additionally, 
 for the $^{12}$C nucleus which has spin 1 and parity $-1$, the pseudoscalar
does not allow a solution that conserves both overall parity and angular
momentum. However, the $^{12}$C excess can be explained by a real 17 MeV scalar. When these observations are combined with the results of of this and previous work on the resolution of the MB and LSND anomalies, an important indicative conclusion  arises, which we note in the concluding  section of the paper.

\subsection {Couplings of the pseudoscalar to quarks}
As shown in the previous section in eqs.~\eqref{eq:sudolag},\eqref{eq:ncoup} can be used to obtain the effective nucleon couplings.
The average nucleon coupling ($\bar{h}^2_N$) of $a'$ with the $^8{\!\,\rm Be}^*$ is given by
\bea
\bar{h}^2_N \equiv \frac{(F_p+F_n)^2}{4} 
\label{eq:hN}\,.
\eea
The next step involves a calculation of the Be decay rates given below,
employing nuclear matrix elements, and for this we use results from~\cite{Ellwanger:2016wfe}
\bea
\frac{{\rm BR}(^8{\!\,\rm Be}^* \rightarrow\, ^8{\!\,\rm Be}\, A)}{{\rm BR}(^8{\!\,\rm Be}^* \rightarrow\, ^8{\!\,\rm Be}\, \gamma)} = \frac{2\bar{h}^2_N}{0.16\,\, e^2} \,\, \frac{|p_A|^3}{|p_\gamma|^3}.
\label{eq:BrRatio}
\eea
We assume ${\rm BR}(A \rightarrow e^+ e^-)=1$ and compare to the experimental value from the $^8{\!\,\rm Be}$ experiment~\cite{Krasznahorkay:2015iga}, which is
\bea
\frac{{\rm BR}(^8{\!\,\rm Be}^* \rightarrow\, ^8{\!\,\rm Be}\, A)}{{\rm BR}(^8{\!\,\rm Be}^* \rightarrow\, ^8{\!\,\rm Be}\, \gamma)} = 5.8\times 10^{-6}
\label{eq:BrRatio2}\,.
\eea
The ratio of momenta depends on the pseudoscalar mass $M_{a'}$ which in our case is 17~MeV. This leads to
\bea
 \frac{|p_A|^3}{|p_\gamma|^3} \sim 0.045\,.
 \label{eq:pRatio}
\eea
%
Solving the above three equations, we get the value of  $\bar{h}^2_N = 9.38 \times 10^{-7}$. We choose the coupling of $a'$ to quarks in table~\ref{tabNeW} such that we get the desired value of $\bar{h}^2_N$. The value $F_p$ ($F_n$) for our benchmark parameters is  $0.0011\,(0.000832)$. The ratio of the absolute values of the effective couplings of $a'$ with proton and neutron (i.e., $F_n/F_p = 0.75 $) is an important factor in reproducing the correct energy and angular distributions in LSND and~MB.
\begin{table}[h!]
\vspace{0.5cm}
\begin{center}\scalebox{0.9}{
 \begin{tabular}{|c|c|c|c|c|c|c|c|c|}
  \hline
  $m_{N_1}$& $m_{N_2}$ & $m_{N_3}$&$y_{u}^{a'}\!\!\times\!\! 10^{6}$ &$y_{e}^{a'}\!\!\times\!\! 10^{5}$&$y_{\mu}^{a'}\!\!\times\!\! 10^{5}$& $M_{H^{\pm}}$&$y_{c}^{a'}\!\!\times\!\! 10^{3}$& $y_{t}^{a'}\!\!\times\!\! 10^{5}$\\
  \hline
   70\,MeV  & $120$\,MeV & $10$\,GeV&$4.613$ &$2.3$& $1$ & 305~GeV & $-6.827$ &$1.0$\\
  \hline
    \hline
  $M_{a'}$& $M_{H}$ &$\sin\xi$&  $y_{d}^{a'}\!\!\times\!\! 10^{5}$&$y_{\nu_{\mu N_2}}^{a'}\!\!\times\!\! 10^{3}$&$\lambda_{N_{\!12}}^{a'}$  & $M_{A}$ & $y_{s}^{a'}\!\!\times\!\! 10^{5}$ & $y_{b}^{a'}$\\[0.05 cm]
  \hline
   17\,MeV  & 300\,GeV &$0.01$& $1.0$&$2.2$&$0.1$ &400~GeV & $1.356$ & $0$\\
  \hline
 \end{tabular}}
\caption{Benchmark parameters used to generate the event spectrum in LSND, MB and for calculating the ATOMKI. The corresponding doublet sector scalar quartic couplings for this benchmark point (BP) are shown in table~\ref{tab:lam} in the Appendix~\ref{eq:loopf}.}
\label{tabNeW}
\end{center}
\end{table}
\subsection {Couplings of the pseudoscalar to the electron}

Our calculation above assumes that the branching fraction of the pseudoscalar
decay to $e^+ e^-$ pairs is very close to 1. Its total width is
\bea
\Gamma(a') =  \frac{(y^{a'}_e)^2}{8 \pi }\,  M_{a'}
\eea
and its decay length is given by
\bea
l_{a'} = \frac{p_{a'}}{M_{a'} \Gamma(a')}.
\eea
The momentum in question can be determined from the energetics of the
ATOMKI $^8$Be results, since in that decay $^8{\rm Be}^* \rightarrow  {\rm^8 Be} \, \,a'$ with $M(^8{\rm Be}^*)-M(^8{\rm Be}) = 18.15$~MeV. 
The size of the ATOMKI detector requires that the pseudoscalar $a'$ decay in about 1 cm.
This gives $y^{a'}_e > 8.3 \times 10^{-6} $. Our benchmark values in table~\ref{tabNeW} satisfy the inequality and give the value $\bar{h}^2_N = 9.38 \times 10^{-7}$.
\section{Combined Results for MB, LSND and ATOMKI}
\label{sec6}
Using the coupling values determined from ATOMKI data in the previous section, we obtain fits to MB and LSND. The fits are identical to those shown in the plots of figure~\ref{fig:MB_LSND_Plot}, for reasons explained towards the end of section~\ref{sec3}, hence we do not display them here.
Our numerical calculations employ the cross-section for the process and the model outlined in section~\ref{sec3}. Fits to LSND and MB, as well as the result of ATOMKI, depend crucially on the couplings of $a'$ to nucleons and the electron. Benchmark values for these couplings, shown in table~\ref{tabNeW}, are obtained from those required to obtain the ATOMKI result in section~\ref{sec5}, and are then fed into the fitting procedure for MB and LSND with other detector specific inputs. Note that the quark couplings to $a'$ are significantly different, and higher, than those required to fit MB and LSND alone.

 The fits shown in figure~\ref{fig:MB_LSND_Plot} are representative, and it is useful to obtain some feeling for the allowed solution space when one considers the three anomalies together. Since obtaining this space is significantly affected by bounds from charged kaon decays, we have relegated its discussion to the next section, which is dedicated to constraints.
\section{Constraints}
\label{sec7}
This section examines constraints on our model from flavour changing meson decays, collider physics, vacuum stability, and electroweak precision data.
%
\subsection{ Constraints on a light (17~MeV) pseudoscalar}
A singlet $17$~MeV pseudoscalar particle is subject to constraints from many sources. For instance, since in our model it predominantly decays to 
 an $e^+e^-$ pair, it is subject to some of the constraints from searches for axions or axion-like particles, such as those summarized in~\cite{Dolan:2014ska,Andreas:2010ms,Essig:2010gu,Proceedings:2012ulb,Dobrich:2015jyk,Liu:2021wap}. However, we note that  $a'$ in our model obtains its couplings to fermions via mixing with the Yukawa couplings of the heavier pseudoscalar $A$ belonging to the second Higgs doublet. These couplings do not follow the usual mass-dependence of the SM Yukawa couplings, and are at present unconstrained in many cases, especially for the heavier fermions. This  weakens the direct applicability of some of the axion and axion-like bounds.  A  discussion of many of the other constraints on a $17$~MeV singlet scalar with couplings in the range required to fit MB and LSND alone is given in Ref.~\cite{Abdallah:2020vgg}. Of special relevance among these are bounds from the electron beam dump experiments E137~\cite{Bjorken:1988as}, E141~\cite{Riordan:1987aw}, ORSAY~\cite{Davier:1989wz} and NA64~\cite{NA64:2018lsq}. While our BP is not in violation of these at present, a more thorough mapping of the allowed regions would be worthwhile.

{\setlength{\parindent}{0pt}
\textbf{\textit{Charged kaon decay constraint and the sample solution space:}}}

 A potentially important class of constraints arises from flavour-violating meson decays, see \eg~\cite{Dolan:2014ska}. We note that in any heavy meson decay that involves $u,d$ quarks, one can radiate an $a'$ which would promptly decay to an $e^+ e^-$ pair via the diagonal couplings between it and the quarks.
 While off-diagonal flavour changing couplings in our model are arbitrarily small, the first generation diagonal quark couplings to the scalars in our model are fixed by the requirements of fitting the ATOMKI, LSND and MB data, and are approximately $\mathcal{O}(10^{-4})$. For this case, the decay of charged kaons leads to constraints which are especially stringent~\cite{Dolan:2014ska,Batell:2018fqo}. These decays are dominated by flavour changing penguin processes, and we  calculate the decay width and branching ratio ${\rm BR}(K^+ \rightarrow \pi^+ a' (\pi^+ + e^+e^-))$  for our model below.

Prior to that, we note certain caveats which may apply to the 
 two experiments which explored the low mass region ($0-50$~MeV) for $a'$, namely a) the $K_{\mu 2}$ experiment E89~\cite{Yamazaki:1984vg} at KEK and b) the BNL-AGS experiment~\cite{Baker:1987gp}. We note certain aspects which could possibly render the bounds less severe than claimed (for a detailed discussion of their limitations see~\cite{Alves:2017avw}). Both experiments, when exploring the low mass region relevant to our model, must confront the difficult background from Dalitz pair production, $K^+ \rightarrow \pi^+ + (\pi^0 \rightarrow e^+e^-\gamma)$. In the case of the KEK experiment, there appears to be significant uncertainty regarding the lower end (in mass) of their sensitivity below 50~MeV, with some papers, for instance~\cite{Yamazaki:1984qx} showing no sensitivity below this value. Additionally, the  method used to infer limits in the range $M_{a'} < 80$~MeV is considered inappropriate for such a low statistics region, in particular  the $\sqrt{N}$ estimation of background error may lead to an overly stringent bound in a Poissonian region~\cite{Alves:2017avw}.

The BNL-AGS experiment has also been criticised for the modelling and subtraction of their background. Their Monte
Carlo apparently  mis-estimates the Dalitz background contamination by
possibly a factor of $\sim 10$~\cite{Baker:1987gp}. This has led to questions on  their Monte Carlo estimation of the signal acceptance as well~\cite{Alves:2017avw}. Overall, there appears to be significant uncertainty in the bounds on $K^+ \rightarrow \pi^+ a'$ in the region relevant to our model. Since the experiments are about four decades old, one may conservatively say that exploring this difficult but very interesting region using a recent-day experiment may be worthwhile. 

Keeping the above considerations in mind, in finding the sample solution space for common solutions to MB, LSND and ATOMKI shown in figure~\ref{fig:BMPregionPlot} we have allowed the possibility that BR$(K^+ \rightarrow \pi^+ a' (\pi^+ + e^+e^-)) < {\cal O}(10^{-6})$. This is about a factor of ~ few to  an order of magnitude less severe than required by~\cite{Yamazaki:1984vg, Baker:1987gp}. Note, however, that in spite of this, figure~\ref{fig:BMPregionPlot} shows a large region (the right panel)  that conforms to the stringent bounds demanded by these experiments. \\

In addition to not violating kaon decay constraints, the derivation of the sample solution spaces shown in figure~\ref{fig:BMPregionPlot} must conform to the following conditions, to a reasonable degree of accuracy:\\
$a)$ the  ratio of the absolute values of the effective couplings of $a'$ with neutron and proton must be $\frac{F_N}{F_P}\simeq 0.77$. This requirement stems from  fits to the MB and LSND distributions.\\
$b)$ The  average nucleon coupling $\bar{h}^2_N$, defined in eq.~\eqref{eq:hN} must be $\simeq 9.38\times 10^{-7}$. This requirement, from ATOMKI data,  was obtained by solving eqs.~\eqref{eq:BrRatio}, \eqref{eq:BrRatio2} and~\eqref{eq:pRatio}.

\begin{figure}[h!]
\centering
\includegraphics[width=0.3\textwidth]{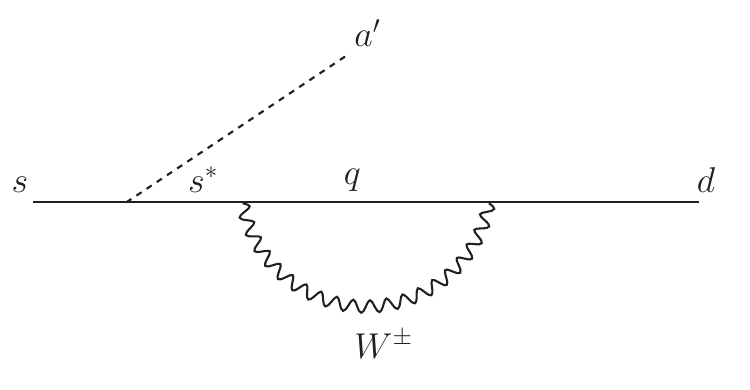}
\includegraphics[width=0.3\textwidth]{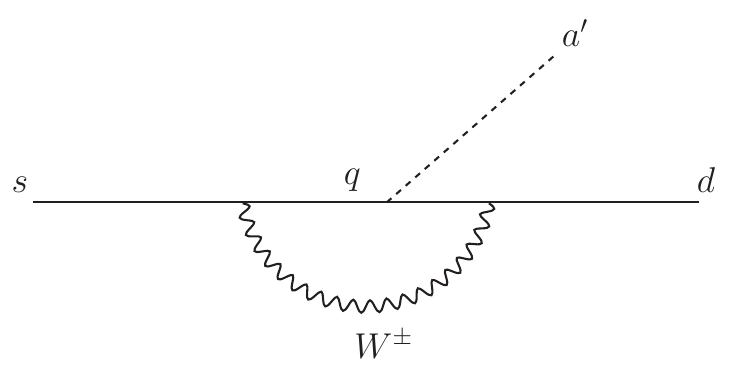}
\includegraphics[width=0.3\textwidth]{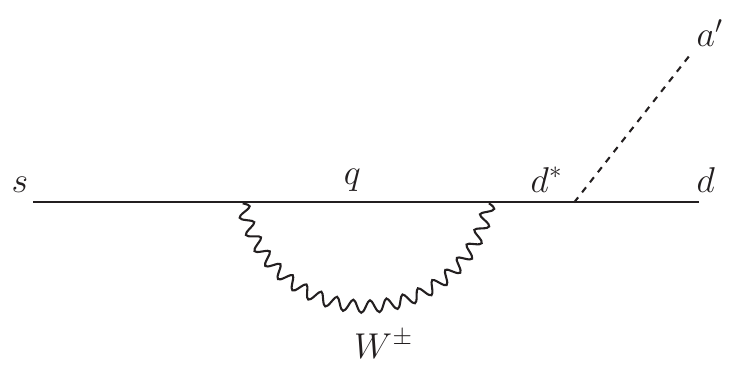}
\caption{ Feynmann diagrams for the flavour-changing transitions such as $s \to d a'$ involving the exchange of quarks and $W^{\pm}$-bosons in the loop. $q$ stands for up-type quarks $u,c,t$.}
\label{fig:1Loopsd}
\end{figure}

\begin{figure}[h!]
	\begin{center}
			{\includegraphics[scale=0.330]{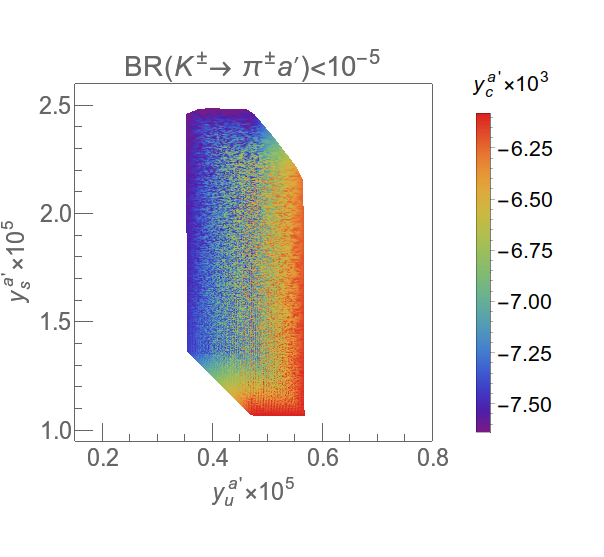}}	
			{\includegraphics[scale=0.330]{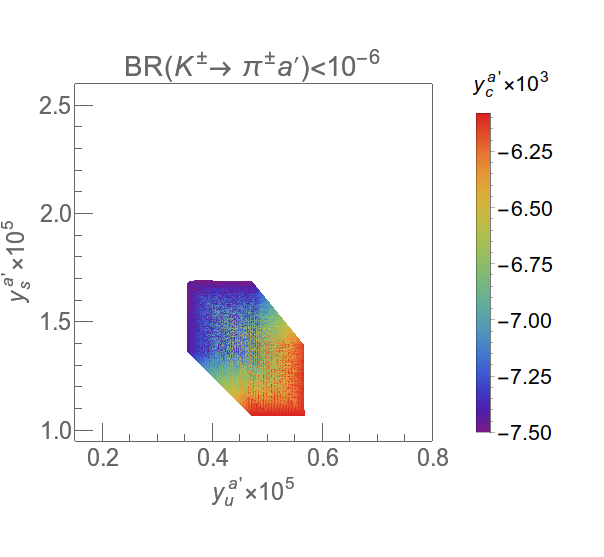}}
	\end{center}
\caption{  The  allowed region in the $y^{a'}_s - y^{a'}_u $ plane (kept $y^{a'}_d=y^{a'}_t=10^{-5}$).  $|\frac{F_N}{F_P}|$=$0.77$ $\pm 0.154$ and $\bar{h}^2_N \equiv $ $\frac{(F_p+F_n)^2}{4}$ = $(9.38\pm 1.876)$  $\times 10^{-7}$. The left plot gives ${\rm BR}(K^\pm \to \pi^\pm a')$ $<10^{-5}$, while the right one maintains ${\rm BR}(K^\pm \to \pi^\pm a')$ $<10^{-6}$. Here we have chosen $y^{a'}_{b}$ $= 0$, and get BR$(B^\pm \to K^\pm a')$ $<5.2 \times 10^{-7}$. For scanning, we keep the branching fraction $1.6 \times 10^{-10}<$ BR $(\pi^+ \rightarrow $ $ e^+ ~\nu_e ~e^+ e^- )< 2.6 \times 10^{-10}$, which is in tension with the SINDRUM bound \cite{SINDRUM:1986klz} at $90 \%$ confidence level.}
\label{fig:BMPregionPlot}
\end{figure}

Noting that while the magnitudes of the couplings $y_u^{a'}$ and $y_d^{a'}$ are largely decided by the fitting requirements above, the couplings of the other quarks to $a'$ are largely free, but do affect the calculation of the BR of the kaon decay through the $W$-mediated penguin diagrams which dominate its value (figure~\ref{fig:1Loopsd})\footnote{Note that, the $H^\pm$-mediated loop diagram is always subdominant in our model, since the corresponding couplings to quarks are not proportional to $m_q$ and are largely free.}. We seek a solution space in terms of the quark couplings to $a'$ and define a tolerance of ($15\%$) in the values defined in $a)$ and $b)$.
 This is sufficient to lead to good fits to the data in all three experiments.

The one-loop contribution from flavour-changing diagrams  shown in figure~\ref{fig:1Loopsd} are in general divergent. The effective coupling for the interaction $h^R_{ds} \, a' \, \overline{d} P_R s$, where $P_R=\frac{1+\gamma_5}{2}$ is given by~\cite{Dolan:2014ska}
\begin{eqnarray}
\hspace{-1.0cm}    h^R_{ds} &=& \sum_{q=u,c,t}
\frac{e^2 \, \left[3 \, m_d \, m_s \, m_q^2 \, y^{a'}_d - (m_d^2 + 2 m_s^2) \, m_q^2 \, y^{a'}_s + 2 \, m_s \, (m_d^2 - m_s^2) \, m_q \, y^{a'}_q \right]}{64 \pi^2 \, m_W^2 \, \sin(\theta_W)^2 \, (m_s^2 - m_d^2)} \, V_{qs} V_{qd}^\ast \times \frac{1}{\epsilon} ,\nonumber\\
\label{eq:1412Ref}
\end{eqnarray}
where  $m_W$ is the $W$-boson mass, $\theta_W$ is the Weinberg angle and $V_{ij}$ are elements of the CKM matrix.   We have used the cutoff regularization replacement $1/\epsilon \rightarrow \log(\Lambda^2 / m_t^2)$ where $\Lambda$ ($= 1$ TeV in our case) is the energy scale at which new physics beyond the Standard Model becomes relevant, and $m_t$ is the mass of the top quark. The corresponding expression (upto a sign) for $h^L_{ds}$  is obtained by exchanging $s$ and $d$ in eq.~\eqref{eq:1412Ref}. The overall effective coupling may then be written as 
\begin{equation}
    h^S_{ds}=\frac{h^R_{ds}+h^L_{ds}}{2},~~~~{\rm and}~~~~ h^P_{ds}=\frac{h^R_{ds}-h^L_{ds}}{2},
\end{equation}
leading to the full pseudoscalar interaction term
\begin{equation}
    i\, \overline{s} (h^S_{ds}+ h^P_{ds} \gamma_5) d.
\end{equation}
 The coupling $h^S_{ds}$ leads to kaon decay via $K^\pm \to \pi^\pm a'$. Note that the part of the coupling proportional to $h^P_{ds}\,\gamma_5$ does not contribute to this decay~\cite{Deshpande:2005mb}. Finally, the partial decay width is given by~\cite{Deshpande:2005mb,Dolan:2014ska}
\begin{equation}
\Gamma(K^+ \rightarrow \pi^+ a') = \frac{1}{16 \pi \, m_{K^+}^3} \lambda^{1/2}(m_{K^+}^2, m_{\pi^+}^2, M_{a'}^2) \left(\frac{m_{K^+}^2 - m_{\pi^+}^2}{m_s - m_d}\right)^2 |h^S_{ds}|^2 \,,
\label{eq:Kwidth}
\end{equation}
where the function $\lambda(a,b,c) =(a-b-c)^2-4\,b\,c$, $m_{K^\pm}$ is the mass of the kaon and $m_{\pi^\pm}$ is the mass of the pion.   
 
For our BP  in table~\ref{tabNeW}, this leads to ${\rm BR}(K^\pm \to \pi^\pm a')=6.13\times 10^{-7}$. The  Yukawa couplings are then varied in accordance with the tolerance mentioned above to explore a larger region of parameter space that can simultaneously explain the MB, LSND, and ATOMKI data while remaining consistent with current kaon decay constraints~\cite{Dolan:2014ska, Batell:2018fqo}.
 There are six independent  couplings, allowing for various solutions.  Our aim is to provide a representative sample of points, rather than undertake a full exploration of the allowed space. Thus, for simplicity, we set $y^{a'}_d=y^{a'}_t=10^{-5}$ and vary $y^{a'}_{u}$, $y^{a'}_s$ and $y^{a'}_c$, showing the allowed region. We also choose $y^{a'}_b=0$ to avoid constraints from the $\Gamma(B^\pm \to K^\pm a')$ decay using a similar loop decay.
 In figure~\ref{fig:BMPregionPlot},  the left plot shows BR$(K^\pm \to \pi^\pm a')$ $<10^{-5}$, while the right one shows BR$(K^\pm \to \pi^\pm a')$ $<10^{-6}$. The color shading indicates the variation of the charm coupling (depicted by the bar on the right side). 
These sample  plots illustrate that a large region of parameter space can explain the observed excesses reported in MB, LSND, and ATOMKI while being consistent with constraints coming from the rare meson decay.

In addition, the constraint ${\rm BR}(K^+ \to \pi^+ 2 (e^+e^-))<{\cal O}(10^{-9})$~\cite{NA62:2023rvm} is also relevant for us, which we discuss in detail in subsection~\ref{Higgsconst}.

\subsection{ Constraints from $\pi^+ \rightarrow e^+ ~\nu_e ~a' \rightarrow e^+ ~\nu_e~e^+ ~ e^- $ }

The above decay channel puts a severe constraint on the $a'$ coupling to the quarks due to the strong bound on the  branching fraction, BR$(\pi^+ \rightarrow  e^+ ~\nu_e ~e^+ e^- )< 0.5 \times 10^{-10}$~\cite{SINDRUM:1986klz} at $90\%$ confidence level. We emulate  the calculation done for axions in  \cite{Krauss:1986bq,Liu:2021wap} and find that the branching fraction for our table \ref{tab} to be almost two orders of magnitude smaller than the current bound. This applies to our solution for MB and LSND alone.  However, when we try to include the solution of Atomki with LSND and MB  as in table \ref{tabNeW}, we face strong  constraints from the above decay channel. We find  the branching fraction for the table \ref{tabNeW} benchmark point to be $2.6 \times 10^{-10} $. This value violates the SINDRUM bound at 90$\%$ confidence level.  Figure \ref{fig:BMPregionPlot} shows a representative sampling of the parameter space.

\subsection{Constraints from the $Z$ decay width}
The LEP experiments measured the $Z$ boson decay width to be $\Gamma(Z\to {\rm All})= 2.4952\pm 0.0023$~GeV~\cite{ParticleDataGroup:2022pthAA}. Hence, any new physics contribution to this width should lie within  $1 \sigma$  of the observed central value.

We thus calculate the contributions to $\Gamma(Z)$ in our model. The $CP$-odd scalar $A$ and $CP$-even scalar $H$ masses are heavy and for the purpose of this calculation  we take them  to be around $\sim 300$~GeV, as shown in table~\ref{tab} and~\ref{tabNeW}.
Thus, while  the two-body decay process $Z \to AH$ is kinematically forbidden, a four-body decay to SM fermions mediated by these heavy scalars via their Yukawa couplings is possible. Similarly $Z \to a' H $ is also kinematically forbidden as $M_Z< M_{a'}+M_H$ (with $M_{a'} = 17$~MeV for our BP). However, the three body decay $Z \rightarrow a' H^* \rightarrow a' f \bar{f}$ is still possible, again via Yukawa couplings of $H$.
Note that the Yukawa couplings of $H, A$ in our model are small, as can be seen from the benchmark values in table~\ref{tabNeW} above. Both the 3-body and 4-body decay widths suffer suppression from heavy mediator masses in the propagator and also from phase space suppression. The 3-body decay width is further reduced by the tiny mixing angle $\sin \xi$.
Thus, any correction to the total width should be small. We have checked this by calculating\footnote{We use {\tt FeynRules}~\cite{Alloul:2013bka} to construct the model UFO files to do the relevant computations using {\tt MadGraph-3.5.3}~\cite{Alwall:2014hca}.} both the three-body and four-body decay widths for our BP. The three-body process yields $\Gamma(Z\to a'\,\bar{f}\,f )\approx 0.001$~MeV, while the four-body decay yields $\Gamma(Z\to 2\,\bar{f} \,2\,f ) \lesssim 10^{-6}$~MeV. Hence, these corrections to the $Z$ width lie comfortably within the presently allowed range.
\subsection{Higgs decay width constraints}\label{Higgsconst}
\begin{figure}[h!]
	\begin{center}
			{\includegraphics[scale=0.320]{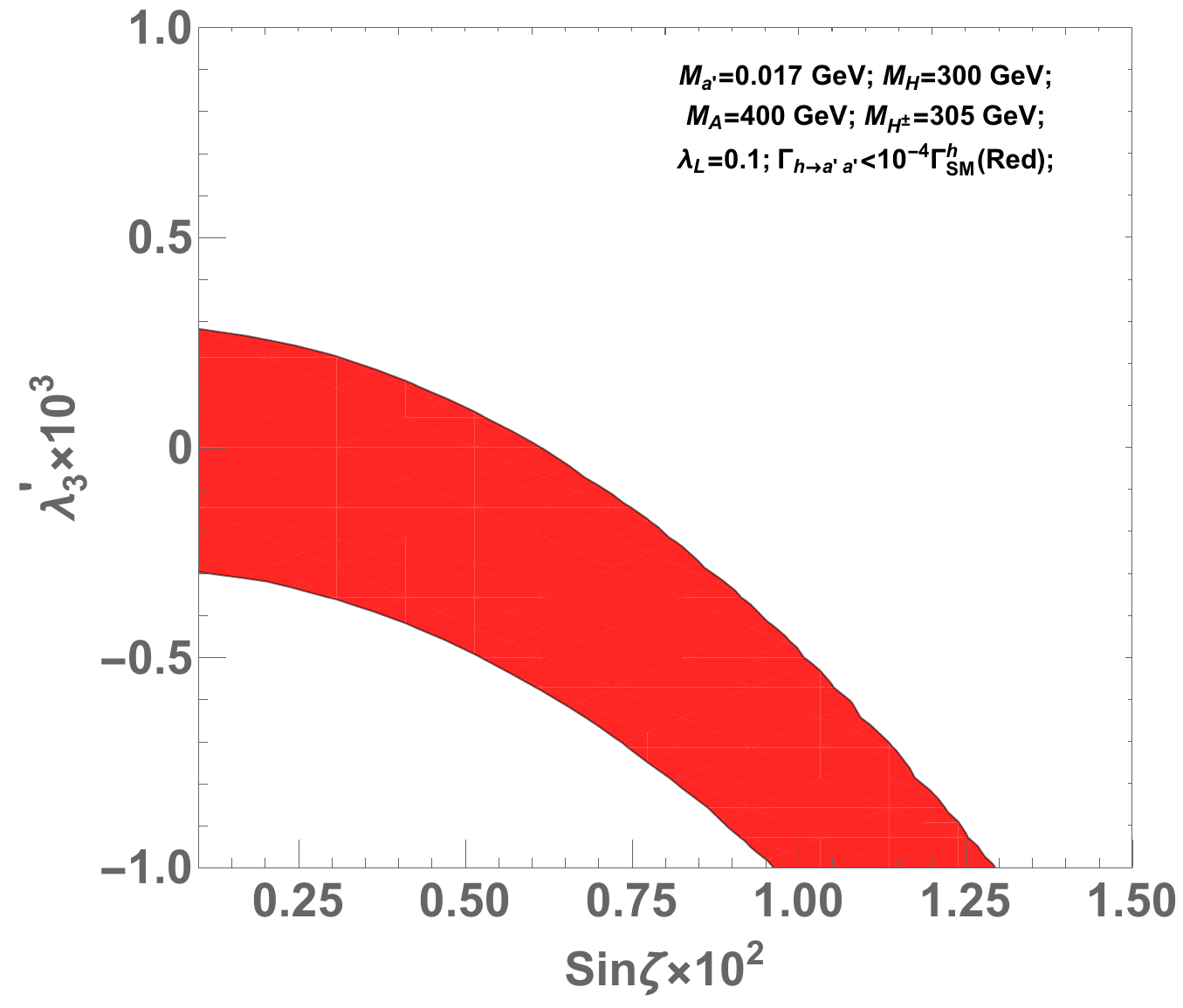}}
	\end{center}
\caption{The region of the parameter space in the $\sin\xi-\lambda^{'}_3$ plane for which the branching ratio ${\rm BR}(h\to a'a')<10^{-4}$ of $\Gamma_{\rm SM}^h$.}
\label{fig:Higg125decayBR}
\end{figure}

The channel ($h \rightarrow a' a'$ with $M_{a'}=17$~MeV)  can alter the total decay width of the SM Higgs. This partial decay width~\cite{Gunion:1989we} is given by
\begin{eqnarray}
\Gamma(h \to a'a')&=& \frac{g_{h a' a'}^{2}}{32 \pi M_h} \, \, \left( 1- 4\, \frac{M_{a'}^2}{M_h^2} \right)^{\frac{1}{2}}.
\end{eqnarray}
The coupling strength $g_{h a' a'}$  is dependent on several factors, including  $\lambda'_3, \sin\xi$, SM VEV $v$ and scalar masses. Detailed expressions for these dependencies can be found in eq.~\eqref{eq:strth1}. Measurements by CMS and ATLAS have put bounds on the Higgs total decay width of $3.2^{+2.4}_{-1.7}$~MeV~\cite{CMS:2022ley} and $4.5^{+3.3}_{-2.4}$~MeV~\cite{ATLAS:2023dnm}, respectively. Given the large error bars, however, stronger bounds result from the $h \rightarrow a' a'$ branching ratio. In our model, the  17~MeV $a'$ will promptly decay to $e^+ e^-$ with a decay length of about one centimeter, a requirement that primarily stems from MB and LSND. For decay lengths in the range  $\mathcal{O}(10^{-1}-10^3~{\rm cm})$, constraints on ${\rm BR}(h \rightarrow a' a' \rightarrow 2 e^+ \, 2e^-)$ can be derived from the CMS search for long-lived particles decaying to leptons~\cite{CMS:2021kdm}. This analysis bounds ${\rm BR}(h \rightarrow a' a') < 10^{-4}$ for $M_{a'}=30$ and 50~GeV with a decay length of $\mathcal{O}(1$ cm). Although our $a'$ is much lighter than the $a'$ masses assumed by CMS, we conservatively choose our BP in such a way that this bound on ${\rm BR}(h \rightarrow a' a')$ is satisfied. We have shown the acceptable region in the $\sin\xi-\lambda_3^\prime$ plane in figure~\ref{fig:Higg125decayBR}. Also, the decay mode $h\to a'a'$ leads to the charged kaon decay through $K^+ \to \pi^+ (h^*\to a'a')$ and for the product ${\rm BR}(K^+ \to \pi^+ (h^*\to a'a'))\times [{\rm BR}(a'\to e^+e^-)]^2$,  an upper limit of $2.1\times 10^{-9}$ is obtained for $M_{a'}=17$~MeV by the NA62 measurements~\cite{NA62:2023rvm}. To evade this constraint, the mixed-quartic coupling $\lambda'_3$ should be smaller than $10^{-2}$~\cite{Hostert:2020xku} which is consistent with our BP, where $\lambda'_3=10^{-3}$. In addition, using figure~3 in Ref.~\cite{He:2020jly}, one can estimate the product ${\rm BR}(K^+ \to \pi^+ (h^*\to a'a'))\times [{\rm BR}(a'\to e^+e^-)]^2$ to be of~${\cal O}(10^{-10})$. Moreover, as shown in figure~\ref{fig:Higg125decayBR}, $\lambda'_3$ can be vanishingly small. Hence, in the limit $\lambda'_3\to 0$, the ${\rm BR}(K^+ \to \pi^+ (h^*\to a'a'))$ is negligible.
\subsection{LHC signal strength constraints}
\label{subsec:diphoton}
In a model such as the one presented in this paper, with one (or more) additional $CP$-even scalar(s), the tree-level couplings of the SM Higgs-like scalar $h$ to SM fermions and gauge bosons are modified due to mixing. However, since we are working in the alignment limit, the mixing between the observed 125~GeV Higgs and the heavier $CP$-even scalar tends to zero, thus any constraint from the LHC experiments on such a mixing angle is satisfied trivially. 

Furthermore, the loop-induced $h \rightarrow \gamma \gamma$ decay is also affected by the presence of the charged scalars ($H^\pm$) of the second $SU(2)$ doublet. 
Despite its small branching fraction of $\sim 0.2 \,\%$~\cite{LHCHiggsCrossSectionWorkingGroup:2013rie}, the $h \to \gamma \gamma$ decay channel provides a clean final-state, leading to a signature that is distinct and easily recognizable against the SM background. Hence, this channel provides stringent constraints in the $g_{hH^+H^-} - M_{H^{\pm}}$ plane, where $g_{hH^+H^-}$ is the coupling of the SM-like Higgs boson with $H^{\pm}$, and in our model is equal to $\lambda_3 v$.

The Higgs to di-photon signal strength, under the narrow width approximation, can be expressed as follows~\cite{Djouadi:2005gj},
\begin{equation}
\mu_{\gamma\gamma}=\, \frac{\sigma(gg\rightarrow h)_{\rm BSM}}{\sigma(gg\rightarrow h)_{\rm SM}} \, \frac{{\rm BR}(h\rightarrow\gamma\gamma)_{\rm BSM}}{{\rm BR}(h\rightarrow\gamma\gamma)_{\rm SM}}.
\end{equation}
In our model, this formula reduces to
\begin{equation}
\mu_{\gamma\gamma} \approx \frac{\Gamma(h\rightarrow\gamma\gamma)_{\rm BSM}}{\Gamma(h\rightarrow\gamma\gamma)_{\rm SM}}.
\end{equation}

At one-loop level, the physical charged Higges $H^{\pm}$ add an extra contribution to the decay width $\Gamma(h\rightarrow\gamma\gamma)_{\rm BSM}$,
\begin{equation}
\Gamma(h\rightarrow \gamma\gamma)_{\rm BSM}=\frac{\alpha^2M_h^3}{256\pi^3 v^2}\Big|Q^2_{H^{\pm}} \, \frac{v \, g_{hH^+H^-}}{2M^2_{H^{\pm}}} \, F_0(\tau_{H^{\pm}})+C_{\rm SM}\Big|,
\end{equation}
where $C_{\rm SM}$ is the contribution from SM particles, $$ C_{\rm SM}=\sum_f N_f^c Q_f^2 y_f F_{1/2}(\tau_f)+y_W F_1(\tau_W) $$ with $\tau_x=\frac{M_h^2}{4 M_x^2}$. $Q$ stands for the electric charge of particles, and $N_f^c$ denotes the color factor. 
The loop functions $F_{(0,1/2,1)}(\tau)$ are  defined in Appendix~\ref{eq:loopf}, which also contains other calculational details.

The relevant coupling strength of $H^{\pm}$ to $h$ at the tree level is given by,
$g_{hH^+H^-} = \lambda_3 v= \, (2 (M_{H^\pm}^2-M_H^2)+v^2 \lambda_L) \, / \, v$, where $\lambda_L=\lambda_3+\lambda_4+\lambda_5$.  
When the mass gap between charged and $CP$-even scalar fields is large, the coupling $\lambda_3$, denoted as $g_{hH^+H^-} / v$, also becomes large. Consequently, this increase in coupling strength can potentially lead to violations of the observed Higgs to di-photon signal strength. It could also affect  unitarity constraints, which we will discuss later.
With a coupling strength of $\lambda_L=\mathcal{O}(0.01)$ and a relatively smaller mass gap between $M_{H^\pm}$ and $M_H$, the parameter $\lambda_3$ decreases. These choices of input parameters make our model consistent with the $1 \sigma$ range of the $h\rightarrow\gamma\gamma$ data.
We note that the quartic couplings $\lambda_3'$ and $\lambda_L$ (which includes $\lambda_{3,4,5}$) also influence the Higgs decay width via ${\rm BR}(h\to a'a')$. This can change the factor $\frac{\Gamma_{\rm SM}^{\rm Total}}{\Gamma_{\rm BSM}^{\rm Total}}$, and hence, $\frac{{\rm BR}(h\rightarrow\gamma\gamma)_{\rm BSM}}{{\rm BR}(h\rightarrow\gamma\gamma)_{\rm SM}}$. In our calculation, we choose ${\rm BR}(h\to a'a')<10^{-4}$, hence the effect is negligibly small as $\frac{\Gamma_{\rm SM}^{\rm Total}}{\Gamma_{\rm BSM}^{\rm Total}}\approx 1$. However, $\lambda_{3,4,5}$ also face significant constraints from stability and unitarity considerations, which we will discuss in the following subsections. 
\subsection{Vacuum stability}
The requirement that  the electroweak vacuum be bounded from below implies that the scalar potential does not tend towards negative infinity along any direction of the scalar field~\cite{Branco:2011iw}. When the field strengths approach infinity, the quartic terms in the scalar potential dominate. From~\cite{Chakrabortty:2013mha,Khan:2022kis}, the  conditions which bound our scalar potential from below  are
\begin{eqnarray}
   && \lambda_{1} \geq 0, \lambda_{2} \geq 0,  \lambda_{2}' \geq 0 ,\, \lambda _3+\lambda _4 - |\lambda _5|>-\sqrt{\lambda _1 \lambda _2}\,,\nonumber\\
    &&\lambda _3'>- \sqrt{2\, \lambda _1 \lambda _2'} , \, \lambda _3>- \sqrt{\lambda _1 \lambda _2}, \, \lambda _4'>- \sqrt{2\, \lambda _1 \lambda _2'}\,, \\
    && \lambda _3+\lambda _4>- \sqrt{\lambda _1 \lambda _2}.\nonumber
    \label{eq:abstb}
\end{eqnarray}
Above conditions are valid when $\lambda_6= \lambda_7 = \lambda_5^\prime =0$. The alignment limit demands $\lambda_6 \rightarrow 0$. For simplicity, we assume $\lambda_6= \lambda_7 = \lambda_5^\prime =0$, which does not affect the phenomenology of the model considered in this paper.

The quartic couplings $\lambda_{3,4,5}$ and $\lambda^\prime_{3,4}$ impose limits on the masses and mixing angles of the scalar sector of the model, particularly affecting the mass differences. For large positive values of $\lambda_{2}$ and $\lambda^\prime_{2}$, these bounds can be relaxed. However, large $\lambda$ values may lead to violation of unitarity bounds, which we  discuss below. Vacuum stability can also lead to strong bounds on Yukawa couplings since they contribute negatively to the beta function of scalar quartic couplings~\cite{Khan:2014kba, Khan:2015ipa, Khan:2022kis}. These considerations are not relevant, however,  for the tiny Yukawa couplings considered in our model.
\subsection{Unitarity}
A set of upper bounds on quartic couplings ($\lambda$'s) of the scalar potential can be imposed by the unitarity of the scattering matrix ($S$-matrix). One can derive the $S$-matrix from various $2 \rightarrow 2$ scattering processes involving gauge bosons, scalar particles, and interactions between gauge and scalar bosons. While deriving the $S$-matrix, one approximates the scatterings involving longitudinal gauge
bosons by Goldstone bosons in the high energy limit by using the Goldstone-boson equivalence theorem~\cite{Kanemura:1993hm, Arhrib:2000is}. In this limit, scatterings are dominated by four-scalar contact interactions.
The procedure for obtaining  unitarity bounds is described in  Appendix~\ref{ap:unitary}. Our model has three sub-S-matrices, and the corresponding eigenvalues are as follows.
{\small
\begin{eqnarray}
    &&\frac{1}{2} \left(\lambda _3-2 \lambda _4\right),  \frac{1}{2} \left(\lambda _3-2 \lambda _5\right), \frac{\lambda _3}{2}+\lambda _4, \frac{\lambda _3}{2}+\lambda _5,\,2 \lambda _2', \frac{1}{4} \left(\lambda _3+\lambda _4-3 \lambda _5\right),\lambda _3+\lambda _4-\frac{\lambda _5}{2}, \nonumber\\
&& \frac{1}{4} \left(\lambda _1+\lambda _2 \pm \sqrt{\lambda _1^2-2 \lambda _2 \lambda _1+\lambda _2^2+16 \lambda _4^2}\right),~\frac{1}{4} \left(\lambda _1+\lambda _2 \pm \sqrt{\lambda _1^2-2 \lambda _2 \lambda _1+\lambda _2^2+16 \lambda _5^2}\right), \nonumber\\
&&\frac{1}{4} \left(\lambda _3'+\lambda _4' \pm \sqrt{\left(\lambda _3'\right){}^2-2 \lambda _4' \lambda _3'+\left(\lambda _4'\right){}^2+4 \left(\lambda _5'\right){}^2}\right),~ \frac{1}{2} \left(\lambda _1+\lambda _2 \pm \sqrt{\left(\lambda _1-\lambda _2\right){}^2+4 \lambda_4^2}\right),\nonumber\\
&& \frac{1}{2} \left(3 \lambda _1+3 \lambda _2 \pm \sqrt{9 \left(\lambda _1-\lambda _2\right){}^2+4 \left(2 \lambda _3+\lambda _4\right){}^2}\right),~\frac{1}{2} \left( \lambda _1+\lambda _2 \pm \sqrt{\left(\lambda _1-\lambda _2\right){}^2+4 \lambda _5^2} \right),\\
&& \frac{1}{8} \left(5 \lambda _3+5 \lambda _4+7 \lambda _5 \pm \sqrt{9 \lambda _3^2+18 \left(\lambda _4-\lambda _5\right) \lambda _3+73 \lambda _4^2+73 \lambda _5^2+110 \lambda _4 \lambda _5}\right),\nonumber\\
&& \frac{1}{8} \left(\lambda _3'+\lambda _4' \pm \sqrt{\left(\lambda _3'-\lambda _4'\right){}^2+16 \left(\lambda _5'\right){}^2}\right).\nn
\end{eqnarray}
}
 Unitarity demands that the eigenvalues of the $S$-matrix must remain below $8\pi$, ensuring the model's consistency and validity at extreme energy scales. The results from unitarity considerations for our model are shown in figure~\ref{fig:Bound}.
\subsection{Constraints from electroweak precision experiments}
New physics phenomena beyond the scale of $W$ and $Z$ masses can significantly influence electroweak precision bounds. If such physics contributes dominantly via virtual loops to precision observables, its effects can be parametrized by three-gauge boson self-energy parameters called the oblique parameters. They were introduced and extensively investigated by Peskin and Takeuchi~\cite{Peskin:1991sw}, Altarelli et al.~\cite{Altarelli:1993bh}, and Baak et al.~\cite{Baak:2014ora} and are denoted by $S$, $T$, and $U$. The $S (S + U)$ parameter describes new physics contributions to neutral (charged) current
processes at different energy scales. The $T$ parameter quantifies differences between the neutral and charged weak currents. The $U$ parameter primarily reflects variations in the mass and width of the $W$-boson. Its contribution is small and subdominant for most new physics cases~\cite{Baak:2011ze, Arhrib:2012ia, Barbieri:2006dq}, including the model presented here, and we neglect it in our following calculations.
All the relevant contributions to the oblique parameters in our model arise from the scalars in the second Higgs doublet, while the singlet scalar contributions are negligible. One may then write
\begin{equation}
S=S_{\rm SM}+\Delta S_{\rm SD}, ~T=T_{\rm SM}+\Delta T_{\rm SD},~U=U_{\rm SM}+\Delta U_{\rm SD}.\nn
\end{equation}
 The notation SD denotes scalar doublet contributions, which, with $\sin\xi \sim 0$, can be expressed as~\cite{Arhrib:2012ia,Barbieri:2006dq},
\bea
\Delta S_{\rm SD}&=&\frac{1}{2\pi}\Big[\frac{1}{6}\ln\frac{M_{H}^2}{M_{H^{\pm}}^2}-\frac{5}{36}+\frac{M_{H}^2M_{A}^2}{3(M_{A}^2-M_{H}^2)^2}+\frac{M_A^4(M_{A}^2-3M_{H}^2)}{6(M_{A}^2-M_{H}^2)^3}\ln \frac{M_{A}^2}{M_{H}^2} \Big],\\
\Delta T_{\rm SD}&=&\frac{1}{32\pi^2\alpha v^2}\Big[F(M^2_{H^{\pm}},M^2_{H})+F(M^2_{H^{\pm}},M_{A}^2)-F(M_{A}^2,M^2_{H})\Big].
\eea
 The two-argument loop function $F$ above is given by
\begin{align*}
F(x,y) =
\begin{cases}
\frac{x+y}{2}-\frac{xy}{x-y}\ln (\frac{x}{y}) & {x \neq y}\\
0 & {x= y}
\end{cases}\,.  
\end{align*}
We now constrain the new parameters by using the next-to-next-to-leading order (NNLO) global electroweak fit results obtained by the Gfitter group~\cite{Haller:2018nnx}. Their study yields significant restrictions on the parameter space. The relevant constraints extracted from the fit at a confidence level of $95\%$ are~\cite{Haller:2018nnx} 
\begin{align*}
 \Delta S_{\rm SD}<0.04\pm0.11, && \Delta T_{\rm SD}<0.09\pm0.14, && \Delta U_{\rm SD}<-0.02\pm0.11  \, .
\end{align*}

\begin{figure}[t]
	\begin{center}
			{\includegraphics[scale=0.40]{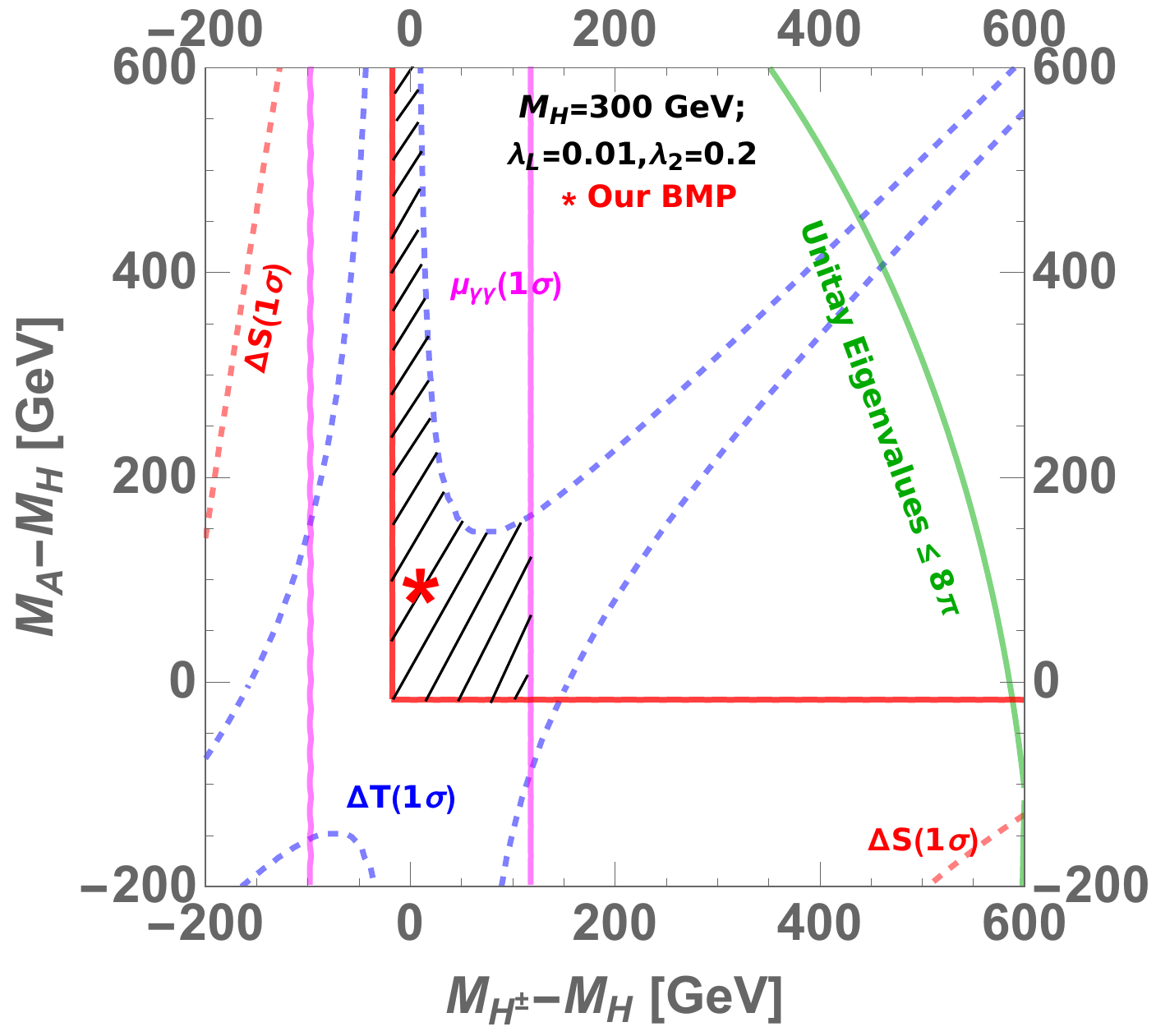}}
	\end{center}
\caption{The contour plot shows the allowed region from $i)$ the $S$ and $T$ parameters, $ii)$ stability and $iii)$  unitarity considerations. The allowed region from the Higgs di-photon signal strength data is  shown  between the magenta lines. For our BP, the most stringent unitarity bound (green line) comes from the eigenvalue: $  \frac{3}{2} (\lambda _1+ \lambda _2) +$ $\frac{1}{2} \sqrt{}\{9 (\lambda _1-\lambda _2)^2+$ $4 (2 \lambda _3+\lambda _4)^2\} $ $\leq 8\pi$. The red horizontal (solid red) line is the stability  bound came from $\lambda_3+\lambda_4-|\lambda_5| \geq$ $ -\sqrt{\lambda_1 \lambda_2}$, whereas the vertical (solid red) line originates from $\lambda_3\geq $ $-\sqrt{\lambda_1 \lambda_2}$. The hatched region is what is allowed after considering all constraints. The red $``star"$ mark is our BP corresponding to table~\ref{tab} and~\ref{tabNeW}.}
\label{fig:Bound}
\end{figure}

The contour plots in figure~\ref{fig:Bound} show the permissible area for certain parameters. We have generated these plots by fixing $M_H = 300$~GeV and varying the masses of $A$ and $H^{\pm}$.
We also assume the quartic coupling $\lambda_2=0.2$ and from the SM we have $\lambda_1=\frac{M_h^2}{2 v^2}$. We also set $\lambda_L=\lambda_3+\lambda_4+\lambda_5=0.01$ for this analysis. The bounds could vary for different choices of $\lambda_2$ and $\lambda_L$.
Within the plot, the space between the blue dashed lines represents the $1 \sigma$ range of the $T$-parameter, while the area between the red dashed lines signifies the $1 \sigma$ range of the $S$-parameter. The region to the left of the green line indicates adherence to unitarity constraints. The entire area, above and to the right of the solid red lines, in the $(M_{H^\pm} - M_H) - (M_A - M_H)$ plane is allowed by stability considerations.
Notably, the most rigorous constraint arises from the Higgs di-photon signal strength data, shown as the region between the two magenta lines.  The hatched area in figure~\ref{fig:Bound} is allowed after considering all constraints. Our chosen benchmark point of table~\ref{tab} and~\ref{tabNeW} is shown by the red star.
\subsection{Bounds on heavy Higgs masses from LHC searches}
Present LHC searches can tightly constrain the singlet and doublet scalar masses via their couplings to the SM fermions. Yukawa couplings relevant to their production and decay are denoted by $\overline{X}$s in eq.~\eqref{eq:YkN}.

For our choice of the benchmark point, as outlined in Table~\ref{tabNeW}, the heavy $CP$-even (odd) scalar, $H \, (A)$, almost entirely decays to $c \bar{c}$. These decays of $H$ and $A$ will manifest as resonant di-jet signals at the LHC. However, in all the resonant di-jet analyses performed by the LHC collaborations, none go down to 300~GeV mass due to large SM backgrounds. The 2016 CMS analysis is the closest but goes down to the di-jet invariant mass of 500 GeV only~\cite{CMS:2016ltu}. Hence, our benchmark point remains unconstrained from resonant di-jet bounds from the LHC. We also checked that for our benchmark point, the production cross-section of the $p \bar{p} \rightarrow H/A \rightarrow jj$ process at the Tevatron is well below the resonant di-jet cross-sections probed by CDF~\cite{CDF:2008ieg}.

Similarly, the heavy charged scalar, $H^{+} \, (H^{-})$, decays to $c \bar{s}$  ($s \bar{c}$) only. Hence, our values for charged scalar masses $M_{H^\pm}$ are consistent with limits obtained from processes such as $pp\to H^\pm t\bar{b} (b\bar{t})$, $H^\pm \to t\bar{b} (b\bar{t})$~\cite{ATLAS:2021upq} at the LHC. The heavy Higgs bosons belonging to a $SU(2)$ doublet can also be produced at the LHC by vector boson fusion (VBF) processes followed by their decay into weak gauge bosons. The bounds from these processes are quite strong~\cite{CMS:2019bnu}. However, these bounds are not relevant for us since the heavy Higgs bosons in our model can not be produced by VBF processes as the VEV of the second Higgs doublet is zero.  

\section {Anomalous magnetic moments of the electron and muon}
\label{sec8}
Since the interaction of $a'$ with electrons plays a significant  role in our explanation of the three low-energy anomalies which are the subject of this work,
it is important to calculate the effect this may have on  $a_e$. Currently there are discrepancies between the SM prediction and the experimentally observed value of this parameter. The latest experimental measurement of $a_e^{\text{Exp}}$ gives the value  $115~965~2180.73\times10^{-12}$~\cite{Hanneke:2008tm}. The theoretical SM estimation of $a_e$~\cite{Aoyama:2012wj, Aoyama:2017uqe}, however,  depends on the experimentally measured value of the fine-structure constant, $\alpha$.
In recent years, three separate measurements of this parameter have been carried out.
Two of these measurements were carried out using the Rubidium (Rb) atomic interferometry~\cite{Bouchendira:2010es, Morel:2020dww}
and one used Caesium (Cs) atomic interferometry~\cite{Parker:2018vye}. The $\alpha$ measurements from  Rb atomic interferometry, carried out in 2010 and 2020, respectively, resulted in  $\Delta a_e$ = $(-13.0 \pm 7.7) \times 10^{-13}$ and $(4.8 \pm 3.0) \times 10^{-13}$. These results had significances of $-1.7\sigma$ and 
$+1.6\sigma$, respectively. The  Cs atomic interferometry experiment (2018)  led to $\Delta a_e$ = $(-8.8 \pm 3.6) \times 10^{-13}$  with a significance of $-2.5\sigma$. Thus, depending on the measured value of $\alpha$ that one chooses, the tension between $a_e^{\text{Exp}}$ and the prediction of the SM can be either positive or negative.

The one-loop contribution from neutral (pseudo)scalars  to $(g_\ell-2)$, with $\ell=e,\mu$ is
\beqa
a_\ell^\textrm{one-loop} = \frac{1}{8  \pi^2} \sum_{\phi = H , A,a'} (y^\phi_\ell)^2  F_\phi \left(  \frac{M_\phi^2}{M_\ell^2} \right),
\label{oneloop}
\eeqa
where,
\beqa
F_H(x) = \int_0^1 dz\frac{(1-z)^2(1+z)}{z x+(1-z)^2} \,, \qquad \qquad 
F_{A,a'}(x)= -\int_0^1 dz\frac{(1-z)^3}{z x+(1-z)^2}. 
\label{oneloopfuc}
\eeqa
 The Barr-Zee two-loop diagram contributions from them~\cite{Chang:2000ii,Larios:2001ma,Ilisie:2015tra} is
\beqa
a_\ell^\textrm{two-loop} = \frac{\alpha}{8\pi^3 } \sum_{\phi= H,A,a'} \sum_{f=t,b,\tau} N^f_c Q_f^2 y_\ell^\phi y_f^\phi \frac{M_\ell M_f}{M_\phi^2}  G_\phi \left(\frac{M_f^2}{M_\phi^2} \right),
\label{twoloop}
\eeqa
where $N_c^f$ is the number of colours with $N_c^f=1(3)$ for leptons (quarks), $Q_f$ the electric charge of fermion and
\beqa
G_H(x) = \int_0^1 d z \frac{2 z(1-z) - 1}{z(1-z)-x} \log \frac{z(1-z)}{x} \,,~~~~~~~G_{A,a'}(x) = \int_0^1 d z \frac{\log \frac{z(1-z)}{x}}{z(1-z)-x}  \,.
\label{twoloopfuc}
\eeqa 
As mentioned above, $\Delta a_{e}$ could be either positive or negative, 
hence the role of contributions coming from both one-loop and two-loop levels could be significant. This is because  at  one-loop, the integrals $F$ become positive (negative) for $CP$-even ($CP$-odd) scalars, as described in eq.~\eqref{oneloopfuc}, while the two-loop integral functions $G$,  described in eq.~\eqref{twoloopfuc}, can become either negative or positive for both $CP$-odd and $CP$-even scalars. Thus, a 17~MeV pseudoscalar will always make a negative contribution to $\Delta a_{e}$ at the one-loop level.  Since the doublet-like scalars are heavy, their contributions can also remain small at the one-loop level, despite their couplings ($y_{e}^{H,A,H^{\pm}}$) not being $\sin\xi$ suppressed.

At the two-loop level, contributions depend linearly on  the product  of $y_\ell^\phi$ and $y_f^\phi$
(whereas the one-loop contributions depend  on  $(y_\ell^\phi)^2$). Thus, depending on the various sign combinations of these two parameters, the  contributions of $CP$-even and $CP$-odd scalars can change signs, as long as they maintain a relative negative sign between them.
 Using the BP values in table~\ref{tabNeW}, we find that the total (one-loop plus two-loop) contribution to the electron $g-2$ is $\Delta a_e \sim - \, {\cal {O}}(10^{-14})$, which is quite small. 

Next, we note that the excesses observed by MB, LSND, and ATOMKI can also be accounted for by considering $y_{e}^{a'} = 1.5 \times 10^{-4}$ (i.e., a value one order of magnitude larger than that in table~\ref{tabNeW}), while keeping all other parameters same as table~\ref{tabNeW}. (This change does not violate  the constraints outlined in section~\ref{sec6}.)
Due to this change, the absolute value of the one-loop contribution, given its quadratic dependence on the coupling,  will exceed  the two-loop correction.
One then obtains  $\Delta a^{\textrm{one-loop}}_{e} \simeq -1.34\times 10^{-12}$. We note that the two-loop contributions of   $M_A$, $M_{a'}$ and $M_H$
 offset each other 
 and yield  a small positive contribution of $\Delta a^{\textrm{two-loop}}_e \simeq 1.11 \times 10^{-14}$  for $y_{\tau}^{H/A} = 0.3$. This demonstrates that 
 it is possible to  align with observations from Rb atomic interferometry  (2010) and Cs atomic interferometry  (2018).
Varying the masses of the $M_H$, $M_A$ and $\sin\xi$, the two-loop correction  can be positive, and of order $10^{-12}$. 
The negative contribution from the one-loop and the positive contribution from the two-loop being of the same order, can partially offset each other and induce a positive $\Delta a_{e} \sim  10^{-13}$ which is close to the Rb atomic interferometry 2020 result.
Thus, in this model it is possible  to address both the positive and negative discrepancies of $\Delta a_{e}$ while concurrently explaining the MB, LSND, and ATOMKI excesses, and satisfying all other constraints.

$a_{\mu}$ can receive contributions from a non-zero  value of $y_{\mu}^{H, A, a'}$. However, there is appreciable  uncertainty in the estimation of its theoretical prediction~\cite{Aoyama:2020ynm,Aoyama:2012wk,Aoyama:2019ryr,Czarnecki:2002nt,Gnendiger:2013pva,Davier:2017zfy,Keshavarzi:2018mgv,Colangelo:2018mtw,Hoferichter:2019mqg,Davier:2019can,Keshavarzi:2019abf,Kurz:2014wya,Melnikov:2003xd,Masjuan:2017tvw,Colangelo:2017fiz,Hoferichter:2018kwz,Gerardin:2019vio,Bijnens:2019ghy,Colangelo:2019uex,Blum:2019ugy,Colangelo:2014qya, Borsanyi:2020mff,Muong-2:2021ojo, Muong-2:2006rrc,Muong-2:2023cdq}.  The hadronic vacuum polarization (HVP) contribution to $a_{\mu}$ obtained using dispersion theory, with its value extracted from the precise experimental measurement of the cross-section of the low-energy
process  ($e^{+} e^{-} \rightarrow$ hadrons)~\cite{Aoyama:2020ynm, Keshavarzi:2018mgv,Colangelo:2018mtw,Hoferichter:2019mqg,Davier:2019can,Keshavarzi:2019abf,Kurz:2014wya,CMD-3:2023alj, CMD-3:2023rfe,BaBar:2012bdw, BESIII:2015equ, KLOE-2:2017fda,Masjuan:2023yam}, differs from the lattice-QCD calculation~\cite{Borsanyi:2020mff,Boccaletti:2024guq}. Thus,  the current discrepancy between $a_\mu^{\rm SM}$ and $a^{\text{avg(Exp)}}_{\mu}$ can vary between a  significance of $5.1 \sigma$ to 0.9$\sigma$ depending on one's choice.\footnote{ If this discrepancy between the cross-section estimation for ($e^{+} e^{-} \rightarrow$ hadrons) from low-energy experiments and the lattice-QCD calculation persists, new physics contributions~\cite{Lehner:2020crt,Crivellin:2020zul,Keshavarzi:2020bfy,deRafael:2020uif, Coyle:2023nmi}  will be necessary to explain it.} In either case,   one needs a positive contribution to $\Delta a_{\mu}$ of order $10^{-9}$ from  new physics to explain it.
In our model, for the choice of $M_H$, $M_A$, $M_{a'}$ and $\sin\xi$ required for our fits to MB and LSND,  $\Delta a^{\textrm{one-loop}}_{\mu}$ always remains negative. 
A positive large $a^{\textrm{two-loop}}_{\mu}$ can be obtained by considering the couplings of the heavy fermion with the scalars to be relatively large. As we discuss in section~\ref{sec5}, constraints coming from the study of the kaon decay restrict these couplings and prevent a  $\Delta a^{\textrm{two-loop}}_{\mu}$ contribution of order $10^{-9}$.
Thus, we always get an overall negative contribution to $\Delta a_{\mu}$. This is small  ($\sim - 10^{-11}$), since  the couplings of the muon with the scalars, being free, can be kept  small as in table~\ref{tabNeW}.

\section{Tests of the model}
\label{sec9}
 Several models proposed to explain the excess in MB lead to the creation of $e^+e^-$ pairs in the detector.  The earliest hints of such pair creation could come from MicroBooNE,  an 85~t active volume Liquid Argon Time-Projection Chamber (LArT-PC)  placed  470~m away from the target.  While no longer running, MicroBooNE is part of the Short Baseline Neutrino (SBN)  program~\cite{MicroBooNE:2015bmn,Machado:2019oxb} at Fermilab, and was fed by its  Booster Neutrino Beam (BNB), which provided  MB with its flux. MicroBooNE has already conducted important tests to restrict or eliminate SM possibilities of the MB excess~\cite{MicroBooNE:2021tya,MicroBooNE:2021pvo,MicroBooNE:2021wad,MicroBooNE:2021nxr}, however, further results from its analysis are expected. Given its superior particle identification capability, its ongoing search for $e^+e^-$ pairs as the origin of the MB excess and its capability to measure their invariant mass, it's new physics analyses may provide the earliest indications for or against the model proposed~here.

 Two additional detectors are part of SBN: the ICARUS detector, currently operating with a fiducial mass of 760 t, located 600 m from the BNB target, and the Short-Baseline Near Detector (SBND), with a 112-t active volume at a distance of 110 m, which has recently started collecting data. Over the next few years, these detectors in conjunction have the capability to test various new physics proposals for the excess, including those involving sterile-active oscillations, additional photons, electrons and $e^+e^-$ pairs. 

 Given the energy of the neutrino beams at MicroBooNE and considering that the $e^{+} e^{-}$ pairs in our model originate from a boosted light pseudoscalar particle with a mass of 17~MeV, the opening angle between the $e^{+} e^{-}$ would likely remain smaller than the detector's resolution threshold for distinguishing them separately. Thus, our events can at first sight appear as single-electron or photon events. However, since   MicroBooNE can measure the rate of the energy loss per unit length ($dE/dx$) for an electromagnetic shower, it may be possible to distinguish such events from single-electron events~\cite{MicroBooNE:2018dfn}. Secondly, if the detector can measure invariant masses of $20$~MeV, it may be possible to separate it from a photon signal. Finally, if a given event has vertex-pointing, a photon gap is expected to be significantly larger in length prior to the onset of pair-production compared to one created by the pseudoscalar, which is of ${\cal O}(1)$~cm or less. With these criteria,  one can possibly also separate these events from single electron or photon backgrounds.

Several experiments are planned to test the ATOMKI result by studying  IPC from nuclear transitions,
as the experiment at ATOMKI did. We mention two of them here, and refer the reader to~\cite{Alves:2023ree} for a much more extensive discussion.
The MEG-II experiment~\cite{MEGII:2018kmf} at the Paul Scherrer Institute in Switzerland, plans to repeat the
$^7$Li$(p, e^+ e^-)$ $^8$Be measurement. The PADME experiment~\cite{Raggi:2014zpa} will explore the region $16.4$~MeV $\leq M_{a'}\leq 17.4$~MeV in its search for the 17~MeV pseudoscalar.
 Additionally, the kaon DAR search, planned at the JSNS$^2$ experiment~\cite{Ajimura:2017fld,JJordan} is in a position to provide a possible test of the proposal presented in this work via its flux of high energy muon neutrinos.

 The mass of $a'$ lies close to, but outside of,  existing bounds from electron beam-dump experiments like E141~\cite{Riordan:1987aw}. Hence it is feasible that this region in mass and coupling may be covered in the near future by other experiments like HPS~\cite{Battaglieri:2014hga}. 
 Finally, we mention an existing experimental hint which could be indicative, \ie, a significant excess in the $10-20$~MeV invariant mass-bin of electron-like FGD1-TPC pairs detected by the T2K ND280 detector, (see figure 11 in~\cite{T2K:2020lrr}).
 \section{Discussion and Conclusions}
 \label{sec10}
 The MB and LSND excesses of electron-like events are long-standing and statistically significant anomalies which have defied attempts to explain them using standard physics, even though stringent  checks of backgrounds and  systematic errors have been conducted. They may be indicative signals of new physics.
 Additionally,  we note that if one seeks a common new physics solution to MB and LSND, scalar mediators do much better than vector ones. This conclusion follows from a comparative study~\cite{Abdallah:2022grs} of the up-scattering cross-section  both above and below MB energies.

 The ATOMKI experiment on rare nuclear transitions  requires a new, 17~MeV mediator to explain the excess seen by it. The possible parity and angular momentum conserving choices are a vector, axial vector or a pseudoscalar, decaying into $e^+e^-$ pairs  to explain the bumps in  its data via new physics. Taking cognizance of this, we have presented a two-Higgs doublet model with a dark singlet  17~MeV pseudoscalar ($a'$).
 We first show that independent of ATOMKI, such a model provides very good fits of MB and LSND excesses. It is  useful to compare  our work in this paper with an earlier attempt with a similar model. It was 
 shown in~\cite{Abdallah:2020vgg} that a second Higgs doublet with a light $(\sim 750$~MeV) $CP$-even scalar, plus  a real, 17~MeV singlet scalar mediator which predominantly decayed to $e^+e^-$ pairs could provide very good fits to datasets of both MB and LSND. At this point it is relevant to note that the observed angular distribution of MB is an important guide to making choices about the new physics model. This distribution is significantly forward, but also has excess events in angular bins in almost all other directions. In~\cite{Abdallah:2020vgg}, such a distribution was achieved by combining the predominantly coherent and forward events generated by the real 17~MeV scalar with a heavier 750~MeV $CP$-even doublet scalar, where the latter helped populate non-forward bins. In this work, the spin dependant couplings and consequent incoherent scattering  of the pseudoscalar $a'$ alone help achieve the necessary angular distribution in MB, and allows us to also fit LSND simultaneously.
All the scalars of the second doublet stay heavy, with masses well above $M_W$ and  $M_Z$, allowing for a natural hierarchy, because the need for the $CP$-even doublet scalar to have a lower mass ($\sim 750$~MeV) is removed. Table~\ref{tab} displays representative benchmark couplings necessary to achieve this solution.

 We next attempt to use the model to fit ATOMKI in addition to MB and LSND.
 We first identify the $u,d,e$ couplings of the pseudoscalar required to satisfy the ATOMKI anomaly, and then use these values as inputs for the interaction (shown in figure~\ref{fig:MB_LSND_Plot}) in MB and LSND, and obtain very good fits of their data, albeit with quark couplings to the $a'$ which are considerably higher than in the previous case described above (see table~\ref{tabNeW}). Relevant constraints which consequently come into play are related to  kaon and pion decay. While we find a large parameter space satisfying kaon decay bounds, our solution (for all three anomalies combined) does remain in tension with the SINDRUM $90\%$ CL bounds on $\pi^+ \rightarrow $ $ e^+ ~\nu_e ~e^+ e^- $.

We compute the contributions to the electron and muon $g-2$ up to two loops. In our scenario, it is possible to explain the currently observed discrepancies in the anomalous magnetic moment of the electron, in addition to addressing the MB, LSND, and ATOMKI excesses, while satisfying all other constraints; however, extending the present model is necessary to explain the observed anomalous magnetic moment of the muon.
  Finally, we have taken into consideration both collider and non-collider constraints on the model and discussed future tests which may validate or invalidate it.

We conclude with a few general comments:

$a)$ The essential elements of the gauge invariant  model presented here are the dark singlet $a'$ and the $N_i$, which are the  sub-GeV ($N_1, N_2$) and GeV-scale ($N_3$) neutral leptons. These  are directly  related to the three low energy anomalies we have made an effort to explain.  As in~\cite{Abdallah:2020vgg}, the 17~MeV singlet provides a portal to the dark sector, and the  three heavy singlet neutrinos help obtain masses for the SM neutrinos via a see-saw mechanism, yielding mass-squared differences in agreement with global oscillation data. Additionally, two of these three singlet neutrinos participate in the interaction used to explain the MB and LSND excess as shown in figure~\ref{FD-SP-LSND-MB}.

$b)$ The heavy sector of the model, $\ie$ the second Higgs doublet, is an important but supplementary part of the solution presented here.  The role of the doublet in this model is to provide a simple mechanism for  the $a'$ to obtain its couplings to the SM fermions in a gauge invariant manner, via its mixing with the heavier pseudoscalar $A$. In nature, it is not inconcievable that this supplementary role is played instead by elements of an extended dark sector interacting feebly with SM particles. 

$c)$ While the excesses seen by MB and LSND are long standing and have been well scrutinized,  our attempt to obtain a common solution for all three anomalies rests on the assumption that the ATOMKI results, which are relatively recent,  are a genuine signal of a  new physics mediator with mass 17 MeV. Further independent tests are clearly necessary to confirm this. In the near future, these include the MEG-II experiment~\cite{MEGII:2018kmf} as well as the PADME experiment~\cite{Raggi:2014zpa}.

$d)$ Finally, we note that if the results of  previous work \cite{Abdallah:2020vgg} on a (real) scalar extension of the SM in order to resolve the MB and LSND anomalies are viewed in conjunction  $a)$ with the pseudoscalar solution for these two anomalies as well as for ATOMKI  presented in this work
and $b)$ with the fact that a real scalar can explain the $^{12}$C nuclei ATOMKI result (as noted in Section \ref{sec5} ) then an important possibile conclusion emerges, \ie that a new complex singlet mediator of mass $17$ MeV, with both real and pseudoscalar parts, could be responsible for an explanation of all three low-energy anomalies considered here (\ie~MB, LSND and the ATOMKI $^8$Be, $^4$He and $^{12}$C results). The specific scalar and pseudoscalar couplings of such a new object to SM quarks and leptons  could then be determined by combining the observations of the these experiments with those of a high statistics LAr-TPC experiment like SBND, where such an object, if it exists, would be copiously produced.

\section*{Acknowledgements} 
Raj Gandhi is thankful to William Louis for his patient help with our many questions on MB, LSND and MicroBooNE. He also acknowledges helpful discussions with Andre de Gouvea, Joachim Kopp, Pedro Machado and Rahul Srivastava. Tathagata Ghosh and Raj Gandhi would like to acknowledge support from the Department of Atomic Energy, Government of India, for Harish-Chandra Research Institute. Najimuddin Khan expresses gratitude to Mohammad Sajjad Athar for providing access to the computational facilities at AMU. Samiran Roy is
supported by the NPDF grant (PDF/2023/001262) from SERB, Government of
India. Subhojit Roy is supported by the U.S. Department of Energy under contracts No. DEAC02 - 06CH11357 at Argonne National Laboratory. He would like to thank Carlos Wagner for various useful discussions on the anomalous magnetic moment of the muon and electron and the issues related to the kaon decay. He also thanks Zelimir Djurcic for useful discussions on MicroBooNE.

\appendix
\section*{Appendices}
\addcontentsline{toc}{section}{\protect\numberline{}Appendices}%
\section{Important vertices}
Vertices that are used in our calculation are shown  below:
\begin{eqnarray}    \label{eq:strth1}
    g_{hHH} &=&  i \,  \lambda_L v, \nonumber\\[0.2cm]
    g_{hAA} &=& \frac{i}{2} \left(\frac{\left(M_{a'}^2-M_A^2\right)}{v}\sin ^2 2\xi   - 2 \,     \lambda'_3 v \sin^2\xi -2  \left(\lambda _L-2\lambda _5\right)   v\cos^2\xi\right), \nonumber\\[0.2cm]
    g_{hAa'} &=& -\frac{i}{2} \left(\frac{\left(M_{a'}^2-M_A^2\right)}{v}\cos 2\xi  \, \sin 2\xi   +      \left(\lambda _L-2\lambda _5- \lambda'_3 \right) v\sin 2\xi  \right), \\[0.2cm]
    g_{ha'a'} &=&  -\frac{i}{2} \left( \frac{\left(M_{a'}^2-M_A^2\right)}{v} \sin ^2 2\xi  +2     \lambda'_3 v\cos^2\xi +2  \left(\lambda _L-2\lambda _5\right)   v\sin^2\xi\right),  \nonumber
\end{eqnarray}
with
\begin{eqnarray*}
\lambda_L&=&\lambda_3+\lambda_4+\lambda_5,\\
\lambda_3&=& \dfrac{1}{v^2}(-2 M_H^2+2 M_{H^\pm}^2+v^2 \lambda _L),\\ 
 \lambda_4&=&  \dfrac{1}{v^2}(M_{{A}}^2 \cos^2\xi+ M_{a'}^2 \sin^2\xi + M_H^2-2 M_{H^\pm}^2),\\
 \lambda_5&=& \dfrac{1}{v^2} (M_H^2-\cos^2\xi M_{A}^2 - \sin^2\xi M_{a'}^2).
\end{eqnarray*}
\begin{eqnarray}
g_{Z_\mu AH}  &=& \frac{\sqrt{g_1^2+g_2^2}}{2}\cos\xi \, (p_\mu^{A} - p_\mu^{H}),~~
g_{Z_\mu a'H} =  \frac{\sqrt{g_1^2+g_2^2}}{2}   \sin\xi\,  (p_\mu^{a'} - p_\mu^{H})    \nonumber\\
g_{W_\mu^\pm H^\mp H} &=&  i\,  \, \frac{g_2}{2}  \, (p_\mu^{H^\pm} - p_\mu^{H}) ,~~ g_{W_\mu^\pm H^\mp A} =   \frac{g_2}{2}  \, (p_\mu^{H^\pm} - p_\mu^{A}), ~~
g_{H^\pm H^\mp h} = i\, \lambda_3 \, v \, .
    \label{eq:strth2}
\end{eqnarray}
\begin{table}[h!]
\begin{center}\scalebox{0.9}{
 \begin{tabular}{|c|c|c|c|c|c|c|c|c|}
  \hline
  $\lambda_1$& $\lambda_2$ & $\lambda_L$&$\lambda_3$ &$\lambda_4$&$\lambda_5$ & $\lambda_6 =\lambda_7 = \lambda_5^{\prime} $&$\lambda_3^\prime$& $\lambda_4^\prime$\\
  \hline
   $\frac{M_h^2}{2 v^2}\approx 0.13$  & $0.2$ & $0.01$ & $0.109$ &$0.553$ & $-0.653$ & $0$ & $0.001$& $0.1$\\[0.1cm]
  \hline
 \end{tabular}}
\caption{The quartic couplings corresponding to our BP as in table~\ref{tab} and~\ref{tabNeW}.}
\label{tab:lam}
\end{center}
\end{table}
\section{Higgs di-photon signal strength}
\label{eq:loopf}
Consider the following ratios of  the production cross-sections  $\sigma(pp\to h)$ and decay widths  $h\to \chi $ in the presence and absence of new physics, 
\begin{equation}
\frac{\sigma(pp\to h)_{\rm BSM}}{\sigma(pp\to h)_{\rm SM}}~~~~~~{\rm and}~~~~~~\frac{\Gamma(h\to \chi)_{\rm BSM}}{\Gamma(h\to \chi)_{\rm SM}},
\end{equation} 
where $\chi$ generically stands for the final states $b\bar{b},\,\tau^-\tau^+,\,W^-W^+,\,ZZ,\,Z\gamma$ and $\gamma\gamma$. 

The Higgs to di-photon strength, under the narrow width approximation, can be expressed as follows,
\begin{equation}
\mu_{\gamma\gamma}=\, \frac{\sigma(gg\rightarrow h)_{\rm BSM}}{\sigma(gg\rightarrow h)_{\rm SM}} \, \frac{{\rm BR}(h\rightarrow\gamma\gamma)_{\rm BSM}}{{\rm BR}(h\rightarrow\gamma\gamma)_{\rm SM}}.
\end{equation}
Here, for our model $\frac{\sigma(gg\rightarrow h)_{\rm BSM}}{\sigma(gg\rightarrow h)_{\rm SM}}= 1$, hence
\begin{equation}
\mu_{\gamma\gamma}=  \frac{\Gamma_{\rm SM}^{\rm Total}}{\Gamma_{\rm BSM}^{\rm Total}}\, \frac{\Gamma(h\rightarrow\gamma\gamma)_{\rm BSM}}{\Gamma(h\rightarrow\gamma\gamma)_{\rm SM}}.
\end{equation}
For the case when the  new particle's mass is greater than $\frac{M_h}{2}$ or small ${\rm BR}(h\to a'a')<10^{-4}$, one can write $\frac{\Gamma_{\rm SM}^{\rm Total}}{\Gamma_{\rm BSM}^{\rm Total}}\approx 1$, hence,
\begin{equation}
\mu_{\gamma\gamma}= \frac{\Gamma(h\rightarrow\gamma\gamma)_{\rm BSM}}{\Gamma(h\rightarrow\gamma\gamma)_{\rm SM}}.
\end{equation}
At one-loop level, the physical charged Higges $H^{\pm}$ add an extra contribution to the decay width $\Gamma(h\rightarrow\gamma\gamma)_{\rm BSM}$~\cite{Djouadi:2005gj}, 
\begin{equation}
\Gamma(h\rightarrow \gamma\gamma)_{\rm BSM}=\frac{\alpha^2M_h^3}{256\pi^3 v^2}\Big|Q^2_{H^{\pm}} \, \frac{v \, g_{hH^+H^-}}{2M^2_{H^{\pm}}} \, F_0(\tau_{H^{\pm}})+C_{\rm SM}\Big|,
\end{equation}
where $C_{\rm SM}$ is the contribution from SM particles, $$ C_{\rm SM}=\sum_f N_f^c Q_f^2 y_f F_{1/2}(\tau_f)+y_W F_1(\tau_W), $$ with $\tau_x=\frac{M_h^2}{4 M_x^2}$. $Q$ stands for the electric charge of particles, and $N_f^c$ denotes the color factor.
The loop functions $F_{(0,1/2,1)}(\tau)$ are defined as~\cite{Djouadi:2005gj}
\begin{align}
F_{0}(\tau)&=-[\tau-f(\tau)]\tau^{-2}\, ,\nn\\
F_{1/2}(\tau)&=2[\tau+(\tau-1)f(\tau)]\tau^{-2}\, ,\nn\\
F_{1}(\tau)&=-[2\tau^2+3\tau+3(2\tau-1)f(\tau)]\tau^{-2}\, , \nn
\label{loopfn}
\end{align}
where 
\beq
f(\tau)= \bigg\{\begin{array}{ll}
(\sin^{-1}\sqrt{\tau})^2\,,\hspace{60pt}& \tau\leq 1\\
-{1\over4}[\ln{1+\sqrt{1-\tau^{-1}}\over1-\sqrt{1-\tau^{-1}}}-i\pi]^2\,,\quad
&\tau>1
\label{ftau}
\end{array}\;.
\eeq
\section{Unitarity of $S$-matrices}
\label{ap:unitary}
The scalar quartic couplings in the physical bases, denoted by $h, G^\pm, G^0, H, H^\pm, A,$ and~$a'$, are intricate functions of the parameters $\lambda$'s and $\xi$. This complexity poses significant challenges when calculating the unitary bounds in these physical bases. An alternative approach involves considering non-physical scalar field bases, namely $H_1^0, G^\pm, G^0, H_2^0, H_2^\pm, A_2^0,$ and $A_3^0$, before Electroweak Symmetry Breaking (EWSB).
A crucial insight is that the $S$-matrix, originally expressed in terms of the physical fields, can be transformed into an $S$-matrix for the non-physical fields by employing a unitary transformation~\cite{Kanemura:1993hm,Arhrib:2000is}. This transformation allows for a more tractable analysis of unitarity bounds and facilitates calculations within this framework.

In this scenario, the complete set of non-physical scalar scattering processes can be represented by a $23\times 23$ $S$-matrix. This matrix comprises three distinct submatrices with dimensions $10 \times 10$, $7 \times 7$, and $6 \times 6$, respectively, corresponding to different initial and final states involved in the scattering processes.

The first Scattering matrix $\mathcal{M}_1$ to scattering processes with one of the following initial and final states: $\{G^0  G^+ , A^0_2  G^+ , A^0_3  G^+ ,  H^0_1  G^+ ,  H^0_2  G^+, G^0  H^+_2 , A^0_2  H^+_2 ,  A^0_3  H^+_2 ,  H^0_1  H^+_2,  H^0_2  H^+_2 \}$ as,
\begin{equation}
\resizebox{0.8\hsize}{!}{ $
  \mathcal{M}_1=  \left(
\begin{array}{cccccccccc}
 \frac{\lambda _1}{2} & 0 & 0 & 0 & 0 & 0 & \frac{\lambda _4}{2}+\frac{\lambda _5}{2} & 0 & 0 & i \left(\frac{\lambda _5}{2}-\frac{\lambda _4}{2}\right) \\
 0 & \frac{\lambda _3}{2} & 0 & 0 & 0 & \frac{\lambda _4}{2}+\frac{\lambda _5}{2} & 0 & 0 & i \left(\frac{\lambda _4}{2}-\frac{\lambda _5}{2}\right) & 0 \\
 0 & 0 & \frac{\lambda _3'}{2} & 0 & 0 & 0 & 0 & \frac{\lambda _5'}{2} & 0 & 0 \\
 0 & 0 & 0 & \frac{\lambda _1}{2} & 0 & 0 & i \left(\frac{\lambda _4}{2}-\frac{\lambda _5}{2}\right) & 0 & 0 & \frac{\lambda _4}{2}+\frac{\lambda _5}{2} \\
 0 & 0 & 0 & 0 & \frac{\lambda _3}{2} & i \left(\frac{\lambda _5}{2}-\frac{\lambda _4}{2}\right) & 0 & 0 & \frac{\lambda _4}{2}+\frac{\lambda _5}{2} & 0 \\
 0 & \frac{\lambda _4}{2}+\frac{\lambda _5}{2} & 0 & 0 & i \left(\frac{\lambda _4}{2}-\frac{\lambda _5}{2}\right) & \frac{\lambda _3}{2} & 0 & 0 & 0 & 0 \\
 \frac{\lambda _4}{2}+\frac{\lambda _5}{2} & 0 & 0 & i \left(\frac{\lambda _5}{2}-\frac{\lambda _4}{2}\right) & 0 & 0 & \frac{\lambda _2}{2} & 0 & 0 & 0 \\
 0 & 0 & \frac{\lambda _5'}{2} & 0 & 0 & 0 & 0 & \frac{\lambda _4'}{2} & 0 & 0 \\
 0 & i \left(\frac{\lambda _5}{2}-\frac{\lambda _4}{2}\right) & 0 & 0 & \frac{\lambda _4}{2}+\frac{\lambda _5}{2} & 0 & 0 & 0 & \frac{\lambda _3}{2} & 0 \\
 i \left(\frac{\lambda _4}{2}-\frac{\lambda _5}{2}\right) & 0 & 0 & \frac{\lambda _4}{2}+\frac{\lambda _5}{2} & 0 & 0 & 0 & 0 & 0 & \frac{\lambda _2}{2} \\
\end{array}
\right)$.}
\end{equation}
The sub-scattering matrix $\mathcal{M}_2$ corresponds to the following initial and final states:
$\{G^+ G^-$,\\$ H^+_2 H^-_2$,$ \frac{(G^0)^2}{\sqrt{2}}$,$ \frac{(A_2^0)^2}{\sqrt{2}}$,$ \frac{(A_3^0)^2}{\sqrt{2}}$,$ \frac{(H^0_1)^2}{\sqrt{2}}$,$ \frac{(H^0_2)^2}{\sqrt{2}}\}$
\begin{equation}
\resizebox{0.8\hsize}{!}{ $
  \mathcal{M}_2= \left(
\begin{array}{ccccccc}
 2 \lambda_1 & \lambda_3+\lambda_4 & \frac{\lambda_1}{\sqrt{2}} & \frac{\lambda_3}{\sqrt{2}} & \frac{\lambda_3'}{\sqrt{2}} & \frac{\lambda_1}{\sqrt{2}} & \frac{\lambda_3}{\sqrt{2}} \\
 \lambda_3+\lambda_4 & 2 \lambda_2 & \frac{\lambda_3}{\sqrt{2}} & \frac{\lambda_2}{\sqrt{2}} & \frac{\lambda_4'}{\sqrt{2}} & \frac{\lambda_3}{\sqrt{2}} & \frac{\lambda_2}{\sqrt{2}} \\
 \frac{\lambda_1}{\sqrt{2}} & \frac{\lambda_3}{\sqrt{2}} & \frac{3 \lambda_1}{2} & \frac{1}{2} \left(\lambda_3+\lambda_4+\lambda_5\right) & \frac{\lambda_3'}{2} & \frac{\lambda_1}{2} & \frac{1}{2} \left(\lambda_3+\lambda_4-\lambda_5\right) \\
 \frac{\lambda_3}{\sqrt{2}} & \frac{\lambda_2}{\sqrt{2}} & \frac{1}{2} \left(\lambda_3+\lambda_4+\lambda_5\right) & \frac{3 \lambda_2}{2} & \frac{\lambda_4'}{2} & \frac{1}{2} \left(\lambda_3+\lambda_4-\lambda_5\right) & \frac{\lambda_2}{2} \\
 \frac{\lambda_3'}{\sqrt{2}} & \frac{\lambda_4'}{\sqrt{2}} & \frac{\lambda_3'}{2} & \frac{\lambda_4'}{2} & 3 \lambda_2' & \frac{\lambda_3'}{2} & \frac{\lambda_4'}{2} \\
 \frac{\lambda_1}{\sqrt{2}} & \frac{\lambda_3}{\sqrt{2}} & \frac{\lambda_1}{2} & \frac{1}{2} \left(\lambda_3+\lambda_4-\lambda_5\right) & \frac{\lambda_3'}{2} & \frac{3 \lambda_1}{2} & \frac{1}{2} \left(\lambda_3+\lambda_4+\lambda_5\right) \\
 \frac{\lambda_3}{\sqrt{2}} & \frac{\lambda_2}{\sqrt{2}} & \frac{1}{2} \left(\lambda_3+\lambda_4-\lambda_5\right) & \frac{\lambda_2}{2} & \frac{\lambda_4'}{2} & \frac{1}{2} \left(\lambda_3+\lambda_4+\lambda_5\right) & \frac{3 \lambda_2}{2} \\
\end{array}
\right)
$.}
\end{equation}
The third Scattering matrix $\mathcal{M}_3$ to scattering processes with one of the following initial and final states: $\{G^+ H^-_2, G^- H^+_2, G^0 A_2^0, G^0 A_3^0, A_2^0 A_3^0, H^0_1 H^0_2\}$
\begin{equation}
\resizebox{0.8\hsize}{!}{ $
  \mathcal{M}_3= \left(
\begin{array}{cccccc}
 \lambda _3+\lambda _4 & \frac{\lambda _5}{2} & \frac{\lambda _4}{2}+\frac{\lambda _5}{2} & 0 & 0 & \frac{\lambda _4}{2}+\frac{\lambda _5}{2} \\
 \frac{\lambda _5}{2} & \lambda _3+\lambda _4 & \frac{\lambda _4}{2}+\frac{\lambda _5}{2} & 0 & 0 & \frac{\lambda _4}{2}+\frac{\lambda _5}{2} \\
 \frac{\lambda _4}{2}+\frac{\lambda _5}{2} & \frac{\lambda _4}{2}+\frac{\lambda _5}{2} & \frac{1}{4} \left(\lambda _3+\lambda _4+\lambda _5\right) & 0 & 0 & \lambda _5 \\
 0 & 0 & 0 & -\frac{\lambda _3'}{4} & -\frac{\lambda _5'}{2} & 0 \\
 0 & 0 & 0 & -\frac{\lambda _5'}{2} & -\frac{\lambda _4'}{4} & 0 \\
 \frac{\lambda _4}{2}+\frac{\lambda _5}{2} & \frac{\lambda _4}{2}+\frac{\lambda _5}{2} & \lambda _5 & 0 & 0 & \frac{1}{4} \left(\lambda _3+\lambda _4+\lambda _5\right) \\
\end{array}
\right)
$.}
\end{equation}

\bibliographystyle{apsrev4-1}
\bibliography{Refs}

\begin{thebibliography}{208}%
\makeatletter
\providecommand \@ifxundefined [1]{%
 \@ifx{#1\undefined}
}%
\providecommand \@ifnum [1]{%
 \ifnum #1\expandafter \@firstoftwo
 \else \expandafter \@secondoftwo
 \fi
}%
\providecommand \@ifx [1]{%
 \ifx #1\expandafter \@firstoftwo
 \else \expandafter \@secondoftwo
 \fi
}%
\providecommand \natexlab [1]{#1}%
\providecommand \enquote  [1]{``#1''}%
\providecommand \bibnamefont  [1]{#1}%
\providecommand \bibfnamefont [1]{#1}%
\providecommand \citenamefont [1]{#1}%
\providecommand \href@noop [0]{\@secondoftwo}%
\providecommand \href [0]{\begingroup \@sanitize@url \@href}%
\providecommand \@href[1]{\@@startlink{#1}\@@href}%
\providecommand \@@href[1]{\endgroup#1\@@endlink}%
\providecommand \@sanitize@url [0]{\catcode `\\12\catcode `\$12\catcode `\&12\catcode `\#12\catcode `\^12\catcode `\_12\catcode `\%12\relax}%
\providecommand \@@startlink[1]{}%
\providecommand \@@endlink[0]{}%
\providecommand \url  [0]{\begingroup\@sanitize@url \@url }%
\providecommand \@url [1]{\endgroup\@href {#1}{\urlprefix }}%
\providecommand \urlprefix  [0]{URL }%
\providecommand \Eprint [0]{\href }%
\providecommand \doibase [0]{http://dx.doi.org/}%
\providecommand \selectlanguage [0]{\@gobble}%
\providecommand \bibinfo  [0]{\@secondoftwo}%
\providecommand \bibfield  [0]{\@secondoftwo}%
\providecommand \translation [1]{[#1]}%
\providecommand \BibitemOpen [0]{}%
\providecommand \bibitemStop [0]{}%
\providecommand \bibitemNoStop [0]{.\EOS\space}%
\providecommand \EOS [0]{\spacefactor3000\relax}%
\providecommand \BibitemShut  [1]{\csname bibitem#1\endcsname}%
\let\auto@bib@innerbib\@empty
\bibitem [{\citenamefont {Martin}(1998)}]{Martin:1997ns}%
  \BibitemOpen
  \bibfield  {author} {\bibinfo {author} {\bibfnamefont {S.~P.}\ \bibnamefont {Martin}},\ }\href {\doibase 10.1142/9789812839657_0001} {\bibfield  {journal} {\bibinfo  {journal} {Adv. Ser. Direct. High Energy Phys.}\ }\textbf {\bibinfo {volume} {18}},\ \bibinfo {pages} {1} (\bibinfo {year} {1998})},\ \Eprint {http://arxiv.org/abs/hep-ph/9709356} {arXiv:hep-ph/9709356} \BibitemShut {NoStop}%
\bibitem [{\citenamefont {Workman}\ \emph {et~al.}(2022)\citenamefont {Workman} \emph {et~al.}}]{ParticleDataGroup:2022pthAA}%
  \BibitemOpen
  \bibfield  {author} {\bibinfo {author} {\bibfnamefont {R.~L.}\ \bibnamefont {Workman}} \emph {et~al.} (\bibinfo {collaboration} {Particle Data Group}),\ }\href {\doibase 10.1093/ptep/ptac097} {\bibfield  {journal} {\bibinfo  {journal} {PTEP}\ }\textbf {\bibinfo {volume} {2022}},\ \bibinfo {pages} {083C01} (\bibinfo {year} {2022})}\BibitemShut {NoStop}%
\bibitem [{\citenamefont {Eichten}\ \emph {et~al.}(1983)\citenamefont {Eichten}, \citenamefont {Lane},\ and\ \citenamefont {Peskin}}]{Eichten:1983hw}%
  \BibitemOpen
  \bibfield  {author} {\bibinfo {author} {\bibfnamefont {E.}~\bibnamefont {Eichten}}, \bibinfo {author} {\bibfnamefont {K.~D.}\ \bibnamefont {Lane}}, \ and\ \bibinfo {author} {\bibfnamefont {M.~E.}\ \bibnamefont {Peskin}},\ }\href {\doibase 10.1103/PhysRevLett.50.811} {\bibfield  {journal} {\bibinfo  {journal} {Phys. Rev. Lett.}\ }\textbf {\bibinfo {volume} {50}},\ \bibinfo {pages} {811} (\bibinfo {year} {1983})}\BibitemShut {NoStop}%
\bibitem [{\citenamefont {Aguilar}\ \emph {et~al.}(2001{\natexlab{a}})\citenamefont {Aguilar} \emph {et~al.}}]{LSND:2001aii}%
  \BibitemOpen
  \bibfield  {author} {\bibinfo {author} {\bibfnamefont {A.}~\bibnamefont {Aguilar}} \emph {et~al.} (\bibinfo {collaboration} {LSND}),\ }\href {\doibase 10.1103/PhysRevD.64.112007} {\bibfield  {journal} {\bibinfo  {journal} {Phys. Rev. D}\ }\textbf {\bibinfo {volume} {64}},\ \bibinfo {pages} {112007} (\bibinfo {year} {2001}{\natexlab{a}})},\ \Eprint {http://arxiv.org/abs/hep-ex/0104049} {arXiv:hep-ex/0104049} \BibitemShut {NoStop}%
\bibitem [{\citenamefont {Aguilar-Arevalo}\ \emph {et~al.}(2007)\citenamefont {Aguilar-Arevalo} \emph {et~al.}}]{MiniBooNE:2007uho}%
  \BibitemOpen
  \bibfield  {author} {\bibinfo {author} {\bibfnamefont {A.~A.}\ \bibnamefont {Aguilar-Arevalo}} \emph {et~al.} (\bibinfo {collaboration} {MiniBooNE}),\ }\href {\doibase 10.1103/PhysRevLett.98.231801} {\bibfield  {journal} {\bibinfo  {journal} {Phys. Rev. Lett.}\ }\textbf {\bibinfo {volume} {98}},\ \bibinfo {pages} {231801} (\bibinfo {year} {2007})},\ \Eprint {http://arxiv.org/abs/0704.1500} {arXiv:0704.1500 [hep-ex]} \BibitemShut {NoStop}%
\bibitem [{\citenamefont {Aguilar-Arevalo}\ \emph {et~al.}(2009)\citenamefont {Aguilar-Arevalo} \emph {et~al.}}]{MiniBooNE:2008yuf}%
  \BibitemOpen
  \bibfield  {author} {\bibinfo {author} {\bibfnamefont {A.~A.}\ \bibnamefont {Aguilar-Arevalo}} \emph {et~al.} (\bibinfo {collaboration} {MiniBooNE}),\ }\href {\doibase 10.1103/PhysRevLett.102.101802} {\bibfield  {journal} {\bibinfo  {journal} {Phys. Rev. Lett.}\ }\textbf {\bibinfo {volume} {102}},\ \bibinfo {pages} {101802} (\bibinfo {year} {2009})},\ \Eprint {http://arxiv.org/abs/0812.2243} {arXiv:0812.2243 [hep-ex]} \BibitemShut {NoStop}%
\bibitem [{\citenamefont {Aguilar-Arevalo}\ \emph {et~al.}(2013)\citenamefont {Aguilar-Arevalo} \emph {et~al.}}]{MiniBooNE:2013uba}%
  \BibitemOpen
  \bibfield  {author} {\bibinfo {author} {\bibfnamefont {A.~A.}\ \bibnamefont {Aguilar-Arevalo}} \emph {et~al.} (\bibinfo {collaboration} {MiniBooNE}),\ }\href {\doibase 10.1103/PhysRevLett.110.161801} {\bibfield  {journal} {\bibinfo  {journal} {Phys. Rev. Lett.}\ }\textbf {\bibinfo {volume} {110}},\ \bibinfo {pages} {161801} (\bibinfo {year} {2013})},\ \Eprint {http://arxiv.org/abs/1303.2588} {arXiv:1303.2588 [hep-ex]} \BibitemShut {NoStop}%
\bibitem [{\citenamefont {Athanassopoulos}\ \emph {et~al.}(1997)\citenamefont {Athanassopoulos} \emph {et~al.}}]{LSND:1996jxj}%
  \BibitemOpen
  \bibfield  {author} {\bibinfo {author} {\bibfnamefont {C.}~\bibnamefont {Athanassopoulos}} \emph {et~al.} (\bibinfo {collaboration} {LSND}),\ }\href {\doibase 10.1016/S0168-9002(96)01155-2} {\bibfield  {journal} {\bibinfo  {journal} {Nucl. Instrum. Meth. A}\ }\textbf {\bibinfo {volume} {388}},\ \bibinfo {pages} {149} (\bibinfo {year} {1997})},\ \Eprint {http://arxiv.org/abs/nucl-ex/9605002} {arXiv:nucl-ex/9605002} \BibitemShut {NoStop}%
\bibitem [{\citenamefont {Katori}(2020)}]{Katori:2020tvv}%
  \BibitemOpen
  \bibfield  {author} {\bibinfo {author} {\bibfnamefont {T.}~\bibnamefont {Katori}} (\bibinfo {collaboration} {MiniBooNE}),\ }in\ \href@noop {} {\emph {\bibinfo {booktitle} {{3rd World Summit on Exploring the Dark Side of the Universe}}}}\ (\bibinfo {year} {2020})\ pp.\ \bibinfo {pages} {139--148},\ \Eprint {http://arxiv.org/abs/2010.06015} {arXiv:2010.06015 [hep-ex]} \BibitemShut {NoStop}%
\bibitem [{\citenamefont {Dasgupta}\ and\ \citenamefont {Kopp}(2021)}]{Dasgupta:2021ies}%
  \BibitemOpen
  \bibfield  {author} {\bibinfo {author} {\bibfnamefont {B.}~\bibnamefont {Dasgupta}}\ and\ \bibinfo {author} {\bibfnamefont {J.}~\bibnamefont {Kopp}},\ }\href {\doibase 10.1016/j.physrep.2021.06.002} {\bibfield  {journal} {\bibinfo  {journal} {Phys. Rept.}\ }\textbf {\bibinfo {volume} {928}},\ \bibinfo {pages} {1} (\bibinfo {year} {2021})},\ \Eprint {http://arxiv.org/abs/2106.05913} {arXiv:2106.05913 [hep-ph]} \BibitemShut {NoStop}%
\bibitem [{\citenamefont {Brdar}\ and\ \citenamefont {Kopp}(2022)}]{Brdar:2021ysi}%
  \BibitemOpen
  \bibfield  {author} {\bibinfo {author} {\bibfnamefont {V.}~\bibnamefont {Brdar}}\ and\ \bibinfo {author} {\bibfnamefont {J.}~\bibnamefont {Kopp}},\ }\href {\doibase 10.1103/PhysRevD.105.115024} {\bibfield  {journal} {\bibinfo  {journal} {Phys. Rev. D}\ }\textbf {\bibinfo {volume} {105}},\ \bibinfo {pages} {115024} (\bibinfo {year} {2022})},\ \Eprint {http://arxiv.org/abs/2109.08157} {arXiv:2109.08157 [hep-ph]} \BibitemShut {NoStop}%
\bibitem [{\citenamefont {Alvarez-Ruso}\ and\ \citenamefont {Saul-Sala}(2021)}]{Alvarez-Ruso:2021dna}%
  \BibitemOpen
  \bibfield  {author} {\bibinfo {author} {\bibfnamefont {L.}~\bibnamefont {Alvarez-Ruso}}\ and\ \bibinfo {author} {\bibfnamefont {E.}~\bibnamefont {Saul-Sala}},\ }\href {\doibase 10.1140/epjs/s11734-021-00293-9} {\bibfield  {journal} {\bibinfo  {journal} {Eur. Phys. J. ST}\ }\textbf {\bibinfo {volume} {230}},\ \bibinfo {pages} {4373} (\bibinfo {year} {2021})},\ \Eprint {http://arxiv.org/abs/2111.02504} {arXiv:2111.02504 [hep-ph]} \BibitemShut {NoStop}%
\bibitem [{\citenamefont {Abratenko}\ \emph {et~al.}(2022{\natexlab{a}})\citenamefont {Abratenko} \emph {et~al.}}]{MicroBooNE:2021zai}%
  \BibitemOpen
  \bibfield  {author} {\bibinfo {author} {\bibfnamefont {P.}~\bibnamefont {Abratenko}} \emph {et~al.} (\bibinfo {collaboration} {MicroBooNE}),\ }\href {\doibase 10.1103/PhysRevLett.128.111801} {\bibfield  {journal} {\bibinfo  {journal} {Phys. Rev. Lett.}\ }\textbf {\bibinfo {volume} {128}},\ \bibinfo {pages} {111801} (\bibinfo {year} {2022}{\natexlab{a}})},\ \Eprint {http://arxiv.org/abs/2110.00409} {arXiv:2110.00409 [hep-ex]} \BibitemShut {NoStop}%
\bibitem [{\citenamefont {Krasznahorkay}\ \emph {et~al.}(2016)\citenamefont {Krasznahorkay} \emph {et~al.}}]{Krasznahorkay:2015iga}%
  \BibitemOpen
  \bibfield  {author} {\bibinfo {author} {\bibfnamefont {A.~J.}\ \bibnamefont {Krasznahorkay}} \emph {et~al.},\ }\href {\doibase 10.1103/PhysRevLett.116.042501} {\bibfield  {journal} {\bibinfo  {journal} {Phys. Rev. Lett.}\ }\textbf {\bibinfo {volume} {116}},\ \bibinfo {pages} {042501} (\bibinfo {year} {2016})},\ \Eprint {http://arxiv.org/abs/1504.01527} {arXiv:1504.01527 [nucl-ex]} \BibitemShut {NoStop}%
\bibitem [{\citenamefont {Krasznahorkay}\ \emph {et~al.}(2019)\citenamefont {Krasznahorkay} \emph {et~al.}}]{Krasznahorkay:2019lyl}%
  \BibitemOpen
  \bibfield  {author} {\bibinfo {author} {\bibfnamefont {A.~J.}\ \bibnamefont {Krasznahorkay}} \emph {et~al.},\ }\href@noop {} {\  (\bibinfo {year} {2019})},\ \Eprint {http://arxiv.org/abs/1910.10459} {arXiv:1910.10459 [nucl-ex]} \BibitemShut {NoStop}%
\bibitem [{\citenamefont {Krasznahorkay}\ \emph {et~al.}(2021)\citenamefont {Krasznahorkay}, \citenamefont {Csatl\'os}, \citenamefont {Csige}, \citenamefont {Guly\'as}, \citenamefont {Krasznahorkay}, \citenamefont {Nyak\'o}, \citenamefont {Rajta}, \citenamefont {Tim\'ar}, \citenamefont {Vajda},\ and\ \citenamefont {Sas}}]{Krasznahorkay:2021joi}%
  \BibitemOpen
  \bibfield  {author} {\bibinfo {author} {\bibfnamefont {A.~J.}\ \bibnamefont {Krasznahorkay}}, \bibinfo {author} {\bibfnamefont {M.}~\bibnamefont {Csatl\'os}}, \bibinfo {author} {\bibfnamefont {L.}~\bibnamefont {Csige}}, \bibinfo {author} {\bibfnamefont {J.}~\bibnamefont {Guly\'as}}, \bibinfo {author} {\bibfnamefont {A.}~\bibnamefont {Krasznahorkay}}, \bibinfo {author} {\bibfnamefont {B.~M.}\ \bibnamefont {Nyak\'o}}, \bibinfo {author} {\bibfnamefont {I.}~\bibnamefont {Rajta}}, \bibinfo {author} {\bibfnamefont {J.}~\bibnamefont {Tim\'ar}}, \bibinfo {author} {\bibfnamefont {I.}~\bibnamefont {Vajda}}, \ and\ \bibinfo {author} {\bibfnamefont {N.~J.}\ \bibnamefont {Sas}},\ }\href {\doibase 10.1103/PhysRevC.104.044003} {\bibfield  {journal} {\bibinfo  {journal} {Phys. Rev. C}\ }\textbf {\bibinfo {volume} {104}},\ \bibinfo {pages} {044003} (\bibinfo {year} {2021})},\ \Eprint {http://arxiv.org/abs/2104.10075} {arXiv:2104.10075 [nucl-ex]} \BibitemShut {NoStop}%
\bibitem [{\citenamefont {Krasznahorkay}\ \emph {et~al.}(2022)\citenamefont {Krasznahorkay} \emph {et~al.}}]{Krasznahorkay:2022pxs}%
  \BibitemOpen
  \bibfield  {author} {\bibinfo {author} {\bibfnamefont {A.~J.}\ \bibnamefont {Krasznahorkay}} \emph {et~al.},\ }\href {\doibase 10.1103/PhysRevC.106.L061601} {\bibfield  {journal} {\bibinfo  {journal} {Phys. Rev. C}\ }\textbf {\bibinfo {volume} {106}},\ \bibinfo {pages} {L061601} (\bibinfo {year} {2022})},\ \Eprint {http://arxiv.org/abs/2209.10795} {arXiv:2209.10795 [nucl-ex]} \BibitemShut {NoStop}%
\bibitem [{\citenamefont {Anselmann}\ \emph {et~al.}(1995)\citenamefont {Anselmann} \emph {et~al.}}]{GALLEX:1994rym}%
  \BibitemOpen
  \bibfield  {author} {\bibinfo {author} {\bibfnamefont {P.}~\bibnamefont {Anselmann}} \emph {et~al.} (\bibinfo {collaboration} {GALLEX}),\ }\href {\doibase 10.1016/0370-2693(94)01586-2} {\bibfield  {journal} {\bibinfo  {journal} {Phys. Lett. B}\ }\textbf {\bibinfo {volume} {342}},\ \bibinfo {pages} {440} (\bibinfo {year} {1995})}\BibitemShut {NoStop}%
\bibitem [{\citenamefont {Hampel}\ \emph {et~al.}(1998)\citenamefont {Hampel} \emph {et~al.}}]{GALLEX:1997lja}%
  \BibitemOpen
  \bibfield  {author} {\bibinfo {author} {\bibfnamefont {W.}~\bibnamefont {Hampel}} \emph {et~al.} (\bibinfo {collaboration} {GALLEX}),\ }\href {\doibase 10.1016/S0370-2693(97)01562-1} {\bibfield  {journal} {\bibinfo  {journal} {Phys. Lett. B}\ }\textbf {\bibinfo {volume} {420}},\ \bibinfo {pages} {114} (\bibinfo {year} {1998})}\BibitemShut {NoStop}%
\bibitem [{\citenamefont {Kaether}\ \emph {et~al.}(2010)\citenamefont {Kaether}, \citenamefont {Hampel}, \citenamefont {Heusser}, \citenamefont {Kiko},\ and\ \citenamefont {Kirsten}}]{Kaether:2010ag}%
  \BibitemOpen
  \bibfield  {author} {\bibinfo {author} {\bibfnamefont {F.}~\bibnamefont {Kaether}}, \bibinfo {author} {\bibfnamefont {W.}~\bibnamefont {Hampel}}, \bibinfo {author} {\bibfnamefont {G.}~\bibnamefont {Heusser}}, \bibinfo {author} {\bibfnamefont {J.}~\bibnamefont {Kiko}}, \ and\ \bibinfo {author} {\bibfnamefont {T.}~\bibnamefont {Kirsten}},\ }\href {\doibase 10.1016/j.physletb.2010.01.030} {\bibfield  {journal} {\bibinfo  {journal} {Phys. Lett. B}\ }\textbf {\bibinfo {volume} {685}},\ \bibinfo {pages} {47} (\bibinfo {year} {2010})},\ \Eprint {http://arxiv.org/abs/1001.2731} {arXiv:1001.2731 [hep-ex]} \BibitemShut {NoStop}%
\bibitem [{\citenamefont {Abdurashitov}\ \emph {et~al.}(1996)\citenamefont {Abdurashitov} \emph {et~al.}}]{Abdurashitov:1996dp}%
  \BibitemOpen
  \bibfield  {author} {\bibinfo {author} {\bibfnamefont {D.~N.}\ \bibnamefont {Abdurashitov}} \emph {et~al.},\ }\href {\doibase 10.1103/PhysRevLett.77.4708} {\bibfield  {journal} {\bibinfo  {journal} {Phys. Rev. Lett.}\ }\textbf {\bibinfo {volume} {77}},\ \bibinfo {pages} {4708} (\bibinfo {year} {1996})}\BibitemShut {NoStop}%
\bibitem [{\citenamefont {Abdurashitov}\ \emph {et~al.}(1999)\citenamefont {Abdurashitov} \emph {et~al.}}]{SAGE:1998fvr}%
  \BibitemOpen
  \bibfield  {author} {\bibinfo {author} {\bibfnamefont {J.~N.}\ \bibnamefont {Abdurashitov}} \emph {et~al.} (\bibinfo {collaboration} {SAGE}),\ }\href {\doibase 10.1103/PhysRevC.59.2246} {\bibfield  {journal} {\bibinfo  {journal} {Phys. Rev. C}\ }\textbf {\bibinfo {volume} {59}},\ \bibinfo {pages} {2246} (\bibinfo {year} {1999})},\ \Eprint {http://arxiv.org/abs/hep-ph/9803418} {arXiv:hep-ph/9803418} \BibitemShut {NoStop}%
\bibitem [{\citenamefont {Abdurashitov}\ \emph {et~al.}(2006)\citenamefont {Abdurashitov} \emph {et~al.}}]{Abdurashitov:2005tb}%
  \BibitemOpen
  \bibfield  {author} {\bibinfo {author} {\bibfnamefont {J.~N.}\ \bibnamefont {Abdurashitov}} \emph {et~al.},\ }\href {\doibase 10.1103/PhysRevC.73.045805} {\bibfield  {journal} {\bibinfo  {journal} {Phys. Rev. C}\ }\textbf {\bibinfo {volume} {73}},\ \bibinfo {pages} {045805} (\bibinfo {year} {2006})},\ \Eprint {http://arxiv.org/abs/nucl-ex/0512041} {arXiv:nucl-ex/0512041} \BibitemShut {NoStop}%
\bibitem [{\citenamefont {Abdurashitov}\ \emph {et~al.}(2009)\citenamefont {Abdurashitov} \emph {et~al.}}]{SAGE:2009eeu}%
  \BibitemOpen
  \bibfield  {author} {\bibinfo {author} {\bibfnamefont {J.~N.}\ \bibnamefont {Abdurashitov}} \emph {et~al.} (\bibinfo {collaboration} {SAGE}),\ }\href {\doibase 10.1103/PhysRevC.80.015807} {\bibfield  {journal} {\bibinfo  {journal} {Phys. Rev. C}\ }\textbf {\bibinfo {volume} {80}},\ \bibinfo {pages} {015807} (\bibinfo {year} {2009})},\ \Eprint {http://arxiv.org/abs/0901.2200} {arXiv:0901.2200 [nucl-ex]} \BibitemShut {NoStop}%
\bibitem [{\citenamefont {Hamann}\ \emph {et~al.}(2011)\citenamefont {Hamann}, \citenamefont {Hannestad}, \citenamefont {Raffelt},\ and\ \citenamefont {Wong}}]{Hamann:2011ge}%
  \BibitemOpen
  \bibfield  {author} {\bibinfo {author} {\bibfnamefont {J.}~\bibnamefont {Hamann}}, \bibinfo {author} {\bibfnamefont {S.}~\bibnamefont {Hannestad}}, \bibinfo {author} {\bibfnamefont {G.~G.}\ \bibnamefont {Raffelt}}, \ and\ \bibinfo {author} {\bibfnamefont {Y.~Y.~Y.}\ \bibnamefont {Wong}},\ }\href {\doibase 10.1088/1475-7516/2011/09/034} {\bibfield  {journal} {\bibinfo  {journal} {JCAP}\ }\textbf {\bibinfo {volume} {09}},\ \bibinfo {pages} {034} (\bibinfo {year} {2011})},\ \Eprint {http://arxiv.org/abs/1108.4136} {arXiv:1108.4136 [astro-ph.CO]} \BibitemShut {NoStop}%
\bibitem [{\citenamefont {Archidiacono}\ \emph {et~al.}(2013)\citenamefont {Archidiacono}, \citenamefont {Fornengo}, \citenamefont {Giunti}, \citenamefont {Hannestad},\ and\ \citenamefont {Melchiorri}}]{Archidiacono:2013xxa}%
  \BibitemOpen
  \bibfield  {author} {\bibinfo {author} {\bibfnamefont {M.}~\bibnamefont {Archidiacono}}, \bibinfo {author} {\bibfnamefont {N.}~\bibnamefont {Fornengo}}, \bibinfo {author} {\bibfnamefont {C.}~\bibnamefont {Giunti}}, \bibinfo {author} {\bibfnamefont {S.}~\bibnamefont {Hannestad}}, \ and\ \bibinfo {author} {\bibfnamefont {A.}~\bibnamefont {Melchiorri}},\ }\href {\doibase 10.1103/PhysRevD.87.125034} {\bibfield  {journal} {\bibinfo  {journal} {Phys. Rev. D}\ }\textbf {\bibinfo {volume} {87}},\ \bibinfo {pages} {125034} (\bibinfo {year} {2013})},\ \Eprint {http://arxiv.org/abs/1302.6720} {arXiv:1302.6720 [astro-ph.CO]} \BibitemShut {NoStop}%
\bibitem [{\citenamefont {Hagstotz}\ \emph {et~al.}(2021)\citenamefont {Hagstotz}, \citenamefont {de~Salas}, \citenamefont {Gariazzo}, \citenamefont {Gerbino}, \citenamefont {Lattanzi}, \citenamefont {Vagnozzi}, \citenamefont {Freese},\ and\ \citenamefont {Pastor}}]{Hagstotz:2020ukm}%
  \BibitemOpen
  \bibfield  {author} {\bibinfo {author} {\bibfnamefont {S.}~\bibnamefont {Hagstotz}}, \bibinfo {author} {\bibfnamefont {P.~F.}\ \bibnamefont {de~Salas}}, \bibinfo {author} {\bibfnamefont {S.}~\bibnamefont {Gariazzo}}, \bibinfo {author} {\bibfnamefont {M.}~\bibnamefont {Gerbino}}, \bibinfo {author} {\bibfnamefont {M.}~\bibnamefont {Lattanzi}}, \bibinfo {author} {\bibfnamefont {S.}~\bibnamefont {Vagnozzi}}, \bibinfo {author} {\bibfnamefont {K.}~\bibnamefont {Freese}}, \ and\ \bibinfo {author} {\bibfnamefont {S.}~\bibnamefont {Pastor}},\ }\href {\doibase 10.1103/PhysRevD.104.123524} {\bibfield  {journal} {\bibinfo  {journal} {Phys. Rev. D}\ }\textbf {\bibinfo {volume} {104}},\ \bibinfo {pages} {123524} (\bibinfo {year} {2021})},\ \Eprint {http://arxiv.org/abs/2003.02289} {arXiv:2003.02289 [astro-ph.CO]} \BibitemShut {NoStop}%
\bibitem [{\citenamefont {Adamson}\ \emph {et~al.}(2020)\citenamefont {Adamson} \emph {et~al.}}]{MINOS:2020iqj}%
  \BibitemOpen
  \bibfield  {author} {\bibinfo {author} {\bibfnamefont {P.}~\bibnamefont {Adamson}} \emph {et~al.} (\bibinfo {collaboration} {MINOS+, Daya Bay}),\ }\href {\doibase 10.1103/PhysRevLett.125.071801} {\bibfield  {journal} {\bibinfo  {journal} {Phys. Rev. Lett.}\ }\textbf {\bibinfo {volume} {125}},\ \bibinfo {pages} {071801} (\bibinfo {year} {2020})},\ \Eprint {http://arxiv.org/abs/2002.00301} {arXiv:2002.00301 [hep-ex]} \BibitemShut {NoStop}%
\bibitem [{\citenamefont {Aartsen}\ \emph {et~al.}(2020{\natexlab{a}})\citenamefont {Aartsen} \emph {et~al.}}]{IceCube:2020phf}%
  \BibitemOpen
  \bibfield  {author} {\bibinfo {author} {\bibfnamefont {M.~G.}\ \bibnamefont {Aartsen}} \emph {et~al.} (\bibinfo {collaboration} {IceCube}),\ }\href {\doibase 10.1103/PhysRevLett.125.141801} {\bibfield  {journal} {\bibinfo  {journal} {Phys. Rev. Lett.}\ }\textbf {\bibinfo {volume} {125}},\ \bibinfo {pages} {141801} (\bibinfo {year} {2020}{\natexlab{a}})},\ \Eprint {http://arxiv.org/abs/2005.12942} {arXiv:2005.12942 [hep-ex]} \BibitemShut {NoStop}%
\bibitem [{\citenamefont {Aartsen}\ \emph {et~al.}(2020{\natexlab{b}})\citenamefont {Aartsen} \emph {et~al.}}]{IceCube:2020tka}%
  \BibitemOpen
  \bibfield  {author} {\bibinfo {author} {\bibfnamefont {M.~G.}\ \bibnamefont {Aartsen}} \emph {et~al.} (\bibinfo {collaboration} {IceCube}),\ }\href {\doibase 10.1103/PhysRevD.102.052009} {\bibfield  {journal} {\bibinfo  {journal} {Phys. Rev. D}\ }\textbf {\bibinfo {volume} {102}},\ \bibinfo {pages} {052009} (\bibinfo {year} {2020}{\natexlab{b}})},\ \Eprint {http://arxiv.org/abs/2005.12943} {arXiv:2005.12943 [hep-ex]} \BibitemShut {NoStop}%
\bibitem [{\citenamefont {Dentler}\ \emph {et~al.}(2018)\citenamefont {Dentler}, \citenamefont {Hern\'andez-Cabezudo}, \citenamefont {Kopp}, \citenamefont {Machado}, \citenamefont {Maltoni}, \citenamefont {Martinez-Soler},\ and\ \citenamefont {Schwetz}}]{Dentler:2018sju}%
  \BibitemOpen
  \bibfield  {author} {\bibinfo {author} {\bibfnamefont {M.}~\bibnamefont {Dentler}}, \bibinfo {author} {\bibfnamefont {A.}~\bibnamefont {Hern\'andez-Cabezudo}}, \bibinfo {author} {\bibfnamefont {J.}~\bibnamefont {Kopp}}, \bibinfo {author} {\bibfnamefont {P.~A.~N.}\ \bibnamefont {Machado}}, \bibinfo {author} {\bibfnamefont {M.}~\bibnamefont {Maltoni}}, \bibinfo {author} {\bibfnamefont {I.}~\bibnamefont {Martinez-Soler}}, \ and\ \bibinfo {author} {\bibfnamefont {T.}~\bibnamefont {Schwetz}},\ }\href {\doibase 10.1007/JHEP08(2018)010} {\bibfield  {journal} {\bibinfo  {journal} {JHEP}\ }\textbf {\bibinfo {volume} {08}},\ \bibinfo {pages} {010} (\bibinfo {year} {2018})},\ \Eprint {http://arxiv.org/abs/1803.10661} {arXiv:1803.10661 [hep-ph]} \BibitemShut {NoStop}%
\bibitem [{\citenamefont {Diaz}\ \emph {et~al.}(2020)\citenamefont {Diaz}, \citenamefont {Arg\"uelles}, \citenamefont {Collin}, \citenamefont {Conrad},\ and\ \citenamefont {Shaevitz}}]{Diaz:2019fwt}%
  \BibitemOpen
  \bibfield  {author} {\bibinfo {author} {\bibfnamefont {A.}~\bibnamefont {Diaz}}, \bibinfo {author} {\bibfnamefont {C.~A.}\ \bibnamefont {Arg\"uelles}}, \bibinfo {author} {\bibfnamefont {G.~H.}\ \bibnamefont {Collin}}, \bibinfo {author} {\bibfnamefont {J.~M.}\ \bibnamefont {Conrad}}, \ and\ \bibinfo {author} {\bibfnamefont {M.~H.}\ \bibnamefont {Shaevitz}},\ }\href {\doibase 10.1016/j.physrep.2020.08.005} {\bibfield  {journal} {\bibinfo  {journal} {Phys. Rept.}\ }\textbf {\bibinfo {volume} {884}},\ \bibinfo {pages} {1} (\bibinfo {year} {2020})},\ \Eprint {http://arxiv.org/abs/1906.00045} {arXiv:1906.00045 [hep-ex]} \BibitemShut {NoStop}%
\bibitem [{\citenamefont {B\"oser}\ \emph {et~al.}(2020)\citenamefont {B\"oser}, \citenamefont {Buck}, \citenamefont {Giunti}, \citenamefont {Lesgourgues}, \citenamefont {Ludhova}, \citenamefont {Mertens}, \citenamefont {Schukraft},\ and\ \citenamefont {Wurm}}]{Boser:2019rta}%
  \BibitemOpen
  \bibfield  {author} {\bibinfo {author} {\bibfnamefont {S.}~\bibnamefont {B\"oser}}, \bibinfo {author} {\bibfnamefont {C.}~\bibnamefont {Buck}}, \bibinfo {author} {\bibfnamefont {C.}~\bibnamefont {Giunti}}, \bibinfo {author} {\bibfnamefont {J.}~\bibnamefont {Lesgourgues}}, \bibinfo {author} {\bibfnamefont {L.}~\bibnamefont {Ludhova}}, \bibinfo {author} {\bibfnamefont {S.}~\bibnamefont {Mertens}}, \bibinfo {author} {\bibfnamefont {A.}~\bibnamefont {Schukraft}}, \ and\ \bibinfo {author} {\bibfnamefont {M.}~\bibnamefont {Wurm}},\ }\href {\doibase 10.1016/j.ppnp.2019.103736} {\bibfield  {journal} {\bibinfo  {journal} {Prog. Part. Nucl. Phys.}\ }\textbf {\bibinfo {volume} {111}},\ \bibinfo {pages} {103736} (\bibinfo {year} {2020})},\ \Eprint {http://arxiv.org/abs/1906.01739} {arXiv:1906.01739 [hep-ex]} \BibitemShut {NoStop}%
\bibitem [{\citenamefont {Acero}\ \emph {et~al.}(2022)\citenamefont {Acero} \emph {et~al.}}]{Acero:2022wqg}%
  \BibitemOpen
  \bibfield  {author} {\bibinfo {author} {\bibfnamefont {M.~A.}\ \bibnamefont {Acero}} \emph {et~al.},\ }\href@noop {} {\  (\bibinfo {year} {2022})},\ \Eprint {http://arxiv.org/abs/2203.07323} {arXiv:2203.07323 [hep-ex]} \BibitemShut {NoStop}%
\bibitem [{\citenamefont {Abdullahi}\ \emph {et~al.}(2023)\citenamefont {Abdullahi}, \citenamefont {Hoefken~Zink}, \citenamefont {Hostert}, \citenamefont {Massaro},\ and\ \citenamefont {Pascoli}}]{Abdullahi:2023ejc}%
  \BibitemOpen
  \bibfield  {author} {\bibinfo {author} {\bibfnamefont {A.~M.}\ \bibnamefont {Abdullahi}}, \bibinfo {author} {\bibfnamefont {J.}~\bibnamefont {Hoefken~Zink}}, \bibinfo {author} {\bibfnamefont {M.}~\bibnamefont {Hostert}}, \bibinfo {author} {\bibfnamefont {D.}~\bibnamefont {Massaro}}, \ and\ \bibinfo {author} {\bibfnamefont {S.}~\bibnamefont {Pascoli}},\ }\href@noop {} {\  (\bibinfo {year} {2023})},\ \Eprint {http://arxiv.org/abs/2308.02543} {arXiv:2308.02543 [hep-ph]} \BibitemShut {NoStop}%
\bibitem [{\citenamefont {Moss}\ \emph {et~al.}(2018)\citenamefont {Moss}, \citenamefont {Moulai}, \citenamefont {Arg\"uelles},\ and\ \citenamefont {Conrad}}]{Moss:2017pur}%
  \BibitemOpen
  \bibfield  {author} {\bibinfo {author} {\bibfnamefont {Z.}~\bibnamefont {Moss}}, \bibinfo {author} {\bibfnamefont {M.~H.}\ \bibnamefont {Moulai}}, \bibinfo {author} {\bibfnamefont {C.~A.}\ \bibnamefont {Arg\"uelles}}, \ and\ \bibinfo {author} {\bibfnamefont {J.~M.}\ \bibnamefont {Conrad}},\ }\href {\doibase 10.1103/PhysRevD.97.055017} {\bibfield  {journal} {\bibinfo  {journal} {Phys. Rev. D}\ }\textbf {\bibinfo {volume} {97}},\ \bibinfo {pages} {055017} (\bibinfo {year} {2018})},\ \Eprint {http://arxiv.org/abs/1711.05921} {arXiv:1711.05921 [hep-ph]} \BibitemShut {NoStop}%
\bibitem [{\citenamefont {Moulai}\ \emph {et~al.}(2020)\citenamefont {Moulai}, \citenamefont {Arg\"uelles}, \citenamefont {Collin}, \citenamefont {Conrad}, \citenamefont {Diaz},\ and\ \citenamefont {Shaevitz}}]{Moulai:2019gpi}%
  \BibitemOpen
  \bibfield  {author} {\bibinfo {author} {\bibfnamefont {M.~H.}\ \bibnamefont {Moulai}}, \bibinfo {author} {\bibfnamefont {C.~A.}\ \bibnamefont {Arg\"uelles}}, \bibinfo {author} {\bibfnamefont {G.~H.}\ \bibnamefont {Collin}}, \bibinfo {author} {\bibfnamefont {J.~M.}\ \bibnamefont {Conrad}}, \bibinfo {author} {\bibfnamefont {A.}~\bibnamefont {Diaz}}, \ and\ \bibinfo {author} {\bibfnamefont {M.~H.}\ \bibnamefont {Shaevitz}},\ }\href {\doibase 10.1103/PhysRevD.101.055020} {\bibfield  {journal} {\bibinfo  {journal} {Phys. Rev. D}\ }\textbf {\bibinfo {volume} {101}},\ \bibinfo {pages} {055020} (\bibinfo {year} {2020})},\ \Eprint {http://arxiv.org/abs/1910.13456} {arXiv:1910.13456 [hep-ph]} \BibitemShut {NoStop}%
\bibitem [{\citenamefont {Akhmedov}\ and\ \citenamefont {Schwetz}(2011)}]{Akhmedov:2011zza}%
  \BibitemOpen
  \bibfield  {author} {\bibinfo {author} {\bibfnamefont {E.~K.}\ \bibnamefont {Akhmedov}}\ and\ \bibinfo {author} {\bibfnamefont {T.}~\bibnamefont {Schwetz}},\ }\href {\doibase 10.1016/j.nuclphysbps.2011.04.106} {\bibfield  {journal} {\bibinfo  {journal} {Nucl. Phys. B Proc. Suppl.}\ }\textbf {\bibinfo {volume} {217}},\ \bibinfo {pages} {217} (\bibinfo {year} {2011})}\BibitemShut {NoStop}%
\bibitem [{\citenamefont {Bramante}(2013)}]{Bramante:2011uu}%
  \BibitemOpen
  \bibfield  {author} {\bibinfo {author} {\bibfnamefont {J.}~\bibnamefont {Bramante}},\ }\href {\doibase 10.1142/S0217751X1350067X} {\bibfield  {journal} {\bibinfo  {journal} {Int. J. Mod. Phys. A}\ }\textbf {\bibinfo {volume} {28}},\ \bibinfo {pages} {1350067} (\bibinfo {year} {2013})},\ \Eprint {http://arxiv.org/abs/1110.4871} {arXiv:1110.4871 [hep-ph]} \BibitemShut {NoStop}%
\bibitem [{\citenamefont {Karagiorgi}\ \emph {et~al.}(2012)\citenamefont {Karagiorgi}, \citenamefont {Shaevitz},\ and\ \citenamefont {Conrad}}]{Karagiorgi:2012kw}%
  \BibitemOpen
  \bibfield  {author} {\bibinfo {author} {\bibfnamefont {G.}~\bibnamefont {Karagiorgi}}, \bibinfo {author} {\bibfnamefont {M.~H.}\ \bibnamefont {Shaevitz}}, \ and\ \bibinfo {author} {\bibfnamefont {J.~M.}\ \bibnamefont {Conrad}},\ }\href@noop {} {\  (\bibinfo {year} {2012})},\ \Eprint {http://arxiv.org/abs/1202.1024} {arXiv:1202.1024 [hep-ph]} \BibitemShut {NoStop}%
\bibitem [{\citenamefont {Asaadi}\ \emph {et~al.}(2018)\citenamefont {Asaadi}, \citenamefont {Church}, \citenamefont {Guenette}, \citenamefont {Jones},\ and\ \citenamefont {Szelc}}]{Asaadi:2017bhx}%
  \BibitemOpen
  \bibfield  {author} {\bibinfo {author} {\bibfnamefont {J.}~\bibnamefont {Asaadi}}, \bibinfo {author} {\bibfnamefont {E.}~\bibnamefont {Church}}, \bibinfo {author} {\bibfnamefont {R.}~\bibnamefont {Guenette}}, \bibinfo {author} {\bibfnamefont {B.~J.~P.}\ \bibnamefont {Jones}}, \ and\ \bibinfo {author} {\bibfnamefont {A.~M.}\ \bibnamefont {Szelc}},\ }\href {\doibase 10.1103/PhysRevD.97.075021} {\bibfield  {journal} {\bibinfo  {journal} {Phys. Rev. D}\ }\textbf {\bibinfo {volume} {97}},\ \bibinfo {pages} {075021} (\bibinfo {year} {2018})},\ \Eprint {http://arxiv.org/abs/1712.08019} {arXiv:1712.08019 [hep-ph]} \BibitemShut {NoStop}%
\bibitem [{\citenamefont {Smirnov}\ and\ \citenamefont {Valera}(2021)}]{Smirnov:2021zgn}%
  \BibitemOpen
  \bibfield  {author} {\bibinfo {author} {\bibfnamefont {A.~Y.}\ \bibnamefont {Smirnov}}\ and\ \bibinfo {author} {\bibfnamefont {V.~B.}\ \bibnamefont {Valera}},\ }\href {\doibase 10.1007/JHEP09(2021)177} {\bibfield  {journal} {\bibinfo  {journal} {JHEP}\ }\textbf {\bibinfo {volume} {09}},\ \bibinfo {pages} {177} (\bibinfo {year} {2021})},\ \Eprint {http://arxiv.org/abs/2106.13829} {arXiv:2106.13829 [hep-ph]} \BibitemShut {NoStop}%
\bibitem [{\citenamefont {Alves}\ \emph {et~al.}(2022)\citenamefont {Alves}, \citenamefont {Louis},\ and\ \citenamefont {deNiverville}}]{Alves:2022vgn}%
  \BibitemOpen
  \bibfield  {author} {\bibinfo {author} {\bibfnamefont {D.~S.~M.}\ \bibnamefont {Alves}}, \bibinfo {author} {\bibfnamefont {W.~C.}\ \bibnamefont {Louis}}, \ and\ \bibinfo {author} {\bibfnamefont {P.~G.}\ \bibnamefont {deNiverville}},\ }\href {\doibase 10.1007/JHEP08(2022)034} {\bibfield  {journal} {\bibinfo  {journal} {JHEP}\ }\textbf {\bibinfo {volume} {08}},\ \bibinfo {pages} {034} (\bibinfo {year} {2022})},\ \Eprint {http://arxiv.org/abs/2201.00876} {arXiv:2201.00876 [hep-ph]} \BibitemShut {NoStop}%
\bibitem [{\citenamefont {Palomares-Ruiz}\ \emph {et~al.}(2005)\citenamefont {Palomares-Ruiz}, \citenamefont {Pascoli},\ and\ \citenamefont {Schwetz}}]{Palomares-Ruiz:2005zbh}%
  \BibitemOpen
  \bibfield  {author} {\bibinfo {author} {\bibfnamefont {S.}~\bibnamefont {Palomares-Ruiz}}, \bibinfo {author} {\bibfnamefont {S.}~\bibnamefont {Pascoli}}, \ and\ \bibinfo {author} {\bibfnamefont {T.}~\bibnamefont {Schwetz}},\ }\href {\doibase 10.1088/1126-6708/2005/09/048} {\bibfield  {journal} {\bibinfo  {journal} {JHEP}\ }\textbf {\bibinfo {volume} {09}},\ \bibinfo {pages} {048} (\bibinfo {year} {2005})},\ \Eprint {http://arxiv.org/abs/hep-ph/0505216} {arXiv:hep-ph/0505216} \BibitemShut {NoStop}%
\bibitem [{\citenamefont {Bai}\ \emph {et~al.}(2016)\citenamefont {Bai}, \citenamefont {Lu}, \citenamefont {Lu}, \citenamefont {Salvado},\ and\ \citenamefont {Stefanek}}]{Bai:2015ztj}%
  \BibitemOpen
  \bibfield  {author} {\bibinfo {author} {\bibfnamefont {Y.}~\bibnamefont {Bai}}, \bibinfo {author} {\bibfnamefont {R.}~\bibnamefont {Lu}}, \bibinfo {author} {\bibfnamefont {S.}~\bibnamefont {Lu}}, \bibinfo {author} {\bibfnamefont {J.}~\bibnamefont {Salvado}}, \ and\ \bibinfo {author} {\bibfnamefont {B.~A.}\ \bibnamefont {Stefanek}},\ }\href {\doibase 10.1103/PhysRevD.93.073004} {\bibfield  {journal} {\bibinfo  {journal} {Phys. Rev. D}\ }\textbf {\bibinfo {volume} {93}},\ \bibinfo {pages} {073004} (\bibinfo {year} {2016})},\ \Eprint {http://arxiv.org/abs/1512.05357} {arXiv:1512.05357 [hep-ph]} \BibitemShut {NoStop}%
\bibitem [{\citenamefont {de~Gouv\^ea}\ \emph {et~al.}(2020)\citenamefont {de~Gouv\^ea}, \citenamefont {Peres}, \citenamefont {Prakash},\ and\ \citenamefont {Stenico}}]{deGouvea:2019qre}%
  \BibitemOpen
  \bibfield  {author} {\bibinfo {author} {\bibfnamefont {A.}~\bibnamefont {de~Gouv\^ea}}, \bibinfo {author} {\bibfnamefont {O.~L.~G.}\ \bibnamefont {Peres}}, \bibinfo {author} {\bibfnamefont {S.}~\bibnamefont {Prakash}}, \ and\ \bibinfo {author} {\bibfnamefont {G.~V.}\ \bibnamefont {Stenico}},\ }\href {\doibase 10.1007/JHEP07(2020)141} {\bibfield  {journal} {\bibinfo  {journal} {JHEP}\ }\textbf {\bibinfo {volume} {07}},\ \bibinfo {pages} {141} (\bibinfo {year} {2020})},\ \Eprint {http://arxiv.org/abs/1911.01447} {arXiv:1911.01447 [hep-ph]} \BibitemShut {NoStop}%
\bibitem [{\citenamefont {Dentler}\ \emph {et~al.}(2020)\citenamefont {Dentler}, \citenamefont {Esteban}, \citenamefont {Kopp},\ and\ \citenamefont {Machado}}]{Dentler:2019dhz}%
  \BibitemOpen
  \bibfield  {author} {\bibinfo {author} {\bibfnamefont {M.}~\bibnamefont {Dentler}}, \bibinfo {author} {\bibfnamefont {I.}~\bibnamefont {Esteban}}, \bibinfo {author} {\bibfnamefont {J.}~\bibnamefont {Kopp}}, \ and\ \bibinfo {author} {\bibfnamefont {P.}~\bibnamefont {Machado}},\ }\href {\doibase 10.1103/PhysRevD.101.115013} {\bibfield  {journal} {\bibinfo  {journal} {Phys. Rev. D}\ }\textbf {\bibinfo {volume} {101}},\ \bibinfo {pages} {115013} (\bibinfo {year} {2020})},\ \Eprint {http://arxiv.org/abs/1911.01427} {arXiv:1911.01427 [hep-ph]} \BibitemShut {NoStop}%
\bibitem [{\citenamefont {Hostert}\ and\ \citenamefont {Pospelov}(2021)}]{Hostert:2020oui}%
  \BibitemOpen
  \bibfield  {author} {\bibinfo {author} {\bibfnamefont {M.}~\bibnamefont {Hostert}}\ and\ \bibinfo {author} {\bibfnamefont {M.}~\bibnamefont {Pospelov}},\ }\href {\doibase 10.1103/PhysRevD.104.055031} {\bibfield  {journal} {\bibinfo  {journal} {Phys. Rev. D}\ }\textbf {\bibinfo {volume} {104}},\ \bibinfo {pages} {055031} (\bibinfo {year} {2021})},\ \Eprint {http://arxiv.org/abs/2008.11851} {arXiv:2008.11851 [hep-ph]} \BibitemShut {NoStop}%
\bibitem [{\citenamefont {Chang}\ \emph {et~al.}(2021)\citenamefont {Chang}, \citenamefont {Chen}, \citenamefont {Ho},\ and\ \citenamefont {Tseng}}]{Chang:2021myh}%
  \BibitemOpen
  \bibfield  {author} {\bibinfo {author} {\bibfnamefont {C.-H.~V.}\ \bibnamefont {Chang}}, \bibinfo {author} {\bibfnamefont {C.-R.}\ \bibnamefont {Chen}}, \bibinfo {author} {\bibfnamefont {S.-Y.}\ \bibnamefont {Ho}}, \ and\ \bibinfo {author} {\bibfnamefont {S.-Y.}\ \bibnamefont {Tseng}},\ }\href {\doibase 10.1103/PhysRevD.104.015030} {\bibfield  {journal} {\bibinfo  {journal} {Phys. Rev. D}\ }\textbf {\bibinfo {volume} {104}},\ \bibinfo {pages} {015030} (\bibinfo {year} {2021})},\ \Eprint {http://arxiv.org/abs/2102.05012} {arXiv:2102.05012 [hep-ph]} \BibitemShut {NoStop}%
\bibitem [{\citenamefont {Gninenko}(2009)}]{Gninenko:2009ks}%
  \BibitemOpen
  \bibfield  {author} {\bibinfo {author} {\bibfnamefont {S.~N.}\ \bibnamefont {Gninenko}},\ }\href {\doibase 10.1103/PhysRevLett.103.241802} {\bibfield  {journal} {\bibinfo  {journal} {Phys. Rev. Lett.}\ }\textbf {\bibinfo {volume} {103}},\ \bibinfo {pages} {241802} (\bibinfo {year} {2009})},\ \Eprint {http://arxiv.org/abs/0902.3802} {arXiv:0902.3802 [hep-ph]} \BibitemShut {NoStop}%
\bibitem [{\citenamefont {Gninenko}(2011)}]{Gninenko:2010pr}%
  \BibitemOpen
  \bibfield  {author} {\bibinfo {author} {\bibfnamefont {S.~N.}\ \bibnamefont {Gninenko}},\ }\href {\doibase 10.1103/PhysRevD.83.015015} {\bibfield  {journal} {\bibinfo  {journal} {Phys. Rev. D}\ }\textbf {\bibinfo {volume} {83}},\ \bibinfo {pages} {015015} (\bibinfo {year} {2011})},\ \Eprint {http://arxiv.org/abs/1009.5536} {arXiv:1009.5536 [hep-ph]} \BibitemShut {NoStop}%
\bibitem [{\citenamefont {Gninenko}(2012)}]{Gninenko:2012rw}%
  \BibitemOpen
  \bibfield  {author} {\bibinfo {author} {\bibfnamefont {S.~N.}\ \bibnamefont {Gninenko}},\ }\href {\doibase 10.1016/j.physletb.2012.02.071} {\bibfield  {journal} {\bibinfo  {journal} {Phys. Lett. B}\ }\textbf {\bibinfo {volume} {710}},\ \bibinfo {pages} {86} (\bibinfo {year} {2012})},\ \Eprint {http://arxiv.org/abs/1201.5194} {arXiv:1201.5194 [hep-ph]} \BibitemShut {NoStop}%
\bibitem [{\citenamefont {Masip}\ \emph {et~al.}(2013)\citenamefont {Masip}, \citenamefont {Masjuan},\ and\ \citenamefont {Meloni}}]{Masip:2012ke}%
  \BibitemOpen
  \bibfield  {author} {\bibinfo {author} {\bibfnamefont {M.}~\bibnamefont {Masip}}, \bibinfo {author} {\bibfnamefont {P.}~\bibnamefont {Masjuan}}, \ and\ \bibinfo {author} {\bibfnamefont {D.}~\bibnamefont {Meloni}},\ }\href {\doibase 10.1007/JHEP01(2013)106} {\bibfield  {journal} {\bibinfo  {journal} {JHEP}\ }\textbf {\bibinfo {volume} {01}},\ \bibinfo {pages} {106} (\bibinfo {year} {2013})},\ \Eprint {http://arxiv.org/abs/1210.1519} {arXiv:1210.1519 [hep-ph]} \BibitemShut {NoStop}%
\bibitem [{\citenamefont {Radionov}(2013)}]{Radionov:2013mca}%
  \BibitemOpen
  \bibfield  {author} {\bibinfo {author} {\bibfnamefont {A.}~\bibnamefont {Radionov}},\ }\href {\doibase 10.1103/PhysRevD.88.015016} {\bibfield  {journal} {\bibinfo  {journal} {Phys. Rev. D}\ }\textbf {\bibinfo {volume} {88}},\ \bibinfo {pages} {015016} (\bibinfo {year} {2013})},\ \Eprint {http://arxiv.org/abs/1303.4587} {arXiv:1303.4587 [hep-ph]} \BibitemShut {NoStop}%
\bibitem [{\citenamefont {Magill}\ \emph {et~al.}(2018)\citenamefont {Magill}, \citenamefont {Plestid}, \citenamefont {Pospelov},\ and\ \citenamefont {Tsai}}]{Magill:2018jla}%
  \BibitemOpen
  \bibfield  {author} {\bibinfo {author} {\bibfnamefont {G.}~\bibnamefont {Magill}}, \bibinfo {author} {\bibfnamefont {R.}~\bibnamefont {Plestid}}, \bibinfo {author} {\bibfnamefont {M.}~\bibnamefont {Pospelov}}, \ and\ \bibinfo {author} {\bibfnamefont {Y.-D.}\ \bibnamefont {Tsai}},\ }\href {\doibase 10.1103/PhysRevD.98.115015} {\bibfield  {journal} {\bibinfo  {journal} {Phys. Rev. D}\ }\textbf {\bibinfo {volume} {98}},\ \bibinfo {pages} {115015} (\bibinfo {year} {2018})},\ \Eprint {http://arxiv.org/abs/1803.03262} {arXiv:1803.03262 [hep-ph]} \BibitemShut {NoStop}%
\bibitem [{\citenamefont {Bertuzzo}\ \emph {et~al.}(2018)\citenamefont {Bertuzzo}, \citenamefont {Jana}, \citenamefont {Machado},\ and\ \citenamefont {Zukanovich~Funchal}}]{Bertuzzo:2018itn}%
  \BibitemOpen
  \bibfield  {author} {\bibinfo {author} {\bibfnamefont {E.}~\bibnamefont {Bertuzzo}}, \bibinfo {author} {\bibfnamefont {S.}~\bibnamefont {Jana}}, \bibinfo {author} {\bibfnamefont {P.~A.~N.}\ \bibnamefont {Machado}}, \ and\ \bibinfo {author} {\bibfnamefont {R.}~\bibnamefont {Zukanovich~Funchal}},\ }\href {\doibase 10.1103/PhysRevLett.121.241801} {\bibfield  {journal} {\bibinfo  {journal} {Phys. Rev. Lett.}\ }\textbf {\bibinfo {volume} {121}},\ \bibinfo {pages} {241801} (\bibinfo {year} {2018})},\ \Eprint {http://arxiv.org/abs/1807.09877} {arXiv:1807.09877 [hep-ph]} \BibitemShut {NoStop}%
\bibitem [{\citenamefont {Ballett}\ \emph {et~al.}(2019)\citenamefont {Ballett}, \citenamefont {Pascoli},\ and\ \citenamefont {Ross-Lonergan}}]{Ballett:2018ynz}%
  \BibitemOpen
  \bibfield  {author} {\bibinfo {author} {\bibfnamefont {P.}~\bibnamefont {Ballett}}, \bibinfo {author} {\bibfnamefont {S.}~\bibnamefont {Pascoli}}, \ and\ \bibinfo {author} {\bibfnamefont {M.}~\bibnamefont {Ross-Lonergan}},\ }\href {\doibase 10.1103/PhysRevD.99.071701} {\bibfield  {journal} {\bibinfo  {journal} {Phys. Rev. D}\ }\textbf {\bibinfo {volume} {99}},\ \bibinfo {pages} {071701} (\bibinfo {year} {2019})},\ \Eprint {http://arxiv.org/abs/1808.02915} {arXiv:1808.02915 [hep-ph]} \BibitemShut {NoStop}%
\bibitem [{\citenamefont {Ballett}\ \emph {et~al.}(2020)\citenamefont {Ballett}, \citenamefont {Hostert},\ and\ \citenamefont {Pascoli}}]{Ballett:2019pyw}%
  \BibitemOpen
  \bibfield  {author} {\bibinfo {author} {\bibfnamefont {P.}~\bibnamefont {Ballett}}, \bibinfo {author} {\bibfnamefont {M.}~\bibnamefont {Hostert}}, \ and\ \bibinfo {author} {\bibfnamefont {S.}~\bibnamefont {Pascoli}},\ }\href {\doibase 10.1103/PhysRevD.101.115025} {\bibfield  {journal} {\bibinfo  {journal} {Phys. Rev. D}\ }\textbf {\bibinfo {volume} {101}},\ \bibinfo {pages} {115025} (\bibinfo {year} {2020})},\ \Eprint {http://arxiv.org/abs/1903.07589} {arXiv:1903.07589 [hep-ph]} \BibitemShut {NoStop}%
\bibitem [{\citenamefont {Datta}\ \emph {et~al.}(2020)\citenamefont {Datta}, \citenamefont {Kamali},\ and\ \citenamefont {Marfatia}}]{Datta:2020auq}%
  \BibitemOpen
  \bibfield  {author} {\bibinfo {author} {\bibfnamefont {A.}~\bibnamefont {Datta}}, \bibinfo {author} {\bibfnamefont {S.}~\bibnamefont {Kamali}}, \ and\ \bibinfo {author} {\bibfnamefont {D.}~\bibnamefont {Marfatia}},\ }\href {\doibase 10.1016/j.physletb.2020.135579} {\bibfield  {journal} {\bibinfo  {journal} {Phys. Lett. B}\ }\textbf {\bibinfo {volume} {807}},\ \bibinfo {pages} {135579} (\bibinfo {year} {2020})},\ \Eprint {http://arxiv.org/abs/2005.08920} {arXiv:2005.08920 [hep-ph]} \BibitemShut {NoStop}%
\bibitem [{\citenamefont {Dutta}\ \emph {et~al.}(2020)\citenamefont {Dutta}, \citenamefont {Ghosh},\ and\ \citenamefont {Li}}]{Dutta:2020scq}%
  \BibitemOpen
  \bibfield  {author} {\bibinfo {author} {\bibfnamefont {B.}~\bibnamefont {Dutta}}, \bibinfo {author} {\bibfnamefont {S.}~\bibnamefont {Ghosh}}, \ and\ \bibinfo {author} {\bibfnamefont {T.}~\bibnamefont {Li}},\ }\href {\doibase 10.1103/PhysRevD.102.055017} {\bibfield  {journal} {\bibinfo  {journal} {Phys. Rev. D}\ }\textbf {\bibinfo {volume} {102}},\ \bibinfo {pages} {055017} (\bibinfo {year} {2020})},\ \Eprint {http://arxiv.org/abs/2006.01319} {arXiv:2006.01319 [hep-ph]} \BibitemShut {NoStop}%
\bibitem [{\citenamefont {Abdallah}\ \emph {et~al.}(2020)\citenamefont {Abdallah}, \citenamefont {Gandhi},\ and\ \citenamefont {Roy}}]{Abdallah:2020biq}%
  \BibitemOpen
  \bibfield  {author} {\bibinfo {author} {\bibfnamefont {W.}~\bibnamefont {Abdallah}}, \bibinfo {author} {\bibfnamefont {R.}~\bibnamefont {Gandhi}}, \ and\ \bibinfo {author} {\bibfnamefont {S.}~\bibnamefont {Roy}},\ }\href {\doibase 10.1007/JHEP12(2020)188} {\bibfield  {journal} {\bibinfo  {journal} {JHEP}\ }\textbf {\bibinfo {volume} {12}},\ \bibinfo {pages} {188} (\bibinfo {year} {2020})},\ \Eprint {http://arxiv.org/abs/2006.01948} {arXiv:2006.01948 [hep-ph]} \BibitemShut {NoStop}%
\bibitem [{\citenamefont {Abdullahi}\ \emph {et~al.}(2021)\citenamefont {Abdullahi}, \citenamefont {Hostert},\ and\ \citenamefont {Pascoli}}]{Abdullahi:2020nyr}%
  \BibitemOpen
  \bibfield  {author} {\bibinfo {author} {\bibfnamefont {A.}~\bibnamefont {Abdullahi}}, \bibinfo {author} {\bibfnamefont {M.}~\bibnamefont {Hostert}}, \ and\ \bibinfo {author} {\bibfnamefont {S.}~\bibnamefont {Pascoli}},\ }\href {\doibase 10.1016/j.physletb.2021.136531} {\bibfield  {journal} {\bibinfo  {journal} {Phys. Lett. B}\ }\textbf {\bibinfo {volume} {820}},\ \bibinfo {pages} {136531} (\bibinfo {year} {2021})},\ \Eprint {http://arxiv.org/abs/2007.11813} {arXiv:2007.11813 [hep-ph]} \BibitemShut {NoStop}%
\bibitem [{\citenamefont {Abdallah}\ \emph {et~al.}(2021)\citenamefont {Abdallah}, \citenamefont {Gandhi},\ and\ \citenamefont {Roy}}]{Abdallah:2020vgg}%
  \BibitemOpen
  \bibfield  {author} {\bibinfo {author} {\bibfnamefont {W.}~\bibnamefont {Abdallah}}, \bibinfo {author} {\bibfnamefont {R.}~\bibnamefont {Gandhi}}, \ and\ \bibinfo {author} {\bibfnamefont {S.}~\bibnamefont {Roy}},\ }\href {\doibase 10.1103/PhysRevD.104.055028} {\bibfield  {journal} {\bibinfo  {journal} {Phys. Rev. D}\ }\textbf {\bibinfo {volume} {104}},\ \bibinfo {pages} {055028} (\bibinfo {year} {2021})},\ \Eprint {http://arxiv.org/abs/2010.06159} {arXiv:2010.06159 [hep-ph]} \BibitemShut {NoStop}%
\bibitem [{\citenamefont {Schwetz}\ \emph {et~al.}(2020)\citenamefont {Schwetz}, \citenamefont {Zhou},\ and\ \citenamefont {Zhu}}]{Schwetz:2020xra}%
  \BibitemOpen
  \bibfield  {author} {\bibinfo {author} {\bibfnamefont {T.}~\bibnamefont {Schwetz}}, \bibinfo {author} {\bibfnamefont {A.}~\bibnamefont {Zhou}}, \ and\ \bibinfo {author} {\bibfnamefont {J.-Y.}\ \bibnamefont {Zhu}},\ }\href {\doibase 10.1007/JHEP07(2021)200} {\bibfield  {journal} {\bibinfo  {journal} {JHEP}\ }\textbf {\bibinfo {volume} {21}},\ \bibinfo {pages} {200} (\bibinfo {year} {2020})},\ \Eprint {http://arxiv.org/abs/2105.09699} {arXiv:2105.09699 [hep-ph]} \BibitemShut {NoStop}%
\bibitem [{\citenamefont {Vergani}\ \emph {et~al.}(2021)\citenamefont {Vergani}, \citenamefont {Kamp}, \citenamefont {Diaz}, \citenamefont {Arg\"uelles}, \citenamefont {Conrad}, \citenamefont {Shaevitz},\ and\ \citenamefont {Uchida}}]{Vergani:2021tgc}%
  \BibitemOpen
  \bibfield  {author} {\bibinfo {author} {\bibfnamefont {S.}~\bibnamefont {Vergani}}, \bibinfo {author} {\bibfnamefont {N.~W.}\ \bibnamefont {Kamp}}, \bibinfo {author} {\bibfnamefont {A.}~\bibnamefont {Diaz}}, \bibinfo {author} {\bibfnamefont {C.~A.}\ \bibnamefont {Arg\"uelles}}, \bibinfo {author} {\bibfnamefont {J.~M.}\ \bibnamefont {Conrad}}, \bibinfo {author} {\bibfnamefont {M.~H.}\ \bibnamefont {Shaevitz}}, \ and\ \bibinfo {author} {\bibfnamefont {M.~A.}\ \bibnamefont {Uchida}},\ }\href {\doibase 10.1103/PhysRevD.104.095005} {\bibfield  {journal} {\bibinfo  {journal} {Phys. Rev. D}\ }\textbf {\bibinfo {volume} {104}},\ \bibinfo {pages} {095005} (\bibinfo {year} {2021})},\ \Eprint {http://arxiv.org/abs/2105.06470} {arXiv:2105.06470 [hep-ph]} \BibitemShut {NoStop}%
\bibitem [{\citenamefont {Hammad}\ \emph {et~al.}(2022)\citenamefont {Hammad}, \citenamefont {Rashed},\ and\ \citenamefont {Moretti}}]{Hammad:2021mpl}%
  \BibitemOpen
  \bibfield  {author} {\bibinfo {author} {\bibfnamefont {A.}~\bibnamefont {Hammad}}, \bibinfo {author} {\bibfnamefont {A.}~\bibnamefont {Rashed}}, \ and\ \bibinfo {author} {\bibfnamefont {S.}~\bibnamefont {Moretti}},\ }\href {\doibase 10.1016/j.physletb.2022.136945} {\bibfield  {journal} {\bibinfo  {journal} {Phys. Lett. B}\ }\textbf {\bibinfo {volume} {827}},\ \bibinfo {pages} {136945} (\bibinfo {year} {2022})},\ \Eprint {http://arxiv.org/abs/2110.08651} {arXiv:2110.08651 [hep-ph]} \BibitemShut {NoStop}%
\bibitem [{\citenamefont {Dutta}\ \emph {et~al.}(2022)\citenamefont {Dutta}, \citenamefont {Kim}, \citenamefont {Thompson}, \citenamefont {Thornton},\ and\ \citenamefont {Van~de Water}}]{Dutta:2021cip}%
  \BibitemOpen
  \bibfield  {author} {\bibinfo {author} {\bibfnamefont {B.}~\bibnamefont {Dutta}}, \bibinfo {author} {\bibfnamefont {D.}~\bibnamefont {Kim}}, \bibinfo {author} {\bibfnamefont {A.}~\bibnamefont {Thompson}}, \bibinfo {author} {\bibfnamefont {R.~T.}\ \bibnamefont {Thornton}}, \ and\ \bibinfo {author} {\bibfnamefont {R.~G.}\ \bibnamefont {Van~de Water}},\ }\href {\doibase 10.1103/PhysRevLett.129.111803} {\bibfield  {journal} {\bibinfo  {journal} {Phys. Rev. Lett.}\ }\textbf {\bibinfo {volume} {129}},\ \bibinfo {pages} {111803} (\bibinfo {year} {2022})},\ \Eprint {http://arxiv.org/abs/2110.11944} {arXiv:2110.11944 [hep-ph]} \BibitemShut {NoStop}%
\bibitem [{\citenamefont {Abdallah}\ \emph {et~al.}(2022)\citenamefont {Abdallah}, \citenamefont {Gandhi},\ and\ \citenamefont {Roy}}]{Abdallah:2022grs}%
  \BibitemOpen
  \bibfield  {author} {\bibinfo {author} {\bibfnamefont {W.}~\bibnamefont {Abdallah}}, \bibinfo {author} {\bibfnamefont {R.}~\bibnamefont {Gandhi}}, \ and\ \bibinfo {author} {\bibfnamefont {S.}~\bibnamefont {Roy}},\ }\href {\doibase 10.1007/JHEP06(2022)160} {\bibfield  {journal} {\bibinfo  {journal} {JHEP}\ }\textbf {\bibinfo {volume} {06}},\ \bibinfo {pages} {160} (\bibinfo {year} {2022})},\ \Eprint {http://arxiv.org/abs/2202.09373} {arXiv:2202.09373 [hep-ph]} \BibitemShut {NoStop}%
\bibitem [{\citenamefont {Kamp}\ \emph {et~al.}(2023)\citenamefont {Kamp}, \citenamefont {Hostert}, \citenamefont {Schneider}, \citenamefont {Vergani}, \citenamefont {Arg\"uelles}, \citenamefont {Conrad}, \citenamefont {Shaevitz},\ and\ \citenamefont {Uchida}}]{Kamp:2022bpt}%
  \BibitemOpen
  \bibfield  {author} {\bibinfo {author} {\bibfnamefont {N.~W.}\ \bibnamefont {Kamp}}, \bibinfo {author} {\bibfnamefont {M.}~\bibnamefont {Hostert}}, \bibinfo {author} {\bibfnamefont {A.}~\bibnamefont {Schneider}}, \bibinfo {author} {\bibfnamefont {S.}~\bibnamefont {Vergani}}, \bibinfo {author} {\bibfnamefont {C.~A.}\ \bibnamefont {Arg\"uelles}}, \bibinfo {author} {\bibfnamefont {J.~M.}\ \bibnamefont {Conrad}}, \bibinfo {author} {\bibfnamefont {M.~H.}\ \bibnamefont {Shaevitz}}, \ and\ \bibinfo {author} {\bibfnamefont {M.~A.}\ \bibnamefont {Uchida}},\ }\href {\doibase 10.1103/PhysRevD.107.055009} {\bibfield  {journal} {\bibinfo  {journal} {Phys. Rev. D}\ }\textbf {\bibinfo {volume} {107}},\ \bibinfo {pages} {055009} (\bibinfo {year} {2023})},\ \Eprint {http://arxiv.org/abs/2206.07100} {arXiv:2206.07100 [hep-ph]} \BibitemShut {NoStop}%
\bibitem [{\citenamefont {Bansal}\ \emph {et~al.}(2023)\citenamefont {Bansal}, \citenamefont {Paz}, \citenamefont {Petrov}, \citenamefont {Tammaro},\ and\ \citenamefont {Zupan}}]{Bansal:2022zpi}%
  \BibitemOpen
  \bibfield  {author} {\bibinfo {author} {\bibfnamefont {S.}~\bibnamefont {Bansal}}, \bibinfo {author} {\bibfnamefont {G.}~\bibnamefont {Paz}}, \bibinfo {author} {\bibfnamefont {A.}~\bibnamefont {Petrov}}, \bibinfo {author} {\bibfnamefont {M.}~\bibnamefont {Tammaro}}, \ and\ \bibinfo {author} {\bibfnamefont {J.}~\bibnamefont {Zupan}},\ }\href {\doibase 10.1007/JHEP05(2023)142} {\bibfield  {journal} {\bibinfo  {journal} {JHEP}\ }\textbf {\bibinfo {volume} {05}},\ \bibinfo {pages} {142} (\bibinfo {year} {2023})},\ \Eprint {http://arxiv.org/abs/2210.05706} {arXiv:2210.05706 [hep-ph]} \BibitemShut {NoStop}%
\bibitem [{\citenamefont {Ghosh}\ and\ \citenamefont {Ko}(2023)}]{Ghosh:2023dgk}%
  \BibitemOpen
  \bibfield  {author} {\bibinfo {author} {\bibfnamefont {S.}~\bibnamefont {Ghosh}}\ and\ \bibinfo {author} {\bibfnamefont {P.}~\bibnamefont {Ko}},\ }\href@noop {} {\  (\bibinfo {year} {2023})},\ \Eprint {http://arxiv.org/abs/2311.14099} {arXiv:2311.14099 [hep-ph]} \BibitemShut {NoStop}%
\bibitem [{\citenamefont {Brdar}\ \emph {et~al.}(2021)\citenamefont {Brdar}, \citenamefont {Fischer},\ and\ \citenamefont {Smirnov}}]{Brdar:2020tle}%
  \BibitemOpen
  \bibfield  {author} {\bibinfo {author} {\bibfnamefont {V.}~\bibnamefont {Brdar}}, \bibinfo {author} {\bibfnamefont {O.}~\bibnamefont {Fischer}}, \ and\ \bibinfo {author} {\bibfnamefont {A.~Y.}\ \bibnamefont {Smirnov}},\ }\href {\doibase 10.1103/PhysRevD.103.075008} {\bibfield  {journal} {\bibinfo  {journal} {Phys. Rev. D}\ }\textbf {\bibinfo {volume} {103}},\ \bibinfo {pages} {075008} (\bibinfo {year} {2021})},\ \Eprint {http://arxiv.org/abs/2007.14411} {arXiv:2007.14411 [hep-ph]} \BibitemShut {NoStop}%
\bibitem [{\citenamefont {Atre}\ \emph {et~al.}(2009)\citenamefont {Atre}, \citenamefont {Han}, \citenamefont {Pascoli},\ and\ \citenamefont {Zhang}}]{Atre:2009rg}%
  \BibitemOpen
  \bibfield  {author} {\bibinfo {author} {\bibfnamefont {A.}~\bibnamefont {Atre}}, \bibinfo {author} {\bibfnamefont {T.}~\bibnamefont {Han}}, \bibinfo {author} {\bibfnamefont {S.}~\bibnamefont {Pascoli}}, \ and\ \bibinfo {author} {\bibfnamefont {B.}~\bibnamefont {Zhang}},\ }\href {\doibase 10.1088/1126-6708/2009/05/030} {\bibfield  {journal} {\bibinfo  {journal} {JHEP}\ }\textbf {\bibinfo {volume} {05}},\ \bibinfo {pages} {030} (\bibinfo {year} {2009})},\ \Eprint {http://arxiv.org/abs/0901.3589} {arXiv:0901.3589 [hep-ph]} \BibitemShut {NoStop}%
\bibitem [{\citenamefont {McKeen}\ and\ \citenamefont {Pospelov}(2010)}]{McKeen:2010rx}%
  \BibitemOpen
  \bibfield  {author} {\bibinfo {author} {\bibfnamefont {D.}~\bibnamefont {McKeen}}\ and\ \bibinfo {author} {\bibfnamefont {M.}~\bibnamefont {Pospelov}},\ }\href {\doibase 10.1103/PhysRevD.82.113018} {\bibfield  {journal} {\bibinfo  {journal} {Phys. Rev. D}\ }\textbf {\bibinfo {volume} {82}},\ \bibinfo {pages} {113018} (\bibinfo {year} {2010})},\ \Eprint {http://arxiv.org/abs/1011.3046} {arXiv:1011.3046 [hep-ph]} \BibitemShut {NoStop}%
\bibitem [{\citenamefont {Duk}\ \emph {et~al.}(2012)\citenamefont {Duk} \emph {et~al.}}]{ISTRA:2011bgc}%
  \BibitemOpen
  \bibfield  {author} {\bibinfo {author} {\bibfnamefont {V.~A.}\ \bibnamefont {Duk}} \emph {et~al.} (\bibinfo {collaboration} {ISTRA+}),\ }\href {\doibase 10.1016/j.physletb.2012.02.087} {\bibfield  {journal} {\bibinfo  {journal} {Phys. Lett. B}\ }\textbf {\bibinfo {volume} {710}},\ \bibinfo {pages} {307} (\bibinfo {year} {2012})},\ \Eprint {http://arxiv.org/abs/1110.1610} {arXiv:1110.1610 [hep-ex]} \BibitemShut {NoStop}%
\bibitem [{\citenamefont {Drewes}\ and\ \citenamefont {Garbrecht}(2017)}]{Drewes:2015iva}%
  \BibitemOpen
  \bibfield  {author} {\bibinfo {author} {\bibfnamefont {M.}~\bibnamefont {Drewes}}\ and\ \bibinfo {author} {\bibfnamefont {B.}~\bibnamefont {Garbrecht}},\ }\href {\doibase 10.1016/j.nuclphysb.2017.05.001} {\bibfield  {journal} {\bibinfo  {journal} {Nucl. Phys. B}\ }\textbf {\bibinfo {volume} {921}},\ \bibinfo {pages} {250} (\bibinfo {year} {2017})},\ \Eprint {http://arxiv.org/abs/1502.00477} {arXiv:1502.00477 [hep-ph]} \BibitemShut {NoStop}%
\bibitem [{\citenamefont {de~Gouv\^ea}\ and\ \citenamefont {Kobach}(2016)}]{deGouvea:2015euy}%
  \BibitemOpen
  \bibfield  {author} {\bibinfo {author} {\bibfnamefont {A.}~\bibnamefont {de~Gouv\^ea}}\ and\ \bibinfo {author} {\bibfnamefont {A.}~\bibnamefont {Kobach}},\ }\href {\doibase 10.1103/PhysRevD.93.033005} {\bibfield  {journal} {\bibinfo  {journal} {Phys. Rev. D}\ }\textbf {\bibinfo {volume} {93}},\ \bibinfo {pages} {033005} (\bibinfo {year} {2016})},\ \Eprint {http://arxiv.org/abs/1511.00683} {arXiv:1511.00683 [hep-ph]} \BibitemShut {NoStop}%
\bibitem [{\citenamefont {Coloma}\ \emph {et~al.}(2017)\citenamefont {Coloma}, \citenamefont {Machado}, \citenamefont {Martinez-Soler},\ and\ \citenamefont {Shoemaker}}]{Coloma:2017ppo}%
  \BibitemOpen
  \bibfield  {author} {\bibinfo {author} {\bibfnamefont {P.}~\bibnamefont {Coloma}}, \bibinfo {author} {\bibfnamefont {P.~A.~N.}\ \bibnamefont {Machado}}, \bibinfo {author} {\bibfnamefont {I.}~\bibnamefont {Martinez-Soler}}, \ and\ \bibinfo {author} {\bibfnamefont {I.~M.}\ \bibnamefont {Shoemaker}},\ }\href {\doibase 10.1103/PhysRevLett.119.201804} {\bibfield  {journal} {\bibinfo  {journal} {Phys. Rev. Lett.}\ }\textbf {\bibinfo {volume} {119}},\ \bibinfo {pages} {201804} (\bibinfo {year} {2017})},\ \Eprint {http://arxiv.org/abs/1707.08573} {arXiv:1707.08573 [hep-ph]} \BibitemShut {NoStop}%
\bibitem [{\citenamefont {Aguilar-Arevalo}\ \emph {et~al.}(2018{\natexlab{a}})\citenamefont {Aguilar-Arevalo} \emph {et~al.}}]{MiniBooNEDM:2018cxm}%
  \BibitemOpen
  \bibfield  {author} {\bibinfo {author} {\bibfnamefont {A.~A.}\ \bibnamefont {Aguilar-Arevalo}} \emph {et~al.} (\bibinfo {collaboration} {MiniBooNE DM}),\ }\href {\doibase 10.1103/PhysRevD.98.112004} {\bibfield  {journal} {\bibinfo  {journal} {Phys. Rev. D}\ }\textbf {\bibinfo {volume} {98}},\ \bibinfo {pages} {112004} (\bibinfo {year} {2018}{\natexlab{a}})},\ \Eprint {http://arxiv.org/abs/1807.06137} {arXiv:1807.06137 [hep-ex]} \BibitemShut {NoStop}%
\bibitem [{\citenamefont {Jordan}\ \emph {et~al.}(2019)\citenamefont {Jordan}, \citenamefont {Kahn}, \citenamefont {Krnjaic}, \citenamefont {Moschella},\ and\ \citenamefont {Spitz}}]{Jordan:2018qiy}%
  \BibitemOpen
  \bibfield  {author} {\bibinfo {author} {\bibfnamefont {J.~R.}\ \bibnamefont {Jordan}}, \bibinfo {author} {\bibfnamefont {Y.}~\bibnamefont {Kahn}}, \bibinfo {author} {\bibfnamefont {G.}~\bibnamefont {Krnjaic}}, \bibinfo {author} {\bibfnamefont {M.}~\bibnamefont {Moschella}}, \ and\ \bibinfo {author} {\bibfnamefont {J.}~\bibnamefont {Spitz}},\ }\href {\doibase 10.1103/PhysRevLett.122.081801} {\bibfield  {journal} {\bibinfo  {journal} {Phys. Rev. Lett.}\ }\textbf {\bibinfo {volume} {122}},\ \bibinfo {pages} {081801} (\bibinfo {year} {2019})},\ \Eprint {http://arxiv.org/abs/1810.07185} {arXiv:1810.07185 [hep-ph]} \BibitemShut {NoStop}%
\bibitem [{\citenamefont {Arg\"uelles}\ \emph {et~al.}(2019)\citenamefont {Arg\"uelles}, \citenamefont {Hostert},\ and\ \citenamefont {Tsai}}]{Arguelles:2018mtc}%
  \BibitemOpen
  \bibfield  {author} {\bibinfo {author} {\bibfnamefont {C.~A.}\ \bibnamefont {Arg\"uelles}}, \bibinfo {author} {\bibfnamefont {M.}~\bibnamefont {Hostert}}, \ and\ \bibinfo {author} {\bibfnamefont {Y.-D.}\ \bibnamefont {Tsai}},\ }\href {\doibase 10.1103/PhysRevLett.123.261801} {\bibfield  {journal} {\bibinfo  {journal} {Phys. Rev. Lett.}\ }\textbf {\bibinfo {volume} {123}},\ \bibinfo {pages} {261801} (\bibinfo {year} {2019})},\ \Eprint {http://arxiv.org/abs/1812.08768} {arXiv:1812.08768 [hep-ph]} \BibitemShut {NoStop}%
\bibitem [{\citenamefont {Bryman}\ and\ \citenamefont {Shrock}(2019{\natexlab{a}})}]{Bryman:2019ssi}%
  \BibitemOpen
  \bibfield  {author} {\bibinfo {author} {\bibfnamefont {D.~A.}\ \bibnamefont {Bryman}}\ and\ \bibinfo {author} {\bibfnamefont {R.}~\bibnamefont {Shrock}},\ }\href {\doibase 10.1103/PhysRevD.100.053006} {\bibfield  {journal} {\bibinfo  {journal} {Phys. Rev. D}\ }\textbf {\bibinfo {volume} {100}},\ \bibinfo {pages} {053006} (\bibinfo {year} {2019}{\natexlab{a}})},\ \Eprint {http://arxiv.org/abs/1904.06787} {arXiv:1904.06787 [hep-ph]} \BibitemShut {NoStop}%
\bibitem [{\citenamefont {Coloma}(2019)}]{Coloma:2019qqj}%
  \BibitemOpen
  \bibfield  {author} {\bibinfo {author} {\bibfnamefont {P.}~\bibnamefont {Coloma}},\ }\href {\doibase 10.1140/epjc/s10052-019-7256-8} {\bibfield  {journal} {\bibinfo  {journal} {Eur. Phys. J. C}\ }\textbf {\bibinfo {volume} {79}},\ \bibinfo {pages} {748} (\bibinfo {year} {2019})},\ \Eprint {http://arxiv.org/abs/1906.02106} {arXiv:1906.02106 [hep-ph]} \BibitemShut {NoStop}%
\bibitem [{\citenamefont {Bryman}\ and\ \citenamefont {Shrock}(2019{\natexlab{b}})}]{Bryman:2019bjg}%
  \BibitemOpen
  \bibfield  {author} {\bibinfo {author} {\bibfnamefont {D.~A.}\ \bibnamefont {Bryman}}\ and\ \bibinfo {author} {\bibfnamefont {R.}~\bibnamefont {Shrock}},\ }\href {\doibase 10.1103/PhysRevD.100.073011} {\bibfield  {journal} {\bibinfo  {journal} {Phys. Rev. D}\ }\textbf {\bibinfo {volume} {100}},\ \bibinfo {pages} {073011} (\bibinfo {year} {2019}{\natexlab{b}})},\ \Eprint {http://arxiv.org/abs/1909.11198} {arXiv:1909.11198 [hep-ph]} \BibitemShut {NoStop}%
\bibitem [{\citenamefont {Aad}\ \emph {et~al.}(2023{\natexlab{a}})\citenamefont {Aad} \emph {et~al.}}]{ATLAS:2022tnm}%
  \BibitemOpen
  \bibfield  {author} {\bibinfo {author} {\bibfnamefont {G.}~\bibnamefont {Aad}} \emph {et~al.} (\bibinfo {collaboration} {ATLAS}),\ }\href {\doibase 10.1007/JHEP07(2023)088} {\bibfield  {journal} {\bibinfo  {journal} {JHEP}\ }\textbf {\bibinfo {volume} {07}},\ \bibinfo {pages} {088} (\bibinfo {year} {2023}{\natexlab{a}})},\ \Eprint {http://arxiv.org/abs/2207.00348} {arXiv:2207.00348 [hep-ex]} \BibitemShut {NoStop}%
\bibitem [{\citenamefont {Ellwanger}\ and\ \citenamefont {Moretti}(2016)}]{Ellwanger:2016wfe}%
  \BibitemOpen
  \bibfield  {author} {\bibinfo {author} {\bibfnamefont {U.}~\bibnamefont {Ellwanger}}\ and\ \bibinfo {author} {\bibfnamefont {S.}~\bibnamefont {Moretti}},\ }\href {\doibase 10.1007/JHEP11(2016)039} {\bibfield  {journal} {\bibinfo  {journal} {JHEP}\ }\textbf {\bibinfo {volume} {11}},\ \bibinfo {pages} {039} (\bibinfo {year} {2016})},\ \Eprint {http://arxiv.org/abs/1609.01669} {arXiv:1609.01669 [hep-ph]} \BibitemShut {NoStop}%
\bibitem [{\citenamefont {Esteban}\ \emph {et~al.}(2020)\citenamefont {Esteban}, \citenamefont {Gonzalez-Garcia}, \citenamefont {Maltoni}, \citenamefont {Schwetz},\ and\ \citenamefont {Zhou}}]{Esteban:2020cvm}%
  \BibitemOpen
  \bibfield  {author} {\bibinfo {author} {\bibfnamefont {I.}~\bibnamefont {Esteban}}, \bibinfo {author} {\bibfnamefont {M.~C.}\ \bibnamefont {Gonzalez-Garcia}}, \bibinfo {author} {\bibfnamefont {M.}~\bibnamefont {Maltoni}}, \bibinfo {author} {\bibfnamefont {T.}~\bibnamefont {Schwetz}}, \ and\ \bibinfo {author} {\bibfnamefont {A.}~\bibnamefont {Zhou}},\ }\href {\doibase 10.1007/JHEP09(2020)178} {\bibfield  {journal} {\bibinfo  {journal} {JHEP}\ }\textbf {\bibinfo {volume} {09}},\ \bibinfo {pages} {178} (\bibinfo {year} {2020})},\ \Eprint {http://arxiv.org/abs/2007.14792} {arXiv:2007.14792 [hep-ph]} \BibitemShut {NoStop}%
\bibitem [{\citenamefont {Cheng}\ and\ \citenamefont {Chiang}(2012)}]{Cheng:2012qr}%
  \BibitemOpen
  \bibfield  {author} {\bibinfo {author} {\bibfnamefont {H.-Y.}\ \bibnamefont {Cheng}}\ and\ \bibinfo {author} {\bibfnamefont {C.-W.}\ \bibnamefont {Chiang}},\ }\href {\doibase 10.1007/JHEP07(2012)009} {\bibfield  {journal} {\bibinfo  {journal} {JHEP}\ }\textbf {\bibinfo {volume} {07}},\ \bibinfo {pages} {009} (\bibinfo {year} {2012})},\ \Eprint {http://arxiv.org/abs/1202.1292} {arXiv:1202.1292 [hep-ph]} \BibitemShut {NoStop}%
\bibitem [{\citenamefont {Arina}\ \emph {et~al.}(2015)\citenamefont {Arina}, \citenamefont {Del~Nobile},\ and\ \citenamefont {Panci}}]{PhysRevLett.114.011301}%
  \BibitemOpen
  \bibfield  {author} {\bibinfo {author} {\bibfnamefont {C.}~\bibnamefont {Arina}}, \bibinfo {author} {\bibfnamefont {E.}~\bibnamefont {Del~Nobile}}, \ and\ \bibinfo {author} {\bibfnamefont {P.}~\bibnamefont {Panci}},\ }\href {\doibase 10.1103/PhysRevLett.114.011301} {\bibfield  {journal} {\bibinfo  {journal} {Phys. Rev. Lett.}\ }\textbf {\bibinfo {volume} {114}},\ \bibinfo {pages} {011301} (\bibinfo {year} {2015})}\BibitemShut {NoStop}%
\bibitem [{\citenamefont {Hikasa}\ \emph {et~al.}(2022)\citenamefont {Hikasa} \emph {et~al.}}]{ParticleDataGroup:2022pth}%
  \BibitemOpen
  \bibfield  {author} {\bibinfo {author} {\bibfnamefont {K.}~\bibnamefont {Hikasa}} \emph {et~al.} (\bibinfo {collaboration} {Particle Data Group}),\ }\href {\doibase 10.1093/ptep/ptac097} {\bibfield  {journal} {\bibinfo  {journal} {PTEP}\ }\textbf {\bibinfo {volume} {2022}},\ \bibinfo {pages} {083C01} (\bibinfo {year} {2022})}\BibitemShut {NoStop}%
\bibitem [{\citenamefont {Aguilar-Arevalo}\ \emph {et~al.}(2021)\citenamefont {Aguilar-Arevalo} \emph {et~al.}}]{MiniBooNE:2020pnu}%
  \BibitemOpen
  \bibfield  {author} {\bibinfo {author} {\bibfnamefont {A.~A.}\ \bibnamefont {Aguilar-Arevalo}} \emph {et~al.} (\bibinfo {collaboration} {MiniBooNE}),\ }\href {\doibase 10.1103/PhysRevD.103.052002} {\bibfield  {journal} {\bibinfo  {journal} {Phys. Rev. D}\ }\textbf {\bibinfo {volume} {103}},\ \bibinfo {pages} {052002} (\bibinfo {year} {2021})},\ \Eprint {http://arxiv.org/abs/2006.16883} {arXiv:2006.16883 [hep-ex]} \BibitemShut {NoStop}%
\bibitem [{\citenamefont {Aguilar-Arevalo}\ \emph {et~al.}(2018{\natexlab{b}})\citenamefont {Aguilar-Arevalo} \emph {et~al.}}]{Aguilar-Arevalo:2018gpe}%
  \BibitemOpen
  \bibfield  {author} {\bibinfo {author} {\bibfnamefont {A.~A.}\ \bibnamefont {Aguilar-Arevalo}} \emph {et~al.} (\bibinfo {collaboration} {MiniBooNE}),\ }\href {\doibase 10.1103/PhysRevLett.121.221801} {\bibfield  {journal} {\bibinfo  {journal} {Phys. Rev. Lett.}\ }\textbf {\bibinfo {volume} {121}},\ \bibinfo {pages} {221801} (\bibinfo {year} {2018}{\natexlab{b}})},\ \Eprint {http://arxiv.org/abs/1805.12028} {arXiv:1805.12028 [hep-ex]} \BibitemShut {NoStop}%
\bibitem [{\citenamefont {Aguilar}\ \emph {et~al.}(2001{\natexlab{b}})\citenamefont {Aguilar} \emph {et~al.}}]{Aguilar:2001ty}%
  \BibitemOpen
  \bibfield  {author} {\bibinfo {author} {\bibfnamefont {A.}~\bibnamefont {Aguilar}} \emph {et~al.} (\bibinfo {collaboration} {LSND}),\ }\href {\doibase 10.1103/PhysRevD.64.112007} {\bibfield  {journal} {\bibinfo  {journal} {Phys. Rev. D}\ }\textbf {\bibinfo {volume} {64}},\ \bibinfo {pages} {112007} (\bibinfo {year} {2001}{\natexlab{b}})},\ \Eprint {http://arxiv.org/abs/hep-ex/0104049} {arXiv:hep-ex/0104049} \BibitemShut {NoStop}%
\bibitem [{\citenamefont {Krasznahorkay}\ \emph {et~al.}(2018)\citenamefont {Krasznahorkay} \emph {et~al.}}]{Krasznahorkay:2018snd}%
  \BibitemOpen
  \bibfield  {author} {\bibinfo {author} {\bibfnamefont {A.~J.}\ \bibnamefont {Krasznahorkay}} \emph {et~al.},\ }\href {\doibase 10.1088/1742-6596/1056/1/012028} {\bibfield  {journal} {\bibinfo  {journal} {J. Phys. Conf. Ser.}\ }\textbf {\bibinfo {volume} {1056}},\ \bibinfo {pages} {012028} (\bibinfo {year} {2018})}\BibitemShut {NoStop}%
\bibitem [{\citenamefont {Barducci}\ and\ \citenamefont {Toni}(2023)}]{Barducci:2022lqd}%
  \BibitemOpen
  \bibfield  {author} {\bibinfo {author} {\bibfnamefont {D.}~\bibnamefont {Barducci}}\ and\ \bibinfo {author} {\bibfnamefont {C.}~\bibnamefont {Toni}},\ }\href {\doibase 10.1007/JHEP02(2023)154} {\bibfield  {journal} {\bibinfo  {journal} {JHEP}\ }\textbf {\bibinfo {volume} {02}},\ \bibinfo {pages} {154} (\bibinfo {year} {2023})},\ \bibinfo {note} {[Erratum: JHEP 07, 168 (2023)]},\ \Eprint {http://arxiv.org/abs/2212.06453} {arXiv:2212.06453 [hep-ph]} \BibitemShut {NoStop}%
\bibitem [{\citenamefont {Tilley}\ \emph {et~al.}(2004)\citenamefont {Tilley}, \citenamefont {Kelley}, \citenamefont {Godwin}, \citenamefont {Millener}, \citenamefont {Purcell}, \citenamefont {Sheu},\ and\ \citenamefont {Weller}}]{Tilley:2004zz}%
  \BibitemOpen
  \bibfield  {author} {\bibinfo {author} {\bibfnamefont {D.~R.}\ \bibnamefont {Tilley}}, \bibinfo {author} {\bibfnamefont {J.~H.}\ \bibnamefont {Kelley}}, \bibinfo {author} {\bibfnamefont {J.~L.}\ \bibnamefont {Godwin}}, \bibinfo {author} {\bibfnamefont {D.~J.}\ \bibnamefont {Millener}}, \bibinfo {author} {\bibfnamefont {J.~E.}\ \bibnamefont {Purcell}}, \bibinfo {author} {\bibfnamefont {C.~G.}\ \bibnamefont {Sheu}}, \ and\ \bibinfo {author} {\bibfnamefont {H.~R.}\ \bibnamefont {Weller}},\ }\href {\doibase 10.1016/j.nuclphysa.2004.09.059} {\bibfield  {journal} {\bibinfo  {journal} {Nucl. Phys. A}\ }\textbf {\bibinfo {volume} {745}},\ \bibinfo {pages} {155} (\bibinfo {year} {2004})}\BibitemShut {NoStop}%
\bibitem [{\citenamefont {Schl\"uter}\ \emph {et~al.}(1981)\citenamefont {Schl\"uter}, \citenamefont {Soff},\ and\ \citenamefont {Greiner}}]{Schluter:1981cjo}%
  \BibitemOpen
  \bibfield  {author} {\bibinfo {author} {\bibfnamefont {P.}~\bibnamefont {Schl\"uter}}, \bibinfo {author} {\bibfnamefont {G.}~\bibnamefont {Soff}}, \ and\ \bibinfo {author} {\bibfnamefont {W.}~\bibnamefont {Greiner}},\ }\href {\doibase 10.1016/0370-1573(81)90166-6} {\bibfield  {journal} {\bibinfo  {journal} {Phys. Rept.}\ }\textbf {\bibinfo {volume} {75}},\ \bibinfo {pages} {327} (\bibinfo {year} {1981})}\BibitemShut {NoStop}%
\bibitem [{\citenamefont {Rose}(1949)}]{Rose:1949zz}%
  \BibitemOpen
  \bibfield  {author} {\bibinfo {author} {\bibfnamefont {M.~E.}\ \bibnamefont {Rose}},\ }\href {\doibase 10.1103/PhysRev.78.184} {\bibfield  {journal} {\bibinfo  {journal} {Phys. Rev.}\ }\textbf {\bibinfo {volume} {76}},\ \bibinfo {pages} {678} (\bibinfo {year} {1949})},\ \bibinfo {note} {[Erratum: Phys.Rev. 78, 184--184 (1950)]}\BibitemShut {NoStop}%
\bibitem [{\citenamefont {Dror}\ \emph {et~al.}(2017{\natexlab{a}})\citenamefont {Dror}, \citenamefont {Lasenby},\ and\ \citenamefont {Pospelov}}]{Dror:2017ehi}%
  \BibitemOpen
  \bibfield  {author} {\bibinfo {author} {\bibfnamefont {J.~A.}\ \bibnamefont {Dror}}, \bibinfo {author} {\bibfnamefont {R.}~\bibnamefont {Lasenby}}, \ and\ \bibinfo {author} {\bibfnamefont {M.}~\bibnamefont {Pospelov}},\ }\href {\doibase 10.1103/PhysRevLett.119.141803} {\bibfield  {journal} {\bibinfo  {journal} {Phys. Rev. Lett.}\ }\textbf {\bibinfo {volume} {119}},\ \bibinfo {pages} {141803} (\bibinfo {year} {2017}{\natexlab{a}})},\ \Eprint {http://arxiv.org/abs/1705.06726} {arXiv:1705.06726 [hep-ph]} \BibitemShut {NoStop}%
\bibitem [{\citenamefont {Dror}\ \emph {et~al.}(2017{\natexlab{b}})\citenamefont {Dror}, \citenamefont {Lasenby},\ and\ \citenamefont {Pospelov}}]{Dror:2017nsg}%
  \BibitemOpen
  \bibfield  {author} {\bibinfo {author} {\bibfnamefont {J.~A.}\ \bibnamefont {Dror}}, \bibinfo {author} {\bibfnamefont {R.}~\bibnamefont {Lasenby}}, \ and\ \bibinfo {author} {\bibfnamefont {M.}~\bibnamefont {Pospelov}},\ }\href {\doibase 10.1103/PhysRevD.96.075036} {\bibfield  {journal} {\bibinfo  {journal} {Phys. Rev. D}\ }\textbf {\bibinfo {volume} {96}},\ \bibinfo {pages} {075036} (\bibinfo {year} {2017}{\natexlab{b}})},\ \Eprint {http://arxiv.org/abs/1707.01503} {arXiv:1707.01503 [hep-ph]} \BibitemShut {NoStop}%
\bibitem [{\citenamefont {Dror}\ \emph {et~al.}(2019)\citenamefont {Dror}, \citenamefont {Lasenby},\ and\ \citenamefont {Pospelov}}]{Dror:2018wfl}%
  \BibitemOpen
  \bibfield  {author} {\bibinfo {author} {\bibfnamefont {J.~A.}\ \bibnamefont {Dror}}, \bibinfo {author} {\bibfnamefont {R.}~\bibnamefont {Lasenby}}, \ and\ \bibinfo {author} {\bibfnamefont {M.}~\bibnamefont {Pospelov}},\ }\href {\doibase 10.1103/PhysRevD.99.055016} {\bibfield  {journal} {\bibinfo  {journal} {Phys. Rev. D}\ }\textbf {\bibinfo {volume} {99}},\ \bibinfo {pages} {055016} (\bibinfo {year} {2019})},\ \Eprint {http://arxiv.org/abs/1811.00595} {arXiv:1811.00595 [hep-ph]} \BibitemShut {NoStop}%
\bibitem [{\citenamefont {Dolan}\ \emph {et~al.}(2015)\citenamefont {Dolan}, \citenamefont {Kahlhoefer}, \citenamefont {McCabe},\ and\ \citenamefont {Schmidt-Hoberg}}]{Dolan:2014ska}%
  \BibitemOpen
  \bibfield  {author} {\bibinfo {author} {\bibfnamefont {M.~J.}\ \bibnamefont {Dolan}}, \bibinfo {author} {\bibfnamefont {F.}~\bibnamefont {Kahlhoefer}}, \bibinfo {author} {\bibfnamefont {C.}~\bibnamefont {McCabe}}, \ and\ \bibinfo {author} {\bibfnamefont {K.}~\bibnamefont {Schmidt-Hoberg}},\ }\href {\doibase 10.1007/JHEP03(2015)171} {\bibfield  {journal} {\bibinfo  {journal} {JHEP}\ }\textbf {\bibinfo {volume} {03}},\ \bibinfo {pages} {171} (\bibinfo {year} {2015})},\ \bibinfo {note} {[Erratum: JHEP 07, 103 (2015)]},\ \Eprint {http://arxiv.org/abs/1412.5174} {arXiv:1412.5174 [hep-ph]} \BibitemShut {NoStop}%
\bibitem [{\citenamefont {Andreas}\ \emph {et~al.}(2010)\citenamefont {Andreas}, \citenamefont {Lebedev}, \citenamefont {Ramos-Sanchez},\ and\ \citenamefont {Ringwald}}]{Andreas:2010ms}%
  \BibitemOpen
  \bibfield  {author} {\bibinfo {author} {\bibfnamefont {S.}~\bibnamefont {Andreas}}, \bibinfo {author} {\bibfnamefont {O.}~\bibnamefont {Lebedev}}, \bibinfo {author} {\bibfnamefont {S.}~\bibnamefont {Ramos-Sanchez}}, \ and\ \bibinfo {author} {\bibfnamefont {A.}~\bibnamefont {Ringwald}},\ }\href {\doibase 10.1007/JHEP08(2010)003} {\bibfield  {journal} {\bibinfo  {journal} {JHEP}\ }\textbf {\bibinfo {volume} {08}},\ \bibinfo {pages} {003} (\bibinfo {year} {2010})},\ \Eprint {http://arxiv.org/abs/1005.3978} {arXiv:1005.3978 [hep-ph]} \BibitemShut {NoStop}%
\bibitem [{\citenamefont {Essig}\ \emph {et~al.}(2010)\citenamefont {Essig}, \citenamefont {Harnik}, \citenamefont {Kaplan},\ and\ \citenamefont {Toro}}]{Essig:2010gu}%
  \BibitemOpen
  \bibfield  {author} {\bibinfo {author} {\bibfnamefont {R.}~\bibnamefont {Essig}}, \bibinfo {author} {\bibfnamefont {R.}~\bibnamefont {Harnik}}, \bibinfo {author} {\bibfnamefont {J.}~\bibnamefont {Kaplan}}, \ and\ \bibinfo {author} {\bibfnamefont {N.}~\bibnamefont {Toro}},\ }\href {\doibase 10.1103/PhysRevD.82.113008} {\bibfield  {journal} {\bibinfo  {journal} {Phys. Rev. D}\ }\textbf {\bibinfo {volume} {82}},\ \bibinfo {pages} {113008} (\bibinfo {year} {2010})},\ \Eprint {http://arxiv.org/abs/1008.0636} {arXiv:1008.0636 [hep-ph]} \BibitemShut {NoStop}%
\bibitem [{Pro(2012)}]{Proceedings:2012ulb}%
  \BibitemOpen
  \href {\doibase 10.2172/1042577} {\emph {\bibinfo {title} {{Fundamental Physics at the Intensity Frontier}}}}\ (\bibinfo {year} {2012})\ \Eprint {http://arxiv.org/abs/1205.2671} {arXiv:1205.2671 [hep-ex]} \BibitemShut {NoStop}%
\bibitem [{\citenamefont {D\"obrich}\ \emph {et~al.}(2016)\citenamefont {D\"obrich}, \citenamefont {Jaeckel}, \citenamefont {Kahlhoefer}, \citenamefont {Ringwald},\ and\ \citenamefont {Schmidt-Hoberg}}]{Dobrich:2015jyk}%
  \BibitemOpen
  \bibfield  {author} {\bibinfo {author} {\bibfnamefont {B.}~\bibnamefont {D\"obrich}}, \bibinfo {author} {\bibfnamefont {J.}~\bibnamefont {Jaeckel}}, \bibinfo {author} {\bibfnamefont {F.}~\bibnamefont {Kahlhoefer}}, \bibinfo {author} {\bibfnamefont {A.}~\bibnamefont {Ringwald}}, \ and\ \bibinfo {author} {\bibfnamefont {K.}~\bibnamefont {Schmidt-Hoberg}},\ }\href {\doibase 10.1007/JHEP02(2016)018} {\bibfield  {journal} {\bibinfo  {journal} {JHEP}\ }\textbf {\bibinfo {volume} {02}},\ \bibinfo {pages} {018} (\bibinfo {year} {2016})},\ \Eprint {http://arxiv.org/abs/1512.03069} {arXiv:1512.03069 [hep-ph]} \BibitemShut {NoStop}%
\bibitem [{\citenamefont {Liu}\ \emph {et~al.}(2021)\citenamefont {Liu}, \citenamefont {McGinnis}, \citenamefont {Wagner},\ and\ \citenamefont {Wang}}]{Liu:2021wap}%
  \BibitemOpen
  \bibfield  {author} {\bibinfo {author} {\bibfnamefont {J.}~\bibnamefont {Liu}}, \bibinfo {author} {\bibfnamefont {N.}~\bibnamefont {McGinnis}}, \bibinfo {author} {\bibfnamefont {C.~E.~M.}\ \bibnamefont {Wagner}}, \ and\ \bibinfo {author} {\bibfnamefont {X.-P.}\ \bibnamefont {Wang}},\ }\href {\doibase 10.1007/JHEP05(2021)138} {\bibfield  {journal} {\bibinfo  {journal} {JHEP}\ }\textbf {\bibinfo {volume} {05}},\ \bibinfo {pages} {138} (\bibinfo {year} {2021})},\ \Eprint {http://arxiv.org/abs/2102.10118} {arXiv:2102.10118 [hep-ph]} \BibitemShut {NoStop}%
\bibitem [{\citenamefont {Bjorken}\ \emph {et~al.}(1988)\citenamefont {Bjorken}, \citenamefont {Ecklund}, \citenamefont {Nelson}, \citenamefont {Abashian}, \citenamefont {Church}, \citenamefont {Lu}, \citenamefont {Mo}, \citenamefont {Nunamaker},\ and\ \citenamefont {Rassmann}}]{Bjorken:1988as}%
  \BibitemOpen
  \bibfield  {author} {\bibinfo {author} {\bibfnamefont {J.~D.}\ \bibnamefont {Bjorken}}, \bibinfo {author} {\bibfnamefont {S.}~\bibnamefont {Ecklund}}, \bibinfo {author} {\bibfnamefont {W.~R.}\ \bibnamefont {Nelson}}, \bibinfo {author} {\bibfnamefont {A.}~\bibnamefont {Abashian}}, \bibinfo {author} {\bibfnamefont {C.}~\bibnamefont {Church}}, \bibinfo {author} {\bibfnamefont {B.}~\bibnamefont {Lu}}, \bibinfo {author} {\bibfnamefont {L.~W.}\ \bibnamefont {Mo}}, \bibinfo {author} {\bibfnamefont {T.~A.}\ \bibnamefont {Nunamaker}}, \ and\ \bibinfo {author} {\bibfnamefont {P.}~\bibnamefont {Rassmann}},\ }\href {\doibase 10.1103/PhysRevD.38.3375} {\bibfield  {journal} {\bibinfo  {journal} {Phys. Rev. D}\ }\textbf {\bibinfo {volume} {38}},\ \bibinfo {pages} {3375} (\bibinfo {year} {1988})}\BibitemShut {NoStop}%
\bibitem [{\citenamefont {Riordan}\ \emph {et~al.}(1987)\citenamefont {Riordan} \emph {et~al.}}]{Riordan:1987aw}%
  \BibitemOpen
  \bibfield  {author} {\bibinfo {author} {\bibfnamefont {E.~M.}\ \bibnamefont {Riordan}} \emph {et~al.},\ }\href {\doibase 10.1103/PhysRevLett.59.755} {\bibfield  {journal} {\bibinfo  {journal} {Phys. Rev. Lett.}\ }\textbf {\bibinfo {volume} {59}},\ \bibinfo {pages} {755} (\bibinfo {year} {1987})}\BibitemShut {NoStop}%
\bibitem [{\citenamefont {Davier}\ and\ \citenamefont {Nguyen~Ngoc}(1989)}]{Davier:1989wz}%
  \BibitemOpen
  \bibfield  {author} {\bibinfo {author} {\bibfnamefont {M.}~\bibnamefont {Davier}}\ and\ \bibinfo {author} {\bibfnamefont {H.}~\bibnamefont {Nguyen~Ngoc}},\ }\href {\doibase 10.1016/0370-2693(89)90174-3} {\bibfield  {journal} {\bibinfo  {journal} {Phys. Lett. B}\ }\textbf {\bibinfo {volume} {229}},\ \bibinfo {pages} {150} (\bibinfo {year} {1989})}\BibitemShut {NoStop}%
\bibitem [{\citenamefont {Banerjee}\ \emph {et~al.}(2018)\citenamefont {Banerjee} \emph {et~al.}}]{NA64:2018lsq}%
  \BibitemOpen
  \bibfield  {author} {\bibinfo {author} {\bibfnamefont {D.}~\bibnamefont {Banerjee}} \emph {et~al.} (\bibinfo {collaboration} {NA64}),\ }\href {\doibase 10.1103/PhysRevLett.120.231802} {\bibfield  {journal} {\bibinfo  {journal} {Phys. Rev. Lett.}\ }\textbf {\bibinfo {volume} {120}},\ \bibinfo {pages} {231802} (\bibinfo {year} {2018})},\ \Eprint {http://arxiv.org/abs/1803.07748} {arXiv:1803.07748 [hep-ex]} \BibitemShut {NoStop}%
\bibitem [{\citenamefont {Batell}\ \emph {et~al.}(2019)\citenamefont {Batell}, \citenamefont {Freitas}, \citenamefont {Ismail},\ and\ \citenamefont {Mckeen}}]{Batell:2018fqo}%
  \BibitemOpen
  \bibfield  {author} {\bibinfo {author} {\bibfnamefont {B.}~\bibnamefont {Batell}}, \bibinfo {author} {\bibfnamefont {A.}~\bibnamefont {Freitas}}, \bibinfo {author} {\bibfnamefont {A.}~\bibnamefont {Ismail}}, \ and\ \bibinfo {author} {\bibfnamefont {D.}~\bibnamefont {Mckeen}},\ }\href {\doibase 10.1103/PhysRevD.100.095020} {\bibfield  {journal} {\bibinfo  {journal} {Phys. Rev. D}\ }\textbf {\bibinfo {volume} {100}},\ \bibinfo {pages} {095020} (\bibinfo {year} {2019})},\ \Eprint {http://arxiv.org/abs/1812.05103} {arXiv:1812.05103 [hep-ph]} \BibitemShut {NoStop}%
\bibitem [{\citenamefont {Yamazaki}\ \emph {et~al.}(1984)\citenamefont {Yamazaki} \emph {et~al.}}]{Yamazaki:1984vg}%
  \BibitemOpen
  \bibfield  {author} {\bibinfo {author} {\bibfnamefont {T.}~\bibnamefont {Yamazaki}} \emph {et~al.},\ }\href {\doibase 10.1103/PhysRevLett.52.1089} {\bibfield  {journal} {\bibinfo  {journal} {Phys. Rev. Lett.}\ }\textbf {\bibinfo {volume} {52}},\ \bibinfo {pages} {1089} (\bibinfo {year} {1984})}\BibitemShut {NoStop}%
\bibitem [{\citenamefont {Baker}\ \emph {et~al.}(1987)\citenamefont {Baker} \emph {et~al.}}]{Baker:1987gp}%
  \BibitemOpen
  \bibfield  {author} {\bibinfo {author} {\bibfnamefont {N.~J.}\ \bibnamefont {Baker}} \emph {et~al.},\ }\href {\doibase 10.1103/PhysRevLett.59.2832} {\bibfield  {journal} {\bibinfo  {journal} {Phys. Rev. Lett.}\ }\textbf {\bibinfo {volume} {59}},\ \bibinfo {pages} {2832} (\bibinfo {year} {1987})}\BibitemShut {NoStop}%
\bibitem [{\citenamefont {Alves}\ and\ \citenamefont {Weiner}(2018)}]{Alves:2017avw}%
  \BibitemOpen
  \bibfield  {author} {\bibinfo {author} {\bibfnamefont {D.~S.~M.}\ \bibnamefont {Alves}}\ and\ \bibinfo {author} {\bibfnamefont {N.}~\bibnamefont {Weiner}},\ }\href {\doibase 10.1007/JHEP07(2018)092} {\bibfield  {journal} {\bibinfo  {journal} {JHEP}\ }\textbf {\bibinfo {volume} {07}},\ \bibinfo {pages} {092} (\bibinfo {year} {2018})},\ \Eprint {http://arxiv.org/abs/1710.03764} {arXiv:1710.03764 [hep-ph]} \BibitemShut {NoStop}%
\bibitem [{\citenamefont {Yamazaki}()}]{Yamazaki:1984qx}%
  \BibitemOpen
  \bibfield  {author} {\bibinfo {author} {\bibfnamefont {T.}~\bibnamefont {Yamazaki}},\ }\href@noop {} {\ }\Eprint {http://arxiv.org/abs/Proceedings of the Second LAMPF II Workshop. Vol. II, Los Alamos U.S.A. (1982), Preprint LA-9572-C, page 413} {Proceedings of the Second LAMPF II Workshop. Vol. II, Los Alamos U.S.A. (1982), Preprint LA-9572-C, page 413} \BibitemShut {NoStop}%
\bibitem [{\citenamefont {Deshpande}\ \emph {et~al.}(2006)\citenamefont {Deshpande}, \citenamefont {Eilam},\ and\ \citenamefont {Jiang}}]{Deshpande:2005mb}%
  \BibitemOpen
  \bibfield  {author} {\bibinfo {author} {\bibfnamefont {N.~G.}\ \bibnamefont {Deshpande}}, \bibinfo {author} {\bibfnamefont {G.}~\bibnamefont {Eilam}}, \ and\ \bibinfo {author} {\bibfnamefont {J.}~\bibnamefont {Jiang}},\ }\href {\doibase 10.1016/j.physletb.2005.10.050} {\bibfield  {journal} {\bibinfo  {journal} {Phys. Lett. B}\ }\textbf {\bibinfo {volume} {632}},\ \bibinfo {pages} {212} (\bibinfo {year} {2006})},\ \Eprint {http://arxiv.org/abs/hep-ph/0509081} {arXiv:hep-ph/0509081} \BibitemShut {NoStop}%
\bibitem [{\citenamefont {Eichler}\ \emph {et~al.}(1986)\citenamefont {Eichler} \emph {et~al.}}]{SINDRUM:1986klz}%
  \BibitemOpen
  \bibfield  {author} {\bibinfo {author} {\bibfnamefont {R.}~\bibnamefont {Eichler}} \emph {et~al.} (\bibinfo {collaboration} {SINDRUM}),\ }\href {\doibase 10.1016/0370-2693(86)90339-4} {\bibfield  {journal} {\bibinfo  {journal} {Phys. Lett. B}\ }\textbf {\bibinfo {volume} {175}},\ \bibinfo {pages} {101} (\bibinfo {year} {1986})}\BibitemShut {NoStop}%
\bibitem [{\citenamefont {Cortina~Gil}\ \emph {et~al.}(2023)\citenamefont {Cortina~Gil} \emph {et~al.}}]{NA62:2023rvm}%
  \BibitemOpen
  \bibfield  {author} {\bibinfo {author} {\bibfnamefont {E.}~\bibnamefont {Cortina~Gil}} \emph {et~al.} (\bibinfo {collaboration} {NA62}),\ }\href {\doibase 10.1016/j.physletb.2023.138193} {\bibfield  {journal} {\bibinfo  {journal} {Phys. Lett. B}\ }\textbf {\bibinfo {volume} {846}},\ \bibinfo {pages} {138193} (\bibinfo {year} {2023})},\ \Eprint {http://arxiv.org/abs/2307.04579} {arXiv:2307.04579 [hep-ex]} \BibitemShut {NoStop}%
\bibitem [{\citenamefont {Krauss}\ and\ \citenamefont {Wise}(1986)}]{Krauss:1986bq}%
  \BibitemOpen
  \bibfield  {author} {\bibinfo {author} {\bibfnamefont {L.~M.}\ \bibnamefont {Krauss}}\ and\ \bibinfo {author} {\bibfnamefont {M.~B.}\ \bibnamefont {Wise}},\ }\href {\doibase 10.1016/0370-2693(86)90201-7} {\bibfield  {journal} {\bibinfo  {journal} {Phys. Lett. B}\ }\textbf {\bibinfo {volume} {176}},\ \bibinfo {pages} {483} (\bibinfo {year} {1986})}\BibitemShut {NoStop}%
\bibitem [{\citenamefont {Alloul}\ \emph {et~al.}(2014)\citenamefont {Alloul}, \citenamefont {Christensen}, \citenamefont {Degrande}, \citenamefont {Duhr},\ and\ \citenamefont {Fuks}}]{Alloul:2013bka}%
  \BibitemOpen
  \bibfield  {author} {\bibinfo {author} {\bibfnamefont {A.}~\bibnamefont {Alloul}}, \bibinfo {author} {\bibfnamefont {N.~D.}\ \bibnamefont {Christensen}}, \bibinfo {author} {\bibfnamefont {C.}~\bibnamefont {Degrande}}, \bibinfo {author} {\bibfnamefont {C.}~\bibnamefont {Duhr}}, \ and\ \bibinfo {author} {\bibfnamefont {B.}~\bibnamefont {Fuks}},\ }\href {\doibase 10.1016/j.cpc.2014.04.012} {\bibfield  {journal} {\bibinfo  {journal} {Comput. Phys. Commun.}\ }\textbf {\bibinfo {volume} {185}},\ \bibinfo {pages} {2250} (\bibinfo {year} {2014})},\ \Eprint {http://arxiv.org/abs/1310.1921} {arXiv:1310.1921 [hep-ph]} \BibitemShut {NoStop}%
\bibitem [{\citenamefont {Alwall}\ \emph {et~al.}(2014)\citenamefont {Alwall}, \citenamefont {Frederix}, \citenamefont {Frixione}, \citenamefont {Hirschi}, \citenamefont {Maltoni}, \citenamefont {Mattelaer}, \citenamefont {Shao}, \citenamefont {Stelzer}, \citenamefont {Torrielli},\ and\ \citenamefont {Zaro}}]{Alwall:2014hca}%
  \BibitemOpen
  \bibfield  {author} {\bibinfo {author} {\bibfnamefont {J.}~\bibnamefont {Alwall}}, \bibinfo {author} {\bibfnamefont {R.}~\bibnamefont {Frederix}}, \bibinfo {author} {\bibfnamefont {S.}~\bibnamefont {Frixione}}, \bibinfo {author} {\bibfnamefont {V.}~\bibnamefont {Hirschi}}, \bibinfo {author} {\bibfnamefont {F.}~\bibnamefont {Maltoni}}, \bibinfo {author} {\bibfnamefont {O.}~\bibnamefont {Mattelaer}}, \bibinfo {author} {\bibfnamefont {H.~S.}\ \bibnamefont {Shao}}, \bibinfo {author} {\bibfnamefont {T.}~\bibnamefont {Stelzer}}, \bibinfo {author} {\bibfnamefont {P.}~\bibnamefont {Torrielli}}, \ and\ \bibinfo {author} {\bibfnamefont {M.}~\bibnamefont {Zaro}},\ }\href {\doibase 10.1007/JHEP07(2014)079} {\bibfield  {journal} {\bibinfo  {journal} {JHEP}\ }\textbf {\bibinfo {volume} {07}},\ \bibinfo {pages} {079} (\bibinfo {year} {2014})},\ \Eprint {http://arxiv.org/abs/1405.0301} {arXiv:1405.0301 [hep-ph]} \BibitemShut {NoStop}%
\bibitem [{\citenamefont {Gunion}\ \emph {et~al.}()\citenamefont {Gunion}, \citenamefont {Haber}, \citenamefont {Kane},\ and\ \citenamefont {Dawson}}]{Gunion:1989we}%
  \BibitemOpen
  \bibfield  {author} {\bibinfo {author} {\bibfnamefont {J.~F.}\ \bibnamefont {Gunion}}, \bibinfo {author} {\bibfnamefont {H.~E.}\ \bibnamefont {Haber}}, \bibinfo {author} {\bibfnamefont {G.~L.}\ \bibnamefont {Kane}}, \ and\ \bibinfo {author} {\bibfnamefont {S.}~\bibnamefont {Dawson}},\ }\href@noop {} {\emph {\bibinfo {title} {{The Higgs Hunter's Guide, {\rm Front. Phys. 80, (2000)}}}}}\BibitemShut {NoStop}%
\bibitem [{\citenamefont {Tumasyan}\ \emph {et~al.}(2022{\natexlab{a}})\citenamefont {Tumasyan} \emph {et~al.}}]{CMS:2022ley}%
  \BibitemOpen
  \bibfield  {author} {\bibinfo {author} {\bibfnamefont {A.}~\bibnamefont {Tumasyan}} \emph {et~al.} (\bibinfo {collaboration} {CMS}),\ }\href {\doibase 10.1038/s41567-022-01682-0} {\bibfield  {journal} {\bibinfo  {journal} {Nature Phys.}\ }\textbf {\bibinfo {volume} {18}},\ \bibinfo {pages} {1329} (\bibinfo {year} {2022}{\natexlab{a}})},\ \Eprint {http://arxiv.org/abs/2202.06923} {arXiv:2202.06923 [hep-ex]} \BibitemShut {NoStop}%
\bibitem [{\citenamefont {Aad}\ \emph {et~al.}(2023{\natexlab{b}})\citenamefont {Aad} \emph {et~al.}}]{ATLAS:2023dnm}%
  \BibitemOpen
  \bibfield  {author} {\bibinfo {author} {\bibfnamefont {G.}~\bibnamefont {Aad}} \emph {et~al.} (\bibinfo {collaboration} {ATLAS}),\ }\href {\doibase 10.1016/j.physletb.2023.138223} {\bibfield  {journal} {\bibinfo  {journal} {Phys. Lett. B}\ }\textbf {\bibinfo {volume} {846}},\ \bibinfo {pages} {138223} (\bibinfo {year} {2023}{\natexlab{b}})},\ \Eprint {http://arxiv.org/abs/2304.01532} {arXiv:2304.01532 [hep-ex]} \BibitemShut {NoStop}%
\bibitem [{\citenamefont {Tumasyan}\ \emph {et~al.}(2022{\natexlab{b}})\citenamefont {Tumasyan} \emph {et~al.}}]{CMS:2021kdm}%
  \BibitemOpen
  \bibfield  {author} {\bibinfo {author} {\bibfnamefont {A.}~\bibnamefont {Tumasyan}} \emph {et~al.} (\bibinfo {collaboration} {CMS}),\ }\href {\doibase 10.1140/epjc/s10052-022-10027-3} {\bibfield  {journal} {\bibinfo  {journal} {Eur. Phys. J. C}\ }\textbf {\bibinfo {volume} {82}},\ \bibinfo {pages} {153} (\bibinfo {year} {2022}{\natexlab{b}})},\ \Eprint {http://arxiv.org/abs/2110.04809} {arXiv:2110.04809 [hep-ex]} \BibitemShut {NoStop}%
\bibitem [{\citenamefont {Hostert}\ and\ \citenamefont {Pospelov}(2022)}]{Hostert:2020xku}%
  \BibitemOpen
  \bibfield  {author} {\bibinfo {author} {\bibfnamefont {M.}~\bibnamefont {Hostert}}\ and\ \bibinfo {author} {\bibfnamefont {M.}~\bibnamefont {Pospelov}},\ }\href {\doibase 10.1103/PhysRevD.105.015017} {\bibfield  {journal} {\bibinfo  {journal} {Phys. Rev. D}\ }\textbf {\bibinfo {volume} {105}},\ \bibinfo {pages} {015017} (\bibinfo {year} {2022})},\ \Eprint {http://arxiv.org/abs/2012.02142} {arXiv:2012.02142 [hep-ph]} \BibitemShut {NoStop}%
\bibitem [{\citenamefont {He}\ \emph {et~al.}(2020)\citenamefont {He}, \citenamefont {Ma}, \citenamefont {Tandean},\ and\ \citenamefont {Valencia}}]{He:2020jly}%
  \BibitemOpen
  \bibfield  {author} {\bibinfo {author} {\bibfnamefont {X.-G.}\ \bibnamefont {He}}, \bibinfo {author} {\bibfnamefont {X.-D.}\ \bibnamefont {Ma}}, \bibinfo {author} {\bibfnamefont {J.}~\bibnamefont {Tandean}}, \ and\ \bibinfo {author} {\bibfnamefont {G.}~\bibnamefont {Valencia}},\ }\href {\doibase 10.1007/JHEP08(2020)034} {\bibfield  {journal} {\bibinfo  {journal} {JHEP}\ }\textbf {\bibinfo {volume} {08}},\ \bibinfo {pages} {034} (\bibinfo {year} {2020})},\ \Eprint {http://arxiv.org/abs/2005.02942} {arXiv:2005.02942 [hep-ph]} \BibitemShut {NoStop}%
\bibitem [{\citenamefont {Andersen}\ \emph {et~al.}(2013)\citenamefont {Andersen} \emph {et~al.}}]{LHCHiggsCrossSectionWorkingGroup:2013rie}%
  \BibitemOpen
  \bibfield  {author} {\bibinfo {author} {\bibfnamefont {J.~R.}\ \bibnamefont {Andersen}} \emph {et~al.} (\bibinfo {collaboration} {LHC Higgs Cross Section Working Group}),\ }\href {\doibase 10.5170/CERN-2013-004} {\  (\bibinfo {year} {2013}),\ 10.5170/CERN-2013-004},\ \Eprint {http://arxiv.org/abs/1307.1347} {arXiv:1307.1347 [hep-ph]} \BibitemShut {NoStop}%
\bibitem [{\citenamefont {Djouadi}(2008)}]{Djouadi:2005gj}%
  \BibitemOpen
  \bibfield  {author} {\bibinfo {author} {\bibfnamefont {A.}~\bibnamefont {Djouadi}},\ }\href {\doibase 10.1016/j.physrep.2007.10.005} {\bibfield  {journal} {\bibinfo  {journal} {Phys. Rept.}\ }\textbf {\bibinfo {volume} {459}},\ \bibinfo {pages} {1} (\bibinfo {year} {2008})},\ \Eprint {http://arxiv.org/abs/hep-ph/0503173} {arXiv:hep-ph/0503173} \BibitemShut {NoStop}%
\bibitem [{\citenamefont {Branco}\ \emph {et~al.}(2012)\citenamefont {Branco}, \citenamefont {Ferreira}, \citenamefont {Lavoura}, \citenamefont {Rebelo}, \citenamefont {Sher},\ and\ \citenamefont {Silva}}]{Branco:2011iw}%
  \BibitemOpen
  \bibfield  {author} {\bibinfo {author} {\bibfnamefont {G.~C.}\ \bibnamefont {Branco}}, \bibinfo {author} {\bibfnamefont {P.~M.}\ \bibnamefont {Ferreira}}, \bibinfo {author} {\bibfnamefont {L.}~\bibnamefont {Lavoura}}, \bibinfo {author} {\bibfnamefont {M.~N.}\ \bibnamefont {Rebelo}}, \bibinfo {author} {\bibfnamefont {M.}~\bibnamefont {Sher}}, \ and\ \bibinfo {author} {\bibfnamefont {J.~P.}\ \bibnamefont {Silva}},\ }\href {\doibase 10.1016/j.physrep.2012.02.002} {\bibfield  {journal} {\bibinfo  {journal} {Phys. Rept.}\ }\textbf {\bibinfo {volume} {516}},\ \bibinfo {pages} {1} (\bibinfo {year} {2012})},\ \Eprint {http://arxiv.org/abs/1106.0034} {arXiv:1106.0034 [hep-ph]} \BibitemShut {NoStop}%
\bibitem [{\citenamefont {Chakrabortty}\ \emph {et~al.}(2014)\citenamefont {Chakrabortty}, \citenamefont {Konar},\ and\ \citenamefont {Mondal}}]{Chakrabortty:2013mha}%
  \BibitemOpen
  \bibfield  {author} {\bibinfo {author} {\bibfnamefont {J.}~\bibnamefont {Chakrabortty}}, \bibinfo {author} {\bibfnamefont {P.}~\bibnamefont {Konar}}, \ and\ \bibinfo {author} {\bibfnamefont {T.}~\bibnamefont {Mondal}},\ }\href {\doibase 10.1103/PhysRevD.89.095008} {\bibfield  {journal} {\bibinfo  {journal} {Phys. Rev. D}\ }\textbf {\bibinfo {volume} {89}},\ \bibinfo {pages} {095008} (\bibinfo {year} {2014})},\ \Eprint {http://arxiv.org/abs/1311.5666} {arXiv:1311.5666 [hep-ph]} \BibitemShut {NoStop}%
\bibitem [{\citenamefont {Khan}(2022)}]{Khan:2022kis}%
  \BibitemOpen
  \bibfield  {author} {\bibinfo {author} {\bibfnamefont {N.}~\bibnamefont {Khan}},\ }\href {\doibase 10.1016/j.nuclphysb.2022.116015} {\bibfield  {journal} {\bibinfo  {journal} {Nucl. Phys. B}\ }\textbf {\bibinfo {volume} {985}},\ \bibinfo {pages} {116015} (\bibinfo {year} {2022})},\ \Eprint {http://arxiv.org/abs/2206.13113} {arXiv:2206.13113 [hep-ph]} \BibitemShut {NoStop}%
\bibitem [{\citenamefont {Khan}\ and\ \citenamefont {Rakshit}(2014)}]{Khan:2014kba}%
  \BibitemOpen
  \bibfield  {author} {\bibinfo {author} {\bibfnamefont {N.}~\bibnamefont {Khan}}\ and\ \bibinfo {author} {\bibfnamefont {S.}~\bibnamefont {Rakshit}},\ }\href {\doibase 10.1103/PhysRevD.90.113008} {\bibfield  {journal} {\bibinfo  {journal} {Phys. Rev. D}\ }\textbf {\bibinfo {volume} {90}},\ \bibinfo {pages} {113008} (\bibinfo {year} {2014})},\ \Eprint {http://arxiv.org/abs/1407.6015} {arXiv:1407.6015 [hep-ph]} \BibitemShut {NoStop}%
\bibitem [{\citenamefont {Khan}\ and\ \citenamefont {Rakshit}(2015)}]{Khan:2015ipa}%
  \BibitemOpen
  \bibfield  {author} {\bibinfo {author} {\bibfnamefont {N.}~\bibnamefont {Khan}}\ and\ \bibinfo {author} {\bibfnamefont {S.}~\bibnamefont {Rakshit}},\ }\href {\doibase 10.1103/PhysRevD.92.055006} {\bibfield  {journal} {\bibinfo  {journal} {Phys. Rev. D}\ }\textbf {\bibinfo {volume} {92}},\ \bibinfo {pages} {055006} (\bibinfo {year} {2015})},\ \Eprint {http://arxiv.org/abs/1503.03085} {arXiv:1503.03085 [hep-ph]} \BibitemShut {NoStop}%
\bibitem [{\citenamefont {Kanemura}\ \emph {et~al.}(1993)\citenamefont {Kanemura}, \citenamefont {Kubota},\ and\ \citenamefont {Takasugi}}]{Kanemura:1993hm}%
  \BibitemOpen
  \bibfield  {author} {\bibinfo {author} {\bibfnamefont {S.}~\bibnamefont {Kanemura}}, \bibinfo {author} {\bibfnamefont {T.}~\bibnamefont {Kubota}}, \ and\ \bibinfo {author} {\bibfnamefont {E.}~\bibnamefont {Takasugi}},\ }\href {\doibase 10.1016/0370-2693(93)91205-2} {\bibfield  {journal} {\bibinfo  {journal} {Phys. Lett. B}\ }\textbf {\bibinfo {volume} {313}},\ \bibinfo {pages} {155} (\bibinfo {year} {1993})},\ \Eprint {http://arxiv.org/abs/hep-ph/9303263} {arXiv:hep-ph/9303263} \BibitemShut {NoStop}%
\bibitem [{\citenamefont {Arhrib}(2000)}]{Arhrib:2000is}%
  \BibitemOpen
  \bibfield  {author} {\bibinfo {author} {\bibfnamefont {A.}~\bibnamefont {Arhrib}},\ }in\ \href@noop {} {\emph {\bibinfo {booktitle} {{Workshop on Noncommutative Geometry, Superstrings and Particle Physics}}}}\ (\bibinfo {year} {2000})\ \Eprint {http://arxiv.org/abs/hep-ph/0012353} {arXiv:hep-ph/0012353} \BibitemShut {NoStop}%
\bibitem [{\citenamefont {Peskin}\ and\ \citenamefont {Takeuchi}(1992)}]{Peskin:1991sw}%
  \BibitemOpen
  \bibfield  {author} {\bibinfo {author} {\bibfnamefont {M.~E.}\ \bibnamefont {Peskin}}\ and\ \bibinfo {author} {\bibfnamefont {T.}~\bibnamefont {Takeuchi}},\ }\href {\doibase 10.1103/PhysRevD.46.381} {\bibfield  {journal} {\bibinfo  {journal} {Phys. Rev. D}\ }\textbf {\bibinfo {volume} {46}},\ \bibinfo {pages} {381} (\bibinfo {year} {1992})}\BibitemShut {NoStop}%
\bibitem [{\citenamefont {Altarelli}\ \emph {et~al.}(1993)\citenamefont {Altarelli}, \citenamefont {Barbieri},\ and\ \citenamefont {Caravaglios}}]{Altarelli:1993bh}%
  \BibitemOpen
  \bibfield  {author} {\bibinfo {author} {\bibfnamefont {G.}~\bibnamefont {Altarelli}}, \bibinfo {author} {\bibfnamefont {R.}~\bibnamefont {Barbieri}}, \ and\ \bibinfo {author} {\bibfnamefont {F.}~\bibnamefont {Caravaglios}},\ }\href {\doibase 10.1016/0370-2693(93)91249-M} {\bibfield  {journal} {\bibinfo  {journal} {Phys. Lett. B}\ }\textbf {\bibinfo {volume} {314}},\ \bibinfo {pages} {357} (\bibinfo {year} {1993})}\BibitemShut {NoStop}%
\bibitem [{\citenamefont {Baak}\ \emph {et~al.}(2014)\citenamefont {Baak}, \citenamefont {Cúth}, \citenamefont {Haller}, \citenamefont {Hoecker}, \citenamefont {Kogler}, \citenamefont {Mönig}, \citenamefont {Schott},\ and\ \citenamefont {Stelzer}}]{Baak:2014ora}%
  \BibitemOpen
  \bibfield  {author} {\bibinfo {author} {\bibfnamefont {M.}~\bibnamefont {Baak}}, \bibinfo {author} {\bibfnamefont {J.}~\bibnamefont {Cúth}}, \bibinfo {author} {\bibfnamefont {J.}~\bibnamefont {Haller}}, \bibinfo {author} {\bibfnamefont {A.}~\bibnamefont {Hoecker}}, \bibinfo {author} {\bibfnamefont {R.}~\bibnamefont {Kogler}}, \bibinfo {author} {\bibfnamefont {K.}~\bibnamefont {Mönig}}, \bibinfo {author} {\bibfnamefont {M.}~\bibnamefont {Schott}}, \ and\ \bibinfo {author} {\bibfnamefont {J.}~\bibnamefont {Stelzer}} (\bibinfo {collaboration} {Gfitter Group}),\ }\href {\doibase 10.1140/epjc/s10052-014-3046-5} {\bibfield  {journal} {\bibinfo  {journal} {Eur. Phys. J.}\ }\textbf {\bibinfo {volume} {C74}},\ \bibinfo {pages} {3046} (\bibinfo {year} {2014})},\ \Eprint {http://arxiv.org/abs/1407.3792} {arXiv:1407.3792 [hep-ph]} \BibitemShut {NoStop}%
\bibitem [{\citenamefont {Baak}\ \emph {et~al.}(2012)\citenamefont {Baak}, \citenamefont {Goebel}, \citenamefont {Haller}, \citenamefont {Hoecker}, \citenamefont {Kennedy}, \citenamefont {Moenig}, \citenamefont {Schott},\ and\ \citenamefont {Stelzer}}]{Baak:2011ze}%
  \BibitemOpen
  \bibfield  {author} {\bibinfo {author} {\bibfnamefont {M.}~\bibnamefont {Baak}}, \bibinfo {author} {\bibfnamefont {M.}~\bibnamefont {Goebel}}, \bibinfo {author} {\bibfnamefont {J.}~\bibnamefont {Haller}}, \bibinfo {author} {\bibfnamefont {A.}~\bibnamefont {Hoecker}}, \bibinfo {author} {\bibfnamefont {D.}~\bibnamefont {Kennedy}}, \bibinfo {author} {\bibfnamefont {K.}~\bibnamefont {Moenig}}, \bibinfo {author} {\bibfnamefont {M.}~\bibnamefont {Schott}}, \ and\ \bibinfo {author} {\bibfnamefont {J.}~\bibnamefont {Stelzer}} (\bibinfo {collaboration} {Gfitter}),\ }\href {\doibase 10.1140/epjc/s10052-012-2003-4} {\bibfield  {journal} {\bibinfo  {journal} {Eur. Phys. J. C}\ }\textbf {\bibinfo {volume} {72}},\ \bibinfo {pages} {2003} (\bibinfo {year} {2012})},\ \Eprint {http://arxiv.org/abs/1107.0975} {arXiv:1107.0975 [hep-ph]} \BibitemShut {NoStop}%
\bibitem [{\citenamefont {Arhrib}\ \emph {et~al.}(2012)\citenamefont {Arhrib}, \citenamefont {Benbrik},\ and\ \citenamefont {Gaur}}]{Arhrib:2012ia}%
  \BibitemOpen
  \bibfield  {author} {\bibinfo {author} {\bibfnamefont {A.}~\bibnamefont {Arhrib}}, \bibinfo {author} {\bibfnamefont {R.}~\bibnamefont {Benbrik}}, \ and\ \bibinfo {author} {\bibfnamefont {N.}~\bibnamefont {Gaur}},\ }\href {\doibase 10.1103/PhysRevD.85.095021} {\bibfield  {journal} {\bibinfo  {journal} {Phys. Rev. D}\ }\textbf {\bibinfo {volume} {85}},\ \bibinfo {pages} {095021} (\bibinfo {year} {2012})},\ \Eprint {http://arxiv.org/abs/1201.2644} {arXiv:1201.2644 [hep-ph]} \BibitemShut {NoStop}%
\bibitem [{\citenamefont {Barbieri}\ \emph {et~al.}(2006)\citenamefont {Barbieri}, \citenamefont {Hall},\ and\ \citenamefont {Rychkov}}]{Barbieri:2006dq}%
  \BibitemOpen
  \bibfield  {author} {\bibinfo {author} {\bibfnamefont {R.}~\bibnamefont {Barbieri}}, \bibinfo {author} {\bibfnamefont {L.~J.}\ \bibnamefont {Hall}}, \ and\ \bibinfo {author} {\bibfnamefont {V.~S.}\ \bibnamefont {Rychkov}},\ }\href {\doibase 10.1103/PhysRevD.74.015007} {\bibfield  {journal} {\bibinfo  {journal} {Phys. Rev. D}\ }\textbf {\bibinfo {volume} {74}},\ \bibinfo {pages} {015007} (\bibinfo {year} {2006})},\ \Eprint {http://arxiv.org/abs/hep-ph/0603188} {arXiv:hep-ph/0603188} \BibitemShut {NoStop}%
\bibitem [{\citenamefont {Haller}\ \emph {et~al.}(2018)\citenamefont {Haller}, \citenamefont {Hoecker}, \citenamefont {Kogler}, \citenamefont {M\"onig}, \citenamefont {Peiffer},\ and\ \citenamefont {Stelzer}}]{Haller:2018nnx}%
  \BibitemOpen
  \bibfield  {author} {\bibinfo {author} {\bibfnamefont {J.}~\bibnamefont {Haller}}, \bibinfo {author} {\bibfnamefont {A.}~\bibnamefont {Hoecker}}, \bibinfo {author} {\bibfnamefont {R.}~\bibnamefont {Kogler}}, \bibinfo {author} {\bibfnamefont {K.}~\bibnamefont {M\"onig}}, \bibinfo {author} {\bibfnamefont {T.}~\bibnamefont {Peiffer}}, \ and\ \bibinfo {author} {\bibfnamefont {J.}~\bibnamefont {Stelzer}},\ }\href {\doibase 10.1140/epjc/s10052-018-6131-3} {\bibfield  {journal} {\bibinfo  {journal} {Eur. Phys. J. C}\ }\textbf {\bibinfo {volume} {78}},\ \bibinfo {pages} {675} (\bibinfo {year} {2018})},\ \Eprint {http://arxiv.org/abs/1803.01853} {arXiv:1803.01853 [hep-ph]} \BibitemShut {NoStop}%
\bibitem [{\citenamefont {Khachatryan}\ \emph {et~al.}(2016)\citenamefont {Khachatryan} \emph {et~al.}}]{CMS:2016ltu}%
  \BibitemOpen
  \bibfield  {author} {\bibinfo {author} {\bibfnamefont {V.}~\bibnamefont {Khachatryan}} \emph {et~al.} (\bibinfo {collaboration} {CMS}),\ }\href {\doibase 10.1103/PhysRevLett.117.031802} {\bibfield  {journal} {\bibinfo  {journal} {Phys. Rev. Lett.}\ }\textbf {\bibinfo {volume} {117}},\ \bibinfo {pages} {031802} (\bibinfo {year} {2016})},\ \Eprint {http://arxiv.org/abs/1604.08907} {arXiv:1604.08907 [hep-ex]} \BibitemShut {NoStop}%
\bibitem [{\citenamefont {Aaltonen}\ \emph {et~al.}(2009)\citenamefont {Aaltonen} \emph {et~al.}}]{CDF:2008ieg}%
  \BibitemOpen
  \bibfield  {author} {\bibinfo {author} {\bibfnamefont {T.}~\bibnamefont {Aaltonen}} \emph {et~al.} (\bibinfo {collaboration} {CDF}),\ }\href {\doibase 10.1103/PhysRevD.79.112002} {\bibfield  {journal} {\bibinfo  {journal} {Phys. Rev. D}\ }\textbf {\bibinfo {volume} {79}},\ \bibinfo {pages} {112002} (\bibinfo {year} {2009})},\ \Eprint {http://arxiv.org/abs/0812.4036} {arXiv:0812.4036 [hep-ex]} \BibitemShut {NoStop}%
\bibitem [{\citenamefont {Aad}\ \emph {et~al.}(2021)\citenamefont {Aad} \emph {et~al.}}]{ATLAS:2021upq}%
  \BibitemOpen
  \bibfield  {author} {\bibinfo {author} {\bibfnamefont {G.}~\bibnamefont {Aad}} \emph {et~al.} (\bibinfo {collaboration} {ATLAS}),\ }\href {\doibase 10.1007/JHEP06(2021)145} {\bibfield  {journal} {\bibinfo  {journal} {JHEP}\ }\textbf {\bibinfo {volume} {06}},\ \bibinfo {pages} {145} (\bibinfo {year} {2021})},\ \Eprint {http://arxiv.org/abs/2102.10076} {arXiv:2102.10076 [hep-ex]} \BibitemShut {NoStop}%
\bibitem [{\citenamefont {Sirunyan}\ \emph {et~al.}(2020)\citenamefont {Sirunyan} \emph {et~al.}}]{CMS:2019bnu}%
  \BibitemOpen
  \bibfield  {author} {\bibinfo {author} {\bibfnamefont {A.~M.}\ \bibnamefont {Sirunyan}} \emph {et~al.} (\bibinfo {collaboration} {CMS}),\ }\href {\doibase 10.1007/JHEP03(2020)034} {\bibfield  {journal} {\bibinfo  {journal} {JHEP}\ }\textbf {\bibinfo {volume} {03}},\ \bibinfo {pages} {034} (\bibinfo {year} {2020})},\ \Eprint {http://arxiv.org/abs/1912.01594} {arXiv:1912.01594 [hep-ex]} \BibitemShut {NoStop}%
\bibitem [{\citenamefont {Hanneke}\ \emph {et~al.}(2008)\citenamefont {Hanneke}, \citenamefont {Fogwell},\ and\ \citenamefont {Gabrielse}}]{Hanneke:2008tm}%
  \BibitemOpen
  \bibfield  {author} {\bibinfo {author} {\bibfnamefont {D.}~\bibnamefont {Hanneke}}, \bibinfo {author} {\bibfnamefont {S.}~\bibnamefont {Fogwell}}, \ and\ \bibinfo {author} {\bibfnamefont {G.}~\bibnamefont {Gabrielse}},\ }\href {\doibase 10.1103/PhysRevLett.100.120801} {\bibfield  {journal} {\bibinfo  {journal} {Phys. Rev. Lett.}\ }\textbf {\bibinfo {volume} {100}},\ \bibinfo {pages} {120801} (\bibinfo {year} {2008})},\ \Eprint {http://arxiv.org/abs/0801.1134} {arXiv:0801.1134 [physics.atom-ph]} \BibitemShut {NoStop}%
\bibitem [{\citenamefont {Aoyama}\ \emph {et~al.}(2012{\natexlab{a}})\citenamefont {Aoyama}, \citenamefont {Hayakawa}, \citenamefont {Kinoshita},\ and\ \citenamefont {Nio}}]{Aoyama:2012wj}%
  \BibitemOpen
  \bibfield  {author} {\bibinfo {author} {\bibfnamefont {T.}~\bibnamefont {Aoyama}}, \bibinfo {author} {\bibfnamefont {M.}~\bibnamefont {Hayakawa}}, \bibinfo {author} {\bibfnamefont {T.}~\bibnamefont {Kinoshita}}, \ and\ \bibinfo {author} {\bibfnamefont {M.}~\bibnamefont {Nio}},\ }\href {\doibase 10.1103/PhysRevLett.109.111807} {\bibfield  {journal} {\bibinfo  {journal} {Phys. Rev. Lett.}\ }\textbf {\bibinfo {volume} {109}},\ \bibinfo {pages} {111807} (\bibinfo {year} {2012}{\natexlab{a}})},\ \Eprint {http://arxiv.org/abs/1205.5368} {arXiv:1205.5368 [hep-ph]} \BibitemShut {NoStop}%
\bibitem [{\citenamefont {Aoyama}\ \emph {et~al.}(2018)\citenamefont {Aoyama}, \citenamefont {Kinoshita},\ and\ \citenamefont {Nio}}]{Aoyama:2017uqe}%
  \BibitemOpen
  \bibfield  {author} {\bibinfo {author} {\bibfnamefont {T.}~\bibnamefont {Aoyama}}, \bibinfo {author} {\bibfnamefont {T.}~\bibnamefont {Kinoshita}}, \ and\ \bibinfo {author} {\bibfnamefont {M.}~\bibnamefont {Nio}},\ }\href {\doibase 10.1103/PhysRevD.97.036001} {\bibfield  {journal} {\bibinfo  {journal} {Phys. Rev. D}\ }\textbf {\bibinfo {volume} {97}},\ \bibinfo {pages} {036001} (\bibinfo {year} {2018})},\ \Eprint {http://arxiv.org/abs/1712.06060} {arXiv:1712.06060 [hep-ph]} \BibitemShut {NoStop}%
\bibitem [{\citenamefont {Bouchendira}\ \emph {et~al.}(2011)\citenamefont {Bouchendira}, \citenamefont {Clade}, \citenamefont {Guellati-Khelifa}, \citenamefont {Nez},\ and\ \citenamefont {Biraben}}]{Bouchendira:2010es}%
  \BibitemOpen
  \bibfield  {author} {\bibinfo {author} {\bibfnamefont {R.}~\bibnamefont {Bouchendira}}, \bibinfo {author} {\bibfnamefont {P.}~\bibnamefont {Clade}}, \bibinfo {author} {\bibfnamefont {S.}~\bibnamefont {Guellati-Khelifa}}, \bibinfo {author} {\bibfnamefont {F.}~\bibnamefont {Nez}}, \ and\ \bibinfo {author} {\bibfnamefont {F.}~\bibnamefont {Biraben}},\ }\href {\doibase 10.1103/PhysRevLett.106.080801} {\bibfield  {journal} {\bibinfo  {journal} {Phys. Rev. Lett.}\ }\textbf {\bibinfo {volume} {106}},\ \bibinfo {pages} {080801} (\bibinfo {year} {2011})},\ \Eprint {http://arxiv.org/abs/1012.3627} {arXiv:1012.3627 [physics.atom-ph]} \BibitemShut {NoStop}%
\bibitem [{\citenamefont {Morel}\ \emph {et~al.}(2020)\citenamefont {Morel}, \citenamefont {Yao}, \citenamefont {Clad\'e},\ and\ \citenamefont {Guellati-Kh\'elifa}}]{Morel:2020dww}%
  \BibitemOpen
  \bibfield  {author} {\bibinfo {author} {\bibfnamefont {L.}~\bibnamefont {Morel}}, \bibinfo {author} {\bibfnamefont {Z.}~\bibnamefont {Yao}}, \bibinfo {author} {\bibfnamefont {P.}~\bibnamefont {Clad\'e}}, \ and\ \bibinfo {author} {\bibfnamefont {S.}~\bibnamefont {Guellati-Kh\'elifa}},\ }\href {\doibase 10.1038/s41586-020-2964-7} {\bibfield  {journal} {\bibinfo  {journal} {Nature}\ }\textbf {\bibinfo {volume} {588}},\ \bibinfo {pages} {61} (\bibinfo {year} {2020})}\BibitemShut {NoStop}%
\bibitem [{\citenamefont {Parker}\ \emph {et~al.}(2018)\citenamefont {Parker}, \citenamefont {Yu}, \citenamefont {Zhong}, \citenamefont {Estey},\ and\ \citenamefont {M\"uller}}]{Parker:2018vye}%
  \BibitemOpen
  \bibfield  {author} {\bibinfo {author} {\bibfnamefont {R.~H.}\ \bibnamefont {Parker}}, \bibinfo {author} {\bibfnamefont {C.}~\bibnamefont {Yu}}, \bibinfo {author} {\bibfnamefont {W.}~\bibnamefont {Zhong}}, \bibinfo {author} {\bibfnamefont {B.}~\bibnamefont {Estey}}, \ and\ \bibinfo {author} {\bibfnamefont {H.}~\bibnamefont {M\"uller}},\ }\href {\doibase 10.1126/science.aap7706} {\bibfield  {journal} {\bibinfo  {journal} {Science}\ }\textbf {\bibinfo {volume} {360}},\ \bibinfo {pages} {191} (\bibinfo {year} {2018})},\ \Eprint {http://arxiv.org/abs/1812.04130} {arXiv:1812.04130 [physics.atom-ph]} \BibitemShut {NoStop}%
\bibitem [{\citenamefont {Chang}\ \emph {et~al.}(2001)\citenamefont {Chang}, \citenamefont {Chang}, \citenamefont {Chou},\ and\ \citenamefont {Keung}}]{Chang:2000ii}%
  \BibitemOpen
  \bibfield  {author} {\bibinfo {author} {\bibfnamefont {D.}~\bibnamefont {Chang}}, \bibinfo {author} {\bibfnamefont {W.-F.}\ \bibnamefont {Chang}}, \bibinfo {author} {\bibfnamefont {C.-H.}\ \bibnamefont {Chou}}, \ and\ \bibinfo {author} {\bibfnamefont {W.-Y.}\ \bibnamefont {Keung}},\ }\href {\doibase 10.1103/PhysRevD.63.091301} {\bibfield  {journal} {\bibinfo  {journal} {Phys. Rev. D}\ }\textbf {\bibinfo {volume} {63}},\ \bibinfo {pages} {091301} (\bibinfo {year} {2001})},\ \Eprint {http://arxiv.org/abs/hep-ph/0009292} {arXiv:hep-ph/0009292} \BibitemShut {NoStop}%
\bibitem [{\citenamefont {Larios}\ \emph {et~al.}(2001)\citenamefont {Larios}, \citenamefont {Tavares-Velasco},\ and\ \citenamefont {Yuan}}]{Larios:2001ma}%
  \BibitemOpen
  \bibfield  {author} {\bibinfo {author} {\bibfnamefont {F.}~\bibnamefont {Larios}}, \bibinfo {author} {\bibfnamefont {G.}~\bibnamefont {Tavares-Velasco}}, \ and\ \bibinfo {author} {\bibfnamefont {C.~P.}\ \bibnamefont {Yuan}},\ }\href {\doibase 10.1103/PhysRevD.64.055004} {\bibfield  {journal} {\bibinfo  {journal} {Phys. Rev. D}\ }\textbf {\bibinfo {volume} {64}},\ \bibinfo {pages} {055004} (\bibinfo {year} {2001})},\ \Eprint {http://arxiv.org/abs/hep-ph/0103292} {arXiv:hep-ph/0103292} \BibitemShut {NoStop}%
\bibitem [{\citenamefont {Ilisie}(2015)}]{Ilisie:2015tra}%
  \BibitemOpen
  \bibfield  {author} {\bibinfo {author} {\bibfnamefont {V.}~\bibnamefont {Ilisie}},\ }\href {\doibase 10.1007/JHEP04(2015)077} {\bibfield  {journal} {\bibinfo  {journal} {JHEP}\ }\textbf {\bibinfo {volume} {04}},\ \bibinfo {pages} {077} (\bibinfo {year} {2015})},\ \Eprint {http://arxiv.org/abs/1502.04199} {arXiv:1502.04199 [hep-ph]} \BibitemShut {NoStop}%
\bibitem [{\citenamefont {Aoyama}\ \emph {et~al.}(2020)\citenamefont {Aoyama} \emph {et~al.}}]{Aoyama:2020ynm}%
  \BibitemOpen
  \bibfield  {author} {\bibinfo {author} {\bibfnamefont {T.}~\bibnamefont {Aoyama}} \emph {et~al.},\ }\href {\doibase 10.1016/j.physrep.2020.07.006} {\bibfield  {journal} {\bibinfo  {journal} {Phys. Rept.}\ }\textbf {\bibinfo {volume} {887}},\ \bibinfo {pages} {1} (\bibinfo {year} {2020})},\ \Eprint {http://arxiv.org/abs/2006.04822} {arXiv:2006.04822 [hep-ph]} \BibitemShut {NoStop}%
\bibitem [{\citenamefont {Aoyama}\ \emph {et~al.}(2012{\natexlab{b}})\citenamefont {Aoyama}, \citenamefont {Hayakawa}, \citenamefont {Kinoshita},\ and\ \citenamefont {Nio}}]{Aoyama:2012wk}%
  \BibitemOpen
  \bibfield  {author} {\bibinfo {author} {\bibfnamefont {T.}~\bibnamefont {Aoyama}}, \bibinfo {author} {\bibfnamefont {M.}~\bibnamefont {Hayakawa}}, \bibinfo {author} {\bibfnamefont {T.}~\bibnamefont {Kinoshita}}, \ and\ \bibinfo {author} {\bibfnamefont {M.}~\bibnamefont {Nio}},\ }\href {\doibase 10.1103/PhysRevLett.109.111808} {\bibfield  {journal} {\bibinfo  {journal} {Phys. Rev. Lett.}\ }\textbf {\bibinfo {volume} {109}},\ \bibinfo {pages} {111808} (\bibinfo {year} {2012}{\natexlab{b}})},\ \Eprint {http://arxiv.org/abs/1205.5370} {arXiv:1205.5370 [hep-ph]} \BibitemShut {NoStop}%
\bibitem [{\citenamefont {Aoyama}\ \emph {et~al.}(2019)\citenamefont {Aoyama}, \citenamefont {Kinoshita},\ and\ \citenamefont {Nio}}]{Aoyama:2019ryr}%
  \BibitemOpen
  \bibfield  {author} {\bibinfo {author} {\bibfnamefont {T.}~\bibnamefont {Aoyama}}, \bibinfo {author} {\bibfnamefont {T.}~\bibnamefont {Kinoshita}}, \ and\ \bibinfo {author} {\bibfnamefont {M.}~\bibnamefont {Nio}},\ }\href {\doibase 10.3390/atoms7010028} {\bibfield  {journal} {\bibinfo  {journal} {Atoms}\ }\textbf {\bibinfo {volume} {7}},\ \bibinfo {pages} {28} (\bibinfo {year} {2019})}\BibitemShut {NoStop}%
\bibitem [{\citenamefont {Czarnecki}\ \emph {et~al.}(2003)\citenamefont {Czarnecki}, \citenamefont {Marciano},\ and\ \citenamefont {Vainshtein}}]{Czarnecki:2002nt}%
  \BibitemOpen
  \bibfield  {author} {\bibinfo {author} {\bibfnamefont {A.}~\bibnamefont {Czarnecki}}, \bibinfo {author} {\bibfnamefont {W.~J.}\ \bibnamefont {Marciano}}, \ and\ \bibinfo {author} {\bibfnamefont {A.}~\bibnamefont {Vainshtein}},\ }\href {\doibase 10.1103/PhysRevD.67.073006} {\bibfield  {journal} {\bibinfo  {journal} {Phys. Rev. D}\ }\textbf {\bibinfo {volume} {67}},\ \bibinfo {pages} {073006} (\bibinfo {year} {2003})},\ \bibinfo {note} {[Erratum: Phys.Rev.D 73, 119901 (2006)]},\ \Eprint {http://arxiv.org/abs/hep-ph/0212229} {arXiv:hep-ph/0212229} \BibitemShut {NoStop}%
\bibitem [{\citenamefont {Gnendiger}\ \emph {et~al.}(2013)\citenamefont {Gnendiger}, \citenamefont {St\"ockinger},\ and\ \citenamefont {St\"ockinger-Kim}}]{Gnendiger:2013pva}%
  \BibitemOpen
  \bibfield  {author} {\bibinfo {author} {\bibfnamefont {C.}~\bibnamefont {Gnendiger}}, \bibinfo {author} {\bibfnamefont {D.}~\bibnamefont {St\"ockinger}}, \ and\ \bibinfo {author} {\bibfnamefont {H.}~\bibnamefont {St\"ockinger-Kim}},\ }\href {\doibase 10.1103/PhysRevD.88.053005} {\bibfield  {journal} {\bibinfo  {journal} {Phys. Rev. D}\ }\textbf {\bibinfo {volume} {88}},\ \bibinfo {pages} {053005} (\bibinfo {year} {2013})},\ \Eprint {http://arxiv.org/abs/1306.5546} {arXiv:1306.5546 [hep-ph]} \BibitemShut {NoStop}%
\bibitem [{\citenamefont {Davier}\ \emph {et~al.}(2017)\citenamefont {Davier}, \citenamefont {Hoecker}, \citenamefont {Malaescu},\ and\ \citenamefont {Zhang}}]{Davier:2017zfy}%
  \BibitemOpen
  \bibfield  {author} {\bibinfo {author} {\bibfnamefont {M.}~\bibnamefont {Davier}}, \bibinfo {author} {\bibfnamefont {A.}~\bibnamefont {Hoecker}}, \bibinfo {author} {\bibfnamefont {B.}~\bibnamefont {Malaescu}}, \ and\ \bibinfo {author} {\bibfnamefont {Z.}~\bibnamefont {Zhang}},\ }\href {\doibase 10.1140/epjc/s10052-017-5161-6} {\bibfield  {journal} {\bibinfo  {journal} {Eur. Phys. J. C}\ }\textbf {\bibinfo {volume} {77}},\ \bibinfo {pages} {827} (\bibinfo {year} {2017})},\ \Eprint {http://arxiv.org/abs/1706.09436} {arXiv:1706.09436 [hep-ph]} \BibitemShut {NoStop}%
\bibitem [{\citenamefont {Keshavarzi}\ \emph {et~al.}(2018)\citenamefont {Keshavarzi}, \citenamefont {Nomura},\ and\ \citenamefont {Teubner}}]{Keshavarzi:2018mgv}%
  \BibitemOpen
  \bibfield  {author} {\bibinfo {author} {\bibfnamefont {A.}~\bibnamefont {Keshavarzi}}, \bibinfo {author} {\bibfnamefont {D.}~\bibnamefont {Nomura}}, \ and\ \bibinfo {author} {\bibfnamefont {T.}~\bibnamefont {Teubner}},\ }\href {\doibase 10.1103/PhysRevD.97.114025} {\bibfield  {journal} {\bibinfo  {journal} {Phys. Rev. D}\ }\textbf {\bibinfo {volume} {97}},\ \bibinfo {pages} {114025} (\bibinfo {year} {2018})},\ \Eprint {http://arxiv.org/abs/1802.02995} {arXiv:1802.02995 [hep-ph]} \BibitemShut {NoStop}%
\bibitem [{\citenamefont {Colangelo}\ \emph {et~al.}(2019)\citenamefont {Colangelo}, \citenamefont {Hoferichter},\ and\ \citenamefont {Stoffer}}]{Colangelo:2018mtw}%
  \BibitemOpen
  \bibfield  {author} {\bibinfo {author} {\bibfnamefont {G.}~\bibnamefont {Colangelo}}, \bibinfo {author} {\bibfnamefont {M.}~\bibnamefont {Hoferichter}}, \ and\ \bibinfo {author} {\bibfnamefont {P.}~\bibnamefont {Stoffer}},\ }\href {\doibase 10.1007/JHEP02(2019)006} {\bibfield  {journal} {\bibinfo  {journal} {JHEP}\ }\textbf {\bibinfo {volume} {02}},\ \bibinfo {pages} {006} (\bibinfo {year} {2019})},\ \Eprint {http://arxiv.org/abs/1810.00007} {arXiv:1810.00007 [hep-ph]} \BibitemShut {NoStop}%
\bibitem [{\citenamefont {Hoferichter}\ \emph {et~al.}(2019)\citenamefont {Hoferichter}, \citenamefont {Hoid},\ and\ \citenamefont {Kubis}}]{Hoferichter:2019mqg}%
  \BibitemOpen
  \bibfield  {author} {\bibinfo {author} {\bibfnamefont {M.}~\bibnamefont {Hoferichter}}, \bibinfo {author} {\bibfnamefont {B.-L.}\ \bibnamefont {Hoid}}, \ and\ \bibinfo {author} {\bibfnamefont {B.}~\bibnamefont {Kubis}},\ }\href {\doibase 10.1007/JHEP08(2019)137} {\bibfield  {journal} {\bibinfo  {journal} {JHEP}\ }\textbf {\bibinfo {volume} {08}},\ \bibinfo {pages} {137} (\bibinfo {year} {2019})},\ \Eprint {http://arxiv.org/abs/1907.01556} {arXiv:1907.01556 [hep-ph]} \BibitemShut {NoStop}%
\bibitem [{\citenamefont {Davier}\ \emph {et~al.}(2020)\citenamefont {Davier}, \citenamefont {Hoecker}, \citenamefont {Malaescu},\ and\ \citenamefont {Zhang}}]{Davier:2019can}%
  \BibitemOpen
  \bibfield  {author} {\bibinfo {author} {\bibfnamefont {M.}~\bibnamefont {Davier}}, \bibinfo {author} {\bibfnamefont {A.}~\bibnamefont {Hoecker}}, \bibinfo {author} {\bibfnamefont {B.}~\bibnamefont {Malaescu}}, \ and\ \bibinfo {author} {\bibfnamefont {Z.}~\bibnamefont {Zhang}},\ }\href {\doibase 10.1140/epjc/s10052-020-7792-2} {\bibfield  {journal} {\bibinfo  {journal} {Eur. Phys. J. C}\ }\textbf {\bibinfo {volume} {80}},\ \bibinfo {pages} {241} (\bibinfo {year} {2020})},\ \bibinfo {note} {[Erratum: Eur.Phys.J.C 80, 410 (2020)]},\ \Eprint {http://arxiv.org/abs/1908.00921} {arXiv:1908.00921 [hep-ph]} \BibitemShut {NoStop}%
\bibitem [{\citenamefont {Keshavarzi}\ \emph {et~al.}(2020{\natexlab{a}})\citenamefont {Keshavarzi}, \citenamefont {Nomura},\ and\ \citenamefont {Teubner}}]{Keshavarzi:2019abf}%
  \BibitemOpen
  \bibfield  {author} {\bibinfo {author} {\bibfnamefont {A.}~\bibnamefont {Keshavarzi}}, \bibinfo {author} {\bibfnamefont {D.}~\bibnamefont {Nomura}}, \ and\ \bibinfo {author} {\bibfnamefont {T.}~\bibnamefont {Teubner}},\ }\href {\doibase 10.1103/PhysRevD.101.014029} {\bibfield  {journal} {\bibinfo  {journal} {Phys. Rev. D}\ }\textbf {\bibinfo {volume} {101}},\ \bibinfo {pages} {014029} (\bibinfo {year} {2020}{\natexlab{a}})},\ \Eprint {http://arxiv.org/abs/1911.00367} {arXiv:1911.00367 [hep-ph]} \BibitemShut {NoStop}%
\bibitem [{\citenamefont {Kurz}\ \emph {et~al.}(2014)\citenamefont {Kurz}, \citenamefont {Liu}, \citenamefont {Marquard},\ and\ \citenamefont {Steinhauser}}]{Kurz:2014wya}%
  \BibitemOpen
  \bibfield  {author} {\bibinfo {author} {\bibfnamefont {A.}~\bibnamefont {Kurz}}, \bibinfo {author} {\bibfnamefont {T.}~\bibnamefont {Liu}}, \bibinfo {author} {\bibfnamefont {P.}~\bibnamefont {Marquard}}, \ and\ \bibinfo {author} {\bibfnamefont {M.}~\bibnamefont {Steinhauser}},\ }\href {\doibase 10.1016/j.physletb.2014.05.043} {\bibfield  {journal} {\bibinfo  {journal} {Phys. Lett. B}\ }\textbf {\bibinfo {volume} {734}},\ \bibinfo {pages} {144} (\bibinfo {year} {2014})},\ \Eprint {http://arxiv.org/abs/1403.6400} {arXiv:1403.6400 [hep-ph]} \BibitemShut {NoStop}%
\bibitem [{\citenamefont {Melnikov}\ and\ \citenamefont {Vainshtein}(2004)}]{Melnikov:2003xd}%
  \BibitemOpen
  \bibfield  {author} {\bibinfo {author} {\bibfnamefont {K.}~\bibnamefont {Melnikov}}\ and\ \bibinfo {author} {\bibfnamefont {A.}~\bibnamefont {Vainshtein}},\ }\href {\doibase 10.1103/PhysRevD.70.113006} {\bibfield  {journal} {\bibinfo  {journal} {Phys. Rev. D}\ }\textbf {\bibinfo {volume} {70}},\ \bibinfo {pages} {113006} (\bibinfo {year} {2004})},\ \Eprint {http://arxiv.org/abs/hep-ph/0312226} {arXiv:hep-ph/0312226} \BibitemShut {NoStop}%
\bibitem [{\citenamefont {Masjuan}\ and\ \citenamefont {Sanchez-Puertas}(2017)}]{Masjuan:2017tvw}%
  \BibitemOpen
  \bibfield  {author} {\bibinfo {author} {\bibfnamefont {P.}~\bibnamefont {Masjuan}}\ and\ \bibinfo {author} {\bibfnamefont {P.}~\bibnamefont {Sanchez-Puertas}},\ }\href {\doibase 10.1103/PhysRevD.95.054026} {\bibfield  {journal} {\bibinfo  {journal} {Phys. Rev. D}\ }\textbf {\bibinfo {volume} {95}},\ \bibinfo {pages} {054026} (\bibinfo {year} {2017})},\ \Eprint {http://arxiv.org/abs/1701.05829} {arXiv:1701.05829 [hep-ph]} \BibitemShut {NoStop}%
\bibitem [{\citenamefont {Colangelo}\ \emph {et~al.}(2017)\citenamefont {Colangelo}, \citenamefont {Hoferichter}, \citenamefont {Procura},\ and\ \citenamefont {Stoffer}}]{Colangelo:2017fiz}%
  \BibitemOpen
  \bibfield  {author} {\bibinfo {author} {\bibfnamefont {G.}~\bibnamefont {Colangelo}}, \bibinfo {author} {\bibfnamefont {M.}~\bibnamefont {Hoferichter}}, \bibinfo {author} {\bibfnamefont {M.}~\bibnamefont {Procura}}, \ and\ \bibinfo {author} {\bibfnamefont {P.}~\bibnamefont {Stoffer}},\ }\href {\doibase 10.1007/JHEP04(2017)161} {\bibfield  {journal} {\bibinfo  {journal} {JHEP}\ }\textbf {\bibinfo {volume} {04}},\ \bibinfo {pages} {161} (\bibinfo {year} {2017})},\ \Eprint {http://arxiv.org/abs/1702.07347} {arXiv:1702.07347 [hep-ph]} \BibitemShut {NoStop}%
\bibitem [{\citenamefont {Hoferichter}\ \emph {et~al.}(2018)\citenamefont {Hoferichter}, \citenamefont {Hoid}, \citenamefont {Kubis}, \citenamefont {Leupold},\ and\ \citenamefont {Schneider}}]{Hoferichter:2018kwz}%
  \BibitemOpen
  \bibfield  {author} {\bibinfo {author} {\bibfnamefont {M.}~\bibnamefont {Hoferichter}}, \bibinfo {author} {\bibfnamefont {B.-L.}\ \bibnamefont {Hoid}}, \bibinfo {author} {\bibfnamefont {B.}~\bibnamefont {Kubis}}, \bibinfo {author} {\bibfnamefont {S.}~\bibnamefont {Leupold}}, \ and\ \bibinfo {author} {\bibfnamefont {S.~P.}\ \bibnamefont {Schneider}},\ }\href {\doibase 10.1007/JHEP10(2018)141} {\bibfield  {journal} {\bibinfo  {journal} {JHEP}\ }\textbf {\bibinfo {volume} {10}},\ \bibinfo {pages} {141} (\bibinfo {year} {2018})},\ \Eprint {http://arxiv.org/abs/1808.04823} {arXiv:1808.04823 [hep-ph]} \BibitemShut {NoStop}%
\bibitem [{\citenamefont {G\'erardin}\ \emph {et~al.}(2019)\citenamefont {G\'erardin}, \citenamefont {Meyer},\ and\ \citenamefont {Nyffeler}}]{Gerardin:2019vio}%
  \BibitemOpen
  \bibfield  {author} {\bibinfo {author} {\bibfnamefont {A.}~\bibnamefont {G\'erardin}}, \bibinfo {author} {\bibfnamefont {H.~B.}\ \bibnamefont {Meyer}}, \ and\ \bibinfo {author} {\bibfnamefont {A.}~\bibnamefont {Nyffeler}},\ }\href {\doibase 10.1103/PhysRevD.100.034520} {\bibfield  {journal} {\bibinfo  {journal} {Phys. Rev. D}\ }\textbf {\bibinfo {volume} {100}},\ \bibinfo {pages} {034520} (\bibinfo {year} {2019})},\ \Eprint {http://arxiv.org/abs/1903.09471} {arXiv:1903.09471 [hep-lat]} \BibitemShut {NoStop}%
\bibitem [{\citenamefont {Bijnens}\ \emph {et~al.}(2019)\citenamefont {Bijnens}, \citenamefont {Hermansson-Truedsson},\ and\ \citenamefont {Rodr\'\i{}guez-S\'anchez}}]{Bijnens:2019ghy}%
  \BibitemOpen
  \bibfield  {author} {\bibinfo {author} {\bibfnamefont {J.}~\bibnamefont {Bijnens}}, \bibinfo {author} {\bibfnamefont {N.}~\bibnamefont {Hermansson-Truedsson}}, \ and\ \bibinfo {author} {\bibfnamefont {A.}~\bibnamefont {Rodr\'\i{}guez-S\'anchez}},\ }\href {\doibase 10.1016/j.physletb.2019.134994} {\bibfield  {journal} {\bibinfo  {journal} {Phys. Lett. B}\ }\textbf {\bibinfo {volume} {798}},\ \bibinfo {pages} {134994} (\bibinfo {year} {2019})},\ \Eprint {http://arxiv.org/abs/1908.03331} {arXiv:1908.03331 [hep-ph]} \BibitemShut {NoStop}%
\bibitem [{\citenamefont {Colangelo}\ \emph {et~al.}(2020)\citenamefont {Colangelo}, \citenamefont {Hagelstein}, \citenamefont {Hoferichter}, \citenamefont {Laub},\ and\ \citenamefont {Stoffer}}]{Colangelo:2019uex}%
  \BibitemOpen
  \bibfield  {author} {\bibinfo {author} {\bibfnamefont {G.}~\bibnamefont {Colangelo}}, \bibinfo {author} {\bibfnamefont {F.}~\bibnamefont {Hagelstein}}, \bibinfo {author} {\bibfnamefont {M.}~\bibnamefont {Hoferichter}}, \bibinfo {author} {\bibfnamefont {L.}~\bibnamefont {Laub}}, \ and\ \bibinfo {author} {\bibfnamefont {P.}~\bibnamefont {Stoffer}},\ }\href {\doibase 10.1007/JHEP03(2020)101} {\bibfield  {journal} {\bibinfo  {journal} {JHEP}\ }\textbf {\bibinfo {volume} {03}},\ \bibinfo {pages} {101} (\bibinfo {year} {2020})},\ \Eprint {http://arxiv.org/abs/1910.13432} {arXiv:1910.13432 [hep-ph]} \BibitemShut {NoStop}%
\bibitem [{\citenamefont {Blum}\ \emph {et~al.}(2020)\citenamefont {Blum}, \citenamefont {Christ}, \citenamefont {Hayakawa}, \citenamefont {Izubuchi}, \citenamefont {Jin}, \citenamefont {Jung},\ and\ \citenamefont {Lehner}}]{Blum:2019ugy}%
  \BibitemOpen
  \bibfield  {author} {\bibinfo {author} {\bibfnamefont {T.}~\bibnamefont {Blum}}, \bibinfo {author} {\bibfnamefont {N.}~\bibnamefont {Christ}}, \bibinfo {author} {\bibfnamefont {M.}~\bibnamefont {Hayakawa}}, \bibinfo {author} {\bibfnamefont {T.}~\bibnamefont {Izubuchi}}, \bibinfo {author} {\bibfnamefont {L.}~\bibnamefont {Jin}}, \bibinfo {author} {\bibfnamefont {C.}~\bibnamefont {Jung}}, \ and\ \bibinfo {author} {\bibfnamefont {C.}~\bibnamefont {Lehner}},\ }\href {\doibase 10.1103/PhysRevLett.124.132002} {\bibfield  {journal} {\bibinfo  {journal} {Phys. Rev. Lett.}\ }\textbf {\bibinfo {volume} {124}},\ \bibinfo {pages} {132002} (\bibinfo {year} {2020})},\ \Eprint {http://arxiv.org/abs/1911.08123} {arXiv:1911.08123 [hep-lat]} \BibitemShut {NoStop}%
\bibitem [{\citenamefont {Colangelo}\ \emph {et~al.}(2014)\citenamefont {Colangelo}, \citenamefont {Hoferichter}, \citenamefont {Nyffeler}, \citenamefont {Passera},\ and\ \citenamefont {Stoffer}}]{Colangelo:2014qya}%
  \BibitemOpen
  \bibfield  {author} {\bibinfo {author} {\bibfnamefont {G.}~\bibnamefont {Colangelo}}, \bibinfo {author} {\bibfnamefont {M.}~\bibnamefont {Hoferichter}}, \bibinfo {author} {\bibfnamefont {A.}~\bibnamefont {Nyffeler}}, \bibinfo {author} {\bibfnamefont {M.}~\bibnamefont {Passera}}, \ and\ \bibinfo {author} {\bibfnamefont {P.}~\bibnamefont {Stoffer}},\ }\href {\doibase 10.1016/j.physletb.2014.06.012} {\bibfield  {journal} {\bibinfo  {journal} {Phys. Lett. B}\ }\textbf {\bibinfo {volume} {735}},\ \bibinfo {pages} {90} (\bibinfo {year} {2014})},\ \Eprint {http://arxiv.org/abs/1403.7512} {arXiv:1403.7512 [hep-ph]} \BibitemShut {NoStop}%
\bibitem [{\citenamefont {Borsanyi}\ \emph {et~al.}(2021)\citenamefont {Borsanyi} \emph {et~al.}}]{Borsanyi:2020mff}%
  \BibitemOpen
  \bibfield  {author} {\bibinfo {author} {\bibfnamefont {S.}~\bibnamefont {Borsanyi}} \emph {et~al.},\ }\href {\doibase 10.1038/s41586-021-03418-1} {\bibfield  {journal} {\bibinfo  {journal} {Nature}\ }\textbf {\bibinfo {volume} {593}},\ \bibinfo {pages} {51} (\bibinfo {year} {2021})},\ \Eprint {http://arxiv.org/abs/2002.12347} {arXiv:2002.12347 [hep-lat]} \BibitemShut {NoStop}%
\bibitem [{\citenamefont {Abi}\ \emph {et~al.}(2021)\citenamefont {Abi} \emph {et~al.}}]{Muong-2:2021ojo}%
  \BibitemOpen
  \bibfield  {author} {\bibinfo {author} {\bibfnamefont {B.}~\bibnamefont {Abi}} \emph {et~al.} (\bibinfo {collaboration} {Muon g-2}),\ }\href {\doibase 10.1103/PhysRevLett.126.141801} {\bibfield  {journal} {\bibinfo  {journal} {Phys. Rev. Lett.}\ }\textbf {\bibinfo {volume} {126}},\ \bibinfo {pages} {141801} (\bibinfo {year} {2021})},\ \Eprint {http://arxiv.org/abs/2104.03281} {arXiv:2104.03281 [hep-ex]} \BibitemShut {NoStop}%
\bibitem [{\citenamefont {Bennett}\ \emph {et~al.}(2006)\citenamefont {Bennett} \emph {et~al.}}]{Muong-2:2006rrc}%
  \BibitemOpen
  \bibfield  {author} {\bibinfo {author} {\bibfnamefont {G.~W.}\ \bibnamefont {Bennett}} \emph {et~al.} (\bibinfo {collaboration} {Muon g-2}),\ }\href {\doibase 10.1103/PhysRevD.73.072003} {\bibfield  {journal} {\bibinfo  {journal} {Phys. Rev. D}\ }\textbf {\bibinfo {volume} {73}},\ \bibinfo {pages} {072003} (\bibinfo {year} {2006})},\ \Eprint {http://arxiv.org/abs/hep-ex/0602035} {arXiv:hep-ex/0602035} \BibitemShut {NoStop}%
\bibitem [{\citenamefont {Aguillard}\ \emph {et~al.}(2023)\citenamefont {Aguillard} \emph {et~al.}}]{Muong-2:2023cdq}%
  \BibitemOpen
  \bibfield  {author} {\bibinfo {author} {\bibfnamefont {D.~P.}\ \bibnamefont {Aguillard}} \emph {et~al.} (\bibinfo {collaboration} {Muon g-2}),\ }\href {\doibase 10.1103/PhysRevLett.131.161802} {\bibfield  {journal} {\bibinfo  {journal} {Phys. Rev. Lett.}\ }\textbf {\bibinfo {volume} {131}},\ \bibinfo {pages} {161802} (\bibinfo {year} {2023})},\ \Eprint {http://arxiv.org/abs/2308.06230} {arXiv:2308.06230 [hep-ex]} \BibitemShut {NoStop}%
\bibitem [{\citenamefont {Ignatov}\ \emph {et~al.}(2024{\natexlab{a}})\citenamefont {Ignatov} \emph {et~al.}}]{CMD-3:2023alj}%
  \BibitemOpen
  \bibfield  {author} {\bibinfo {author} {\bibfnamefont {F.~V.}\ \bibnamefont {Ignatov}} \emph {et~al.} (\bibinfo {collaboration} {CMD-3}),\ }\href {\doibase 10.1103/PhysRevD.109.112002} {\bibfield  {journal} {\bibinfo  {journal} {Phys. Rev. D}\ }\textbf {\bibinfo {volume} {109}},\ \bibinfo {pages} {112002} (\bibinfo {year} {2024}{\natexlab{a}})},\ \Eprint {http://arxiv.org/abs/2302.08834} {arXiv:2302.08834 [hep-ex]} \BibitemShut {NoStop}%
\bibitem [{\citenamefont {Ignatov}\ \emph {et~al.}(2024{\natexlab{b}})\citenamefont {Ignatov} \emph {et~al.}}]{CMD-3:2023rfe}%
  \BibitemOpen
  \bibfield  {author} {\bibinfo {author} {\bibfnamefont {F.~V.}\ \bibnamefont {Ignatov}} \emph {et~al.} (\bibinfo {collaboration} {CMD-3}),\ }\href {\doibase 10.1103/PhysRevLett.132.231903} {\bibfield  {journal} {\bibinfo  {journal} {Phys. Rev. Lett.}\ }\textbf {\bibinfo {volume} {132}},\ \bibinfo {pages} {231903} (\bibinfo {year} {2024}{\natexlab{b}})},\ \Eprint {http://arxiv.org/abs/2309.12910} {arXiv:2309.12910 [hep-ex]} \BibitemShut {NoStop}%
\bibitem [{\citenamefont {Lees}\ \emph {et~al.}(2012)\citenamefont {Lees} \emph {et~al.}}]{BaBar:2012bdw}%
  \BibitemOpen
  \bibfield  {author} {\bibinfo {author} {\bibfnamefont {J.~P.}\ \bibnamefont {Lees}} \emph {et~al.} (\bibinfo {collaboration} {BaBar}),\ }\href {\doibase 10.1103/PhysRevD.86.032013} {\bibfield  {journal} {\bibinfo  {journal} {Phys. Rev. D}\ }\textbf {\bibinfo {volume} {86}},\ \bibinfo {pages} {032013} (\bibinfo {year} {2012})},\ \Eprint {http://arxiv.org/abs/1205.2228} {arXiv:1205.2228 [hep-ex]} \BibitemShut {NoStop}%
\bibitem [{\citenamefont {Ablikim}\ \emph {et~al.}(2016)\citenamefont {Ablikim} \emph {et~al.}}]{BESIII:2015equ}%
  \BibitemOpen
  \bibfield  {author} {\bibinfo {author} {\bibfnamefont {M.}~\bibnamefont {Ablikim}} \emph {et~al.} (\bibinfo {collaboration} {BESIII}),\ }\href {\doibase 10.1016/j.physletb.2015.11.043} {\bibfield  {journal} {\bibinfo  {journal} {Phys. Lett. B}\ }\textbf {\bibinfo {volume} {753}},\ \bibinfo {pages} {629} (\bibinfo {year} {2016})},\ \bibinfo {note} {[Erratum: Phys.Lett.B 812, 135982 (2021)]},\ \Eprint {http://arxiv.org/abs/1507.08188} {arXiv:1507.08188 [hep-ex]} \BibitemShut {NoStop}%
\bibitem [{\citenamefont {Anastasi}\ \emph {et~al.}(2018)\citenamefont {Anastasi} \emph {et~al.}}]{KLOE-2:2017fda}%
  \BibitemOpen
  \bibfield  {author} {\bibinfo {author} {\bibfnamefont {A.}~\bibnamefont {Anastasi}} \emph {et~al.} (\bibinfo {collaboration} {KLOE-2}),\ }\href {\doibase 10.1007/JHEP03(2018)173} {\bibfield  {journal} {\bibinfo  {journal} {JHEP}\ }\textbf {\bibinfo {volume} {03}},\ \bibinfo {pages} {173} (\bibinfo {year} {2018})},\ \Eprint {http://arxiv.org/abs/1711.03085} {arXiv:1711.03085 [hep-ex]} \BibitemShut {NoStop}%
\bibitem [{\citenamefont {Masjuan}\ \emph {et~al.}(2024)\citenamefont {Masjuan}, \citenamefont {Miranda},\ and\ \citenamefont {Roig}}]{Masjuan:2023yam}%
  \BibitemOpen
  \bibfield  {author} {\bibinfo {author} {\bibfnamefont {P.}~\bibnamefont {Masjuan}}, \bibinfo {author} {\bibfnamefont {A.}~\bibnamefont {Miranda}}, \ and\ \bibinfo {author} {\bibfnamefont {P.}~\bibnamefont {Roig}},\ }\href {\doibase 10.1016/j.nuclphysbps.2023.12.001} {\bibfield  {journal} {\bibinfo  {journal} {Nucl. Part. Phys. Proc.}\ }\textbf {\bibinfo {volume} {343}},\ \bibinfo {pages} {99} (\bibinfo {year} {2024})},\ \Eprint {http://arxiv.org/abs/2310.14102} {arXiv:2310.14102 [hep-ph]} \BibitemShut {NoStop}%
\bibitem [{\citenamefont {Boccaletti}\ \emph {et~al.}(2024)\citenamefont {Boccaletti} \emph {et~al.}}]{Boccaletti:2024guq}%
  \BibitemOpen
  \bibfield  {author} {\bibinfo {author} {\bibfnamefont {A.}~\bibnamefont {Boccaletti}} \emph {et~al.},\ }\href@noop {} {\  (\bibinfo {year} {2024})},\ \Eprint {http://arxiv.org/abs/2407.10913} {arXiv:2407.10913 [hep-lat]} \BibitemShut {NoStop}%
\bibitem [{\citenamefont {Lehner}\ and\ \citenamefont {Meyer}(2020)}]{Lehner:2020crt}%
  \BibitemOpen
  \bibfield  {author} {\bibinfo {author} {\bibfnamefont {C.}~\bibnamefont {Lehner}}\ and\ \bibinfo {author} {\bibfnamefont {A.~S.}\ \bibnamefont {Meyer}},\ }\href {\doibase 10.1103/PhysRevD.101.074515} {\bibfield  {journal} {\bibinfo  {journal} {Phys. Rev. D}\ }\textbf {\bibinfo {volume} {101}},\ \bibinfo {pages} {074515} (\bibinfo {year} {2020})},\ \Eprint {http://arxiv.org/abs/2003.04177} {arXiv:2003.04177 [hep-lat]} \BibitemShut {NoStop}%
\bibitem [{\citenamefont {Crivellin}\ \emph {et~al.}(2020)\citenamefont {Crivellin}, \citenamefont {Hoferichter}, \citenamefont {Manzari},\ and\ \citenamefont {Montull}}]{Crivellin:2020zul}%
  \BibitemOpen
  \bibfield  {author} {\bibinfo {author} {\bibfnamefont {A.}~\bibnamefont {Crivellin}}, \bibinfo {author} {\bibfnamefont {M.}~\bibnamefont {Hoferichter}}, \bibinfo {author} {\bibfnamefont {C.~A.}\ \bibnamefont {Manzari}}, \ and\ \bibinfo {author} {\bibfnamefont {M.}~\bibnamefont {Montull}},\ }\href {\doibase 10.1103/PhysRevLett.125.091801} {\bibfield  {journal} {\bibinfo  {journal} {Phys. Rev. Lett.}\ }\textbf {\bibinfo {volume} {125}},\ \bibinfo {pages} {091801} (\bibinfo {year} {2020})},\ \Eprint {http://arxiv.org/abs/2003.04886} {arXiv:2003.04886 [hep-ph]} \BibitemShut {NoStop}%
\bibitem [{\citenamefont {Keshavarzi}\ \emph {et~al.}(2020{\natexlab{b}})\citenamefont {Keshavarzi}, \citenamefont {Marciano}, \citenamefont {Passera},\ and\ \citenamefont {Sirlin}}]{Keshavarzi:2020bfy}%
  \BibitemOpen
  \bibfield  {author} {\bibinfo {author} {\bibfnamefont {A.}~\bibnamefont {Keshavarzi}}, \bibinfo {author} {\bibfnamefont {W.~J.}\ \bibnamefont {Marciano}}, \bibinfo {author} {\bibfnamefont {M.}~\bibnamefont {Passera}}, \ and\ \bibinfo {author} {\bibfnamefont {A.}~\bibnamefont {Sirlin}},\ }\href {\doibase 10.1103/PhysRevD.102.033002} {\bibfield  {journal} {\bibinfo  {journal} {Phys. Rev. D}\ }\textbf {\bibinfo {volume} {102}},\ \bibinfo {pages} {033002} (\bibinfo {year} {2020}{\natexlab{b}})},\ \Eprint {http://arxiv.org/abs/2006.12666} {arXiv:2006.12666 [hep-ph]} \BibitemShut {NoStop}%
\bibitem [{\citenamefont {de~Rafael}(2020)}]{deRafael:2020uif}%
  \BibitemOpen
  \bibfield  {author} {\bibinfo {author} {\bibfnamefont {E.}~\bibnamefont {de~Rafael}},\ }\href {\doibase 10.1103/PhysRevD.102.056025} {\bibfield  {journal} {\bibinfo  {journal} {Phys. Rev. D}\ }\textbf {\bibinfo {volume} {102}},\ \bibinfo {pages} {056025} (\bibinfo {year} {2020})},\ \Eprint {http://arxiv.org/abs/2006.13880} {arXiv:2006.13880 [hep-ph]} \BibitemShut {NoStop}%
\bibitem [{\citenamefont {Coyle}\ and\ \citenamefont {Wagner}(2023)}]{Coyle:2023nmi}%
  \BibitemOpen
  \bibfield  {author} {\bibinfo {author} {\bibfnamefont {N.~M.}\ \bibnamefont {Coyle}}\ and\ \bibinfo {author} {\bibfnamefont {C.~E.~M.}\ \bibnamefont {Wagner}},\ }\href {\doibase 10.1007/JHEP12(2023)071} {\bibfield  {journal} {\bibinfo  {journal} {JHEP}\ }\textbf {\bibinfo {volume} {12}},\ \bibinfo {pages} {071} (\bibinfo {year} {2023})},\ \Eprint {http://arxiv.org/abs/2305.02354} {arXiv:2305.02354 [hep-ph]} \BibitemShut {NoStop}%
\bibitem [{\citenamefont {Acciarri}\ \emph {et~al.}(2015)\citenamefont {Acciarri} \emph {et~al.}}]{MicroBooNE:2015bmn}%
  \BibitemOpen
  \bibfield  {author} {\bibinfo {author} {\bibfnamefont {R.}~\bibnamefont {Acciarri}} \emph {et~al.} (\bibinfo {collaboration} {MicroBooNE, LAr1-ND, ICARUS-WA104}),\ }\href@noop {} {\  (\bibinfo {year} {2015})},\ \Eprint {http://arxiv.org/abs/1503.01520} {arXiv:1503.01520 [physics.ins-det]} \BibitemShut {NoStop}%
\bibitem [{\citenamefont {Machado}\ \emph {et~al.}(2019)\citenamefont {Machado}, \citenamefont {Palamara},\ and\ \citenamefont {Schmitz}}]{Machado:2019oxb}%
  \BibitemOpen
  \bibfield  {author} {\bibinfo {author} {\bibfnamefont {P.~A.}\ \bibnamefont {Machado}}, \bibinfo {author} {\bibfnamefont {O.}~\bibnamefont {Palamara}}, \ and\ \bibinfo {author} {\bibfnamefont {D.~W.}\ \bibnamefont {Schmitz}},\ }\href {\doibase 10.1146/annurev-nucl-101917-020949} {\bibfield  {journal} {\bibinfo  {journal} {Ann. Rev. Nucl. Part. Sci.}\ }\textbf {\bibinfo {volume} {69}},\ \bibinfo {pages} {363} (\bibinfo {year} {2019})},\ \Eprint {http://arxiv.org/abs/1903.04608} {arXiv:1903.04608 [hep-ex]} \BibitemShut {NoStop}%
\bibitem [{\citenamefont {Abratenko}\ \emph {et~al.}(2022{\natexlab{b}})\citenamefont {Abratenko} \emph {et~al.}}]{MicroBooNE:2021tya}%
  \BibitemOpen
  \bibfield  {author} {\bibinfo {author} {\bibfnamefont {P.}~\bibnamefont {Abratenko}} \emph {et~al.} (\bibinfo {collaboration} {MicroBooNE}),\ }\href {\doibase 10.1103/PhysRevLett.128.241801} {\bibfield  {journal} {\bibinfo  {journal} {Phys. Rev. Lett.}\ }\textbf {\bibinfo {volume} {128}},\ \bibinfo {pages} {241801} (\bibinfo {year} {2022}{\natexlab{b}})},\ \Eprint {http://arxiv.org/abs/2110.14054} {arXiv:2110.14054 [hep-ex]} \BibitemShut {NoStop}%
\bibitem [{\citenamefont {Abratenko}\ \emph {et~al.}(2022{\natexlab{c}})\citenamefont {Abratenko} \emph {et~al.}}]{MicroBooNE:2021pvo}%
  \BibitemOpen
  \bibfield  {author} {\bibinfo {author} {\bibfnamefont {P.}~\bibnamefont {Abratenko}} \emph {et~al.} (\bibinfo {collaboration} {MicroBooNE}),\ }\href {\doibase 10.1103/PhysRevD.105.112003} {\bibfield  {journal} {\bibinfo  {journal} {Phys. Rev. D}\ }\textbf {\bibinfo {volume} {105}},\ \bibinfo {pages} {112003} (\bibinfo {year} {2022}{\natexlab{c}})},\ \Eprint {http://arxiv.org/abs/2110.14080} {arXiv:2110.14080 [hep-ex]} \BibitemShut {NoStop}%
\bibitem [{\citenamefont {Abratenko}\ \emph {et~al.}(2022{\natexlab{d}})\citenamefont {Abratenko} \emph {et~al.}}]{MicroBooNE:2021wad}%
  \BibitemOpen
  \bibfield  {author} {\bibinfo {author} {\bibfnamefont {P.}~\bibnamefont {Abratenko}} \emph {et~al.} (\bibinfo {collaboration} {MicroBooNE}),\ }\href {\doibase 10.1103/PhysRevD.105.112004} {\bibfield  {journal} {\bibinfo  {journal} {Phys. Rev. D}\ }\textbf {\bibinfo {volume} {105}},\ \bibinfo {pages} {112004} (\bibinfo {year} {2022}{\natexlab{d}})},\ \Eprint {http://arxiv.org/abs/2110.14065} {arXiv:2110.14065 [hep-ex]} \BibitemShut {NoStop}%
\bibitem [{\citenamefont {Abratenko}\ \emph {et~al.}(2022{\natexlab{e}})\citenamefont {Abratenko} \emph {et~al.}}]{MicroBooNE:2021nxr}%
  \BibitemOpen
  \bibfield  {author} {\bibinfo {author} {\bibfnamefont {P.}~\bibnamefont {Abratenko}} \emph {et~al.} (\bibinfo {collaboration} {MicroBooNE}),\ }\href {\doibase 10.1103/PhysRevD.105.112005} {\bibfield  {journal} {\bibinfo  {journal} {Phys. Rev. D}\ }\textbf {\bibinfo {volume} {105}},\ \bibinfo {pages} {112005} (\bibinfo {year} {2022}{\natexlab{e}})},\ \Eprint {http://arxiv.org/abs/2110.13978} {arXiv:2110.13978 [hep-ex]} \BibitemShut {NoStop}%
\bibitem [{Mic(2018)}]{MicroBooNE:2018dfn}%
  \BibitemOpen
  \href {\doibase 10.2172/1573219} {\  (\bibinfo {year} {2018}),\ 10.2172/1573219}\BibitemShut {NoStop}%
\bibitem [{\citenamefont {Alves}\ \emph {et~al.}(2023)\citenamefont {Alves} \emph {et~al.}}]{Alves:2023ree}%
  \BibitemOpen
  \bibfield  {author} {\bibinfo {author} {\bibfnamefont {D.~S.~M.}\ \bibnamefont {Alves}} \emph {et~al.},\ }\href {\doibase 10.1140/epjc/s10052-023-11271-x} {\bibfield  {journal} {\bibinfo  {journal} {Eur. Phys. J. C}\ }\textbf {\bibinfo {volume} {83}},\ \bibinfo {pages} {230} (\bibinfo {year} {2023})}\BibitemShut {NoStop}%
\bibitem [{\citenamefont {Baldini}\ \emph {et~al.}(2018)\citenamefont {Baldini} \emph {et~al.}}]{MEGII:2018kmf}%
  \BibitemOpen
  \bibfield  {author} {\bibinfo {author} {\bibfnamefont {A.~M.}\ \bibnamefont {Baldini}} \emph {et~al.} (\bibinfo {collaboration} {MEG II}),\ }\href {\doibase 10.1140/epjc/s10052-018-5845-6} {\bibfield  {journal} {\bibinfo  {journal} {Eur. Phys. J. C}\ }\textbf {\bibinfo {volume} {78}},\ \bibinfo {pages} {380} (\bibinfo {year} {2018})},\ \Eprint {http://arxiv.org/abs/1801.04688} {arXiv:1801.04688 [physics.ins-det]} \BibitemShut {NoStop}%
\bibitem [{\citenamefont {Raggi}\ and\ \citenamefont {Kozhuharov}(2014)}]{Raggi:2014zpa}%
  \BibitemOpen
  \bibfield  {author} {\bibinfo {author} {\bibfnamefont {M.}~\bibnamefont {Raggi}}\ and\ \bibinfo {author} {\bibfnamefont {V.}~\bibnamefont {Kozhuharov}},\ }\href {\doibase 10.1155/2014/959802} {\bibfield  {journal} {\bibinfo  {journal} {Adv. High Energy Phys.}\ }\textbf {\bibinfo {volume} {2014}},\ \bibinfo {pages} {959802} (\bibinfo {year} {2014})},\ \Eprint {http://arxiv.org/abs/1403.3041} {arXiv:1403.3041 [physics.ins-det]} \BibitemShut {NoStop}%
\bibitem [{\citenamefont {Ajimura}\ \emph {et~al.}(2017)\citenamefont {Ajimura} \emph {et~al.}}]{Ajimura:2017fld}%
  \BibitemOpen
  \bibfield  {author} {\bibinfo {author} {\bibfnamefont {S.}~\bibnamefont {Ajimura}} \emph {et~al.},\ }\href@noop {} {\  (\bibinfo {year} {2017})},\ \Eprint {http://arxiv.org/abs/1705.08629} {arXiv:1705.08629 [physics.ins-det]} \BibitemShut {NoStop}%
\bibitem [{\citenamefont {Jordan}(2020)}]{JJordan}%
  \BibitemOpen
  \bibfield  {author} {\bibinfo {author} {\bibfnamefont {J.}~\bibnamefont {Jordan}},\ }\href {\doibase 10.5281/zenodo.4122990} {\bibfield  {journal} {\bibinfo  {journal} {Neutrino 2020 - Virtual Meeting (posters)}\ ,\ \bibinfo {pages} {482}} (\bibinfo {year} {2020})},\ \Eprint {http://arxiv.org/abs/https://zenodo.org/records/4122990} {https://zenodo.org/records/4122990} \BibitemShut {NoStop}%
\bibitem [{\citenamefont {Battaglieri}\ \emph {et~al.}(2015)\citenamefont {Battaglieri} \emph {et~al.}}]{Battaglieri:2014hga}%
  \BibitemOpen
  \bibfield  {author} {\bibinfo {author} {\bibfnamefont {M.}~\bibnamefont {Battaglieri}} \emph {et~al.},\ }\href {\doibase 10.1016/j.nima.2014.12.017} {\bibfield  {journal} {\bibinfo  {journal} {Nucl. Instrum. Meth. A}\ }\textbf {\bibinfo {volume} {777}},\ \bibinfo {pages} {91} (\bibinfo {year} {2015})},\ \Eprint {http://arxiv.org/abs/1406.6115} {arXiv:1406.6115 [physics.ins-det]} \BibitemShut {NoStop}%
\bibitem [{\citenamefont {Abe}\ \emph {et~al.}(2020)\citenamefont {Abe} \emph {et~al.}}]{T2K:2020lrr}%
  \BibitemOpen
  \bibfield  {author} {\bibinfo {author} {\bibfnamefont {K.}~\bibnamefont {Abe}} \emph {et~al.} (\bibinfo {collaboration} {T2K}),\ }\href {\doibase 10.1007/JHEP10(2020)114} {\bibfield  {journal} {\bibinfo  {journal} {JHEP}\ }\textbf {\bibinfo {volume} {10}},\ \bibinfo {pages} {114} (\bibinfo {year} {2020})},\ \Eprint {http://arxiv.org/abs/2002.11986} {arXiv:2002.11986 [hep-ex]} \BibitemShut {NoStop}%
\end{thebibliography}%
\end{document}